\documentclass[aps,pra,reprint,groupedaddress,nofootinbib,longbibliography,showpacs]{revtex4-1}
\usepackage{graphicx}
\usepackage{latexsym,amssymb,amsmath}
\usepackage{bm}
\usepackage[caption=false]{subfig}
\usepackage{xcolor}
\usepackage{mathrsfs}
\usepackage[colorlinks=true,
urlcolor=blue,
linkcolor=blue,
citecolor=blue
]{hyperref}

\begin{document}
\title{Topological phonon polaritons in one-dimensional non-Hermitian silicon carbide nanoparticle chains}

\author{B. X. Wang}
\author{C. Y. Zhao}
\email{Changying.zhao@sjtu.edu.cn}

\affiliation{Institute of Engineering Thermophysics, Shanghai Jiao Tong University,
	Shanghai, 200240, China}
\date{\today}

\begin{abstract}
Topological phonon polaritons (TPhPs) are highly protected and localized edge modes that are capable of achieving a strong confinement of electromagnetic waves and immune to impurities and disorder. Here we realize TPhPs by constructing one-dimensional dimerized silicon carbide nanoparticle chains, which mimic the topological property of the well-known Su-Schrieffer-Heeger (SSH) model. We analytically calculate the complex band structure of such chains by taking all near-field and far-field dipole-dipole interactions into account. For longitudinal modes, we demonstrate that, despite the non-Hermiticity and breaking of the chiral symmetry, the band topology can be still characterized by the complex Zak phase, which is quantized and indicates a topological phase transition when the dimerization parameter $\beta$ changes from less than 0.5 to larger than 0.5, like the conventional Hermitian SSH model. By calculating the eigenmodes of a finite chain, we find such a dimerized chain with $\beta>0.5$ supports nontrivial topological eigenmodes localized over both of its edges, indicating the validity of the bulk-boundary correspondence. On the other hand, for transverse modes, we discover a topological phase transition by increasing the lattice constant, which is due to the presence of strong long-range far-field dipole-dipole interactions decaying with the distance $r$ as $1/r$ for an infinitely long chain. However, we surprisingly find the emergence of non-Hermitian skin effect in a finite chain, which leads to the breakdown of the bulk-boundary correspondence. Furthermore, by incorporating the effect of localized bulk eigenmodes and proposing a modified complex Zak phase for a finite lattice, we still recover the topological behavior of the conventional SSH model. Our comprehensive study provides profound implications to the fields of non-Hermitian topological physics and quantum mechanical models with long-range interactions. In addition to the theoretical analysis, we demonstrate the excitation of the topological phonon polaritons and show their enhancement to the photonic LDOS. These TPhPs offer an efficient tool for enhancing light-matter interaction in the mid-infrared. 
\end{abstract}

\maketitle
\section{Introduction}
Phonon polaritons (PhPs) are bosonic quasi-particles originating from the strong coupling of phonons and photons \cite{mutschkeAandA1999,hillenbrandNature2002}. They are able to confine light into the deep subwavelength scale \cite{caldwellNL2013,shiACSPhoton2015,alfaroNaturecomms2017,gubbinPRB2017,duanAM2017,liScience2018}, and thus show a great potential in enhancing light-matter interactions in the infrared and terahertz (THz) range. As a consequence, phonon polaritons are very promising for nanophotonic applications like nanoscale thermal radiation heat transfer \cite{LeGallPRB1997,francoeurOE2011,tervoPRMater2018}, enhanced infrared molecular nano-spectroscopy \cite{autoreLSA2018}, enhanced infrared absorption \cite{zhaoOE2017} and radiative cooling \cite{zhaiScience2017}, etc. Moreover, compared to plasmon polaritons that suffer a lot from the inherent optical losses of metals, phonon polaritons, which are usually excited in low-loss polar dielectrics, like silicon carbide (SiC) \cite{mutschkeAandA1999,hillenbrandNature2002,caldwellNL2013} as well as hexagonal boron nitride (hBN) \cite{shiACSPhoton2015,alfaroNaturecomms2017,duanAM2017,liScience2018}, can have very high quality factors \cite{caldwellNL2013}, even though their frequencies are generally much lower. On the other hand, the rise of topological photonic systems provides great opportunities to create topological states of light, which are highly localized and can propagate unidirectionally without any backscattering processes, even in the presence of disorder and impurities \cite{luNPhoton2014,khanikaevNPhoton2017,ozawa2018topological}. These topological states can be utilized to realize novel photonic devices, such as unidirectional waveguides \cite{poliNComms2015}, optical isolators \cite{el-GanainyOL2015} as well topological lasers \cite{stjeanNaturephoton2017,partoPRL2018,zhaoNaturecomms2018}. Therefore, on the basis of the inherent advantages of phonon polaritons, it is then of great interest to create topologically protected phonon polaritons (TPhPs) to further enhance the light-matter interaction in the infrared and THz range, achieve more extreme light confinement and reduce the consequences of undesirable intrinsic or external perturbations. Moreover, in this context, TPhPs, due to their bosonic quasiparticle nature, may also potentially act as one of the fundamental building blocks of low-loss phonon-polariton based nanoscale circuits, following from their rapid developing counterpart of quantum plasmonics \cite{tameNaturephys2013,gubbinPRB2016,gubbinPRL2016,passlerNL2018}. 

Our interest for realizing TPhPs is also stimulated by recent works on topological plasmon polaritons \cite{lingOE2015,downingPRB2017,pocockArxiv2017,downing2018topological}. The simplest nanostructure supporting topological plasmon polaritons is based on an optical analogy of the Su-Schrieffer-Heeger (SSH) model, by using one-dimensional (1D) periodic, dimerized plasmonic nanoparticle (NP) chains \cite{lingOE2015,downingPRB2017,pocockArxiv2017,downing2018topological}. Hence, similarly in this paper, we aim to realize topologically protected phonon polaritons in 1D dimerized silicon carbide (SiC) NP chains also by mimicking the SSH model. Since a single SiC NP is capable to sustain localized phonon polaritons, a 1D periodic SiC NP chain can also support collective PhPs bound along the chain due to the collective excitation of these localized PhPs \cite{LeGallPRB1997,gubbinPRL2016}. If an appropriate dimerization parameter can be found to make the system fall into the topologically nontrivial phase, it is then possible to observe collective PhPs that spatially localize at the boundaries between topologically non-trivial and trivial phases, and spectrally reside at the gap of the optical band structure. These collective PhPs thus can be regarded as TPhPs.

However, as the Hamiltonian of conventional SSH model for electron transport in a diatomic chain is Hermitian, and only accounts for the nearest-neighbor electron hoppings \cite{asboth2016short}, we are not able to directly draw conclusions about the topological properties of our system based on it. More specifically, this is because the long-range dipole-dipole interactions and retardation effect are natural ingredients of electromagnetic (EM) interactions between the localized PhPs, and in this circumstance our system becomes non-Hermitian. The non-Hermiticity introduces difficulties in studying the topological properties of this system. Recently, there are extensive discussions on the topological properties of non-Hermitian Hamiltonians \cite{huPRB2011,esakiPRB2011,liangPRA2013,schomerusOL2013,leePRL2016,lingSR2016,leykamPRL2017,jinPRA2017,weimannNaturemat2017,lieuPRB2018,yucePRA2018,xiongJPC2018,shenPRL2018,yao2018edge,yinPRA2018,alvarez2018topologicalreview,dangel2018topological,kunst2018biorthogonal,gong2018topological,kawabata2018nonhermitian,wang2018topological,jin2018bulk}, since traditional topological physics is mainly developed on the basis of Hermitian Hamiltonians.  In fact, there are still several crucial open questions not fully solved about the topological properties of non-Hermitian systems. For example, can appropriate topological invariants be defined in non-Hermitian systems to describe the band topology \cite{esakiPRB2011,liangPRA2013,leykamPRL2017,ozawa2018topological,alvarez2018topologicalreview,shenPRL2018}? And if the answer is positive, then is the bulk-boundary correspondence that allows us to predict the number of topological boundary (edge) states solely based on the (Bloch) bulk band topology still valid \cite{ozawa2018topological,alvarez2018topologicalreview,xiongJPC2018}?  Since only until very recently there was a periodic table of topological insulators for non-Hermitian systems \cite{gong2018topological}, like the Altland-Zirnbauer (AZ) classification for Hermitian systems \cite{ryuNJP2010},  so far most of the analyses on topological properties in this field remains case by case. Moreover, long-range dipole-dipole interactions in the dimerized chain can possibly induce a significant dependence on the boundary conditions, for example, the non-Hermitian skin effect in a finite system \cite{yao2018edge,yao2018nonhermitian}, which can break the Bloch theorem and corresponding bulk-boundary correspondence. In that sense, a systematic investigation on the non-Hermitian topological properties of the present system is still necessary.

In a recent paper \cite{wang2018topological}, we studied the topological optical states in 1D dimerized ultracold atomic chains, which are typically described by an effective non-Hermitian Hamiltonian if the photonic degrees of freedom in the reservoir (namely the quantized EM field) are integrated out. It was shown that the complex Zak phase is still quantized and becomes nontrivial when the dimerization parameter $\beta>0.5$, despite the non-Hermiticity. We also verified the bulk-boundary correspondence for that system by analyzing the eigenstate distributions. 

In this paper, in a similar way, we also aim to demonstrate whether above conclusions are still valid for the present non-Hermitian system involving phonon polaritons supported by SiC NP chains. For small enough SiC NPs, we treat them as electric dipoles. The complex band structure of such chains is analytically calculated by taking all near-field and far-field dipole-dipole interactions into account. For longitudinal modes, we find the band topology of this system can still be characterized by the complex Zak phase, which indicates a topological phase transition when the dimerization parameter changes from less than 0.5 to larger than 0.5. By analyzing the eigenmodes of a finite chain as well as their inverse participation ratios (IPRs), we demonstrate that the chain indeed supports nontrivial topological eigenmodes localized over the edges, as long as the dimerization parameter $\beta>0.5$, implying the validity of the bulk-boundary correspondence. The situation is drastically different for the transverse modes. By analyzing the transverse Bloch band, we show the increase of lattice constant can lead to a topological phase transition in an infinitely long chain, due to the presence of strong long-range far-field dipole-dipole interactions decaying with the distance $r$ as $1/r$. However, for a finite chain, we discover that the non-Hermitian skin effect, which indicates the emergence of localized (non-Bloch) bulk modes, indeed violates the bulk-boundary correspondence. This phenomenon necessitates a careful treatment for the effects brought by the strong non-Hermiticity. Furthermore, by incorporating the effect of localized bulk eigenmodes and proposing a modified complex Zak phase for a finite lattice, we recover the topological behavior of the conventional SSH model. In addition, we exhibit the excitation of TPhPs and show their enhancement to the photonic local density of states (LDOS). These TPhPs can offer an efficient interface for enhancing light-matter interaction in the mid-infrared. We also envision our findings can be beneficial to the study of non-Hermitian topological physics and quantum mechanical models with long-range interactions.

\section{Model}\label{model}
In this section, we briefly describe the analytical and numerical models used in this study. We restrict our model in the classical electrodynamics and do not take any nonlinear and quantum effects into account. We first present the detailed calculation of the Bloch band structure for an infinite chain and then show how to determine the band structure or eigenmode distribution of a finite chain. The former calculation can be dubbed the periodic boundary condition (PBC) and the later the open boundary condition (OBC). These two boundaries conditions, in certain circumstances, may show a qualitative difference in investigating the topological properties of our system, especially when long-range far-field dipole-dipole interactions are significant, as will be discussed in Section \ref{transmode_sec}. 
\subsection{The electric dipole approximation and coupled-dipole model}
\begin{figure}[htbp]
	\flushleft
	\subfloat{
		\includegraphics[width=0.46\linewidth]{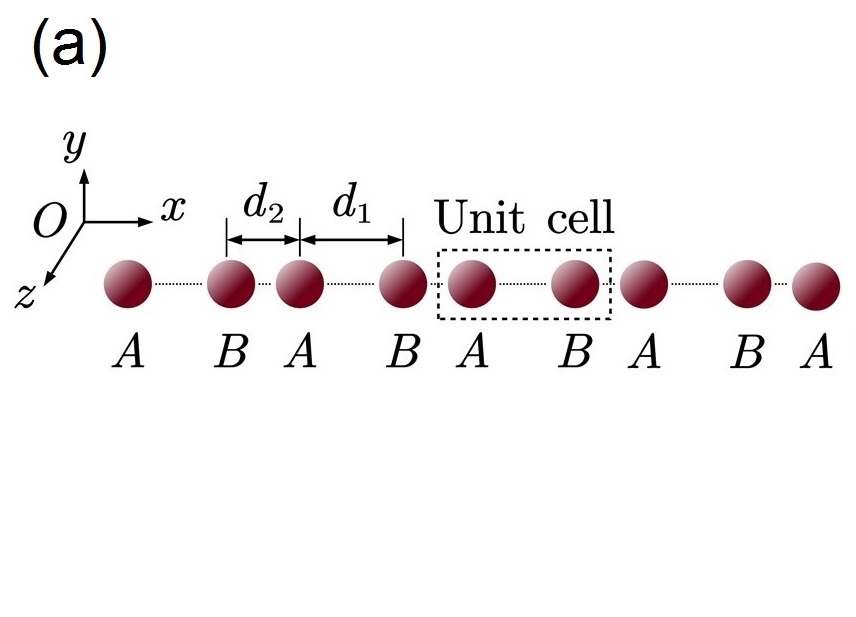}\label{schematic}
	}
	\hspace{0.01in}
	\subfloat{
		\includegraphics[width=0.46\linewidth]{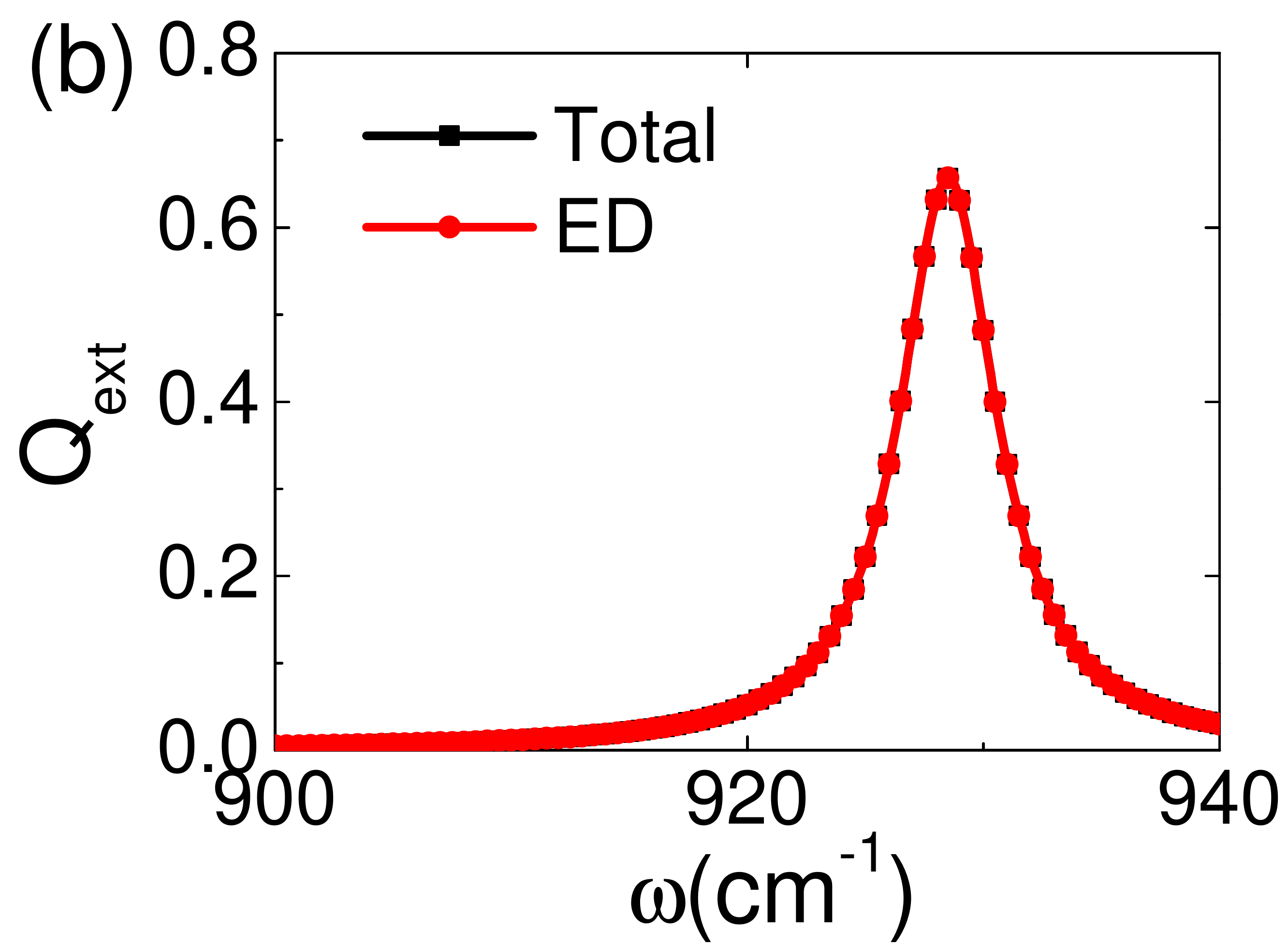}\label{qextsingle}
	}
	\caption{(a) Schematic of the dimerized SiC nanoparticle chain. The nanoparticles are identical and each sublattice with different lattice constants are denoted by $A$ and $B$ respectively. The inequivalent inter-particle spacings are represented by $d_1$ and $d_2$. (b) Extinction efficiency $Q_\mathrm{ext}$ of a single SiC NP with a radius of $a=0.1\mathrm{\mu m}$ calculated from the Mie theory, compared with the result under the electric dipole (ED) approximation.}\label{schematicandmie}
	
\end{figure} 

Consider a 1D dimerized chain composed of spherical $\alpha-$ (hexagonal) SiC NPs schematically shown in Fig.\ref{schematic}. The chain is well aligned along the $x$-axis, where the dimerization is introduced by using inequivalent spacings $d_1$ and $d_2$ for the two sublattices, denoted by $A$ and $B$. Here we define the dimerization parameter as $\beta=d_1/d$ where $d=d_1+d_2$ is the overall lattice constant. SiC NPs support strong localized phonon polaritonic resonances in the infrared region around $11\mathrm{\mu m}$ due to excitation of longitudinal optical phonons, and the permittivity function is modeled by a Lorentz model as \cite{wheelerPRB2009}
\begin{equation}\label{permittivity}  \varepsilon_p(\omega)=\varepsilon_\infty\Big(1+\frac{\omega_L^2-\omega_T^2}{\omega_T^2-\omega^2-i\omega\gamma}\Big),
\end{equation}
where $\omega$ is the angular frequency of the driving field in the unit of $\mathrm{cm}^{-1}$, $\varepsilon_\infty=6.7$ is the high-frequency limit of the permittivity, $\omega_T=790 \mathrm{cm}^{-1}$ is the transverse optical phonon frequency,
$\omega_L=966 \mathrm{cm}^{-1}$ is the longitudinal optical phonon frequency, and $\gamma=5\mathrm{cm}^{-1}$ is the (non-radiative) damping coefficient \cite{wheelerPRB2009}. Herein we set the radius of the spherical SiC NP as $a=0.1\mathrm{\mu m}$. In this situation, the single particle extinction efficiency $Q_\mathrm{ext}=C_\mathrm{ext}/(\pi a^2)$ is calculated using the Mie theory as shown in Fig.\ref{qextsingle}, where $C_\mathrm{ext}$ is the extinction cross section. Actually, such a small NP can be well modeled by the electric dipole (ED) approximation. By considering electric dipole excitations only, the EM response of an individual SiC NP is described by the dipole polarizability with the radiative correction, which is given as \cite{tervoPRMater2018,markelPRB2007,parkPRB2004}
\begin{equation}\label{radiativecorrection}
\alpha(\omega)=\frac{4\pi a^3\alpha_0}{1-2i\alpha_0(ka)^3/3}
\end{equation}
where 
\begin{equation}
\alpha_0(\omega)=\frac{\varepsilon_p(\omega)-1}{\varepsilon_p(\omega)+2}.
\end{equation}

The extinction efficiency calculated under the ED approximation using the polarizability with radiative correction is also shown in Fig.\ref{qextsingle}, and a good agreement with the exact Mie theory is observed. An extra examination of the validity of radiative correction for different particle sizes is given in Appendix \ref{rad_correct_append}.

Furthermore, when the distance between the centers of different spherical NPs is less than $3a$, the EM response of such a chain is described by the well-known set of coupled-dipole equations \cite{parkPRB2004,markelPRB2007,tervoPRMater2018}:
\begin{equation}\label{coupled_dipole_eq}
\mathbf{p}_j(\omega)=\alpha(\omega)\left[\mathbf{E}_\mathrm{inc}(\mathbf{r}_j)+\frac{\omega^2}{c^2}\sum_{i=1,i\neq j}^{\infty}\mathbf{G}_0(\omega,\mathbf{r}_j,\mathbf{r}_i)\mathbf{p}_i(\omega)\right],
\end{equation}
where $c$ is the speed of light in vacuum. $\mathbf{E}_\mathrm{inc}(\mathbf{r})$ is the external incident field and $\mathbf{p}_j(\omega)$ is the excited electric dipole moment of the $j$-th NP. $\mathbf{G}_{0}(\omega,\mathbf{r}_j,\mathbf{r}_i)$ is the free-space dyadic Green's function describing the propagation of field emitting from the $i$-th NP to $j$-th NP \cite{markelPRB2007}. This model takes all types of near-field and far-field dipole-dipole interactions into account and is thus beyond the traditional nearest-neighbor approximation, which is implemented in the SSH model. 

\subsection{Infinite chains}\label{model_infinite}
For 1D chains, there are two types of electromagnetic eigenmodes, including the transverse and longitudinal ones \cite{weberPRB2004}. For the longitudinal eigenmodes, the dipole moments of the NPs are aligned to the $x$-axis. By applying the Bloch theorem in Eq.(\ref{coupled_dipole_eq}) for an infinitely periodic chain with zero incident field, we can analytically solve the longitudinal Bloch eigenmode with a momentum $k_x$ along the $x$-axis. Such an eigenmode, in which the dipole moment of $i$-th NP can be expressed as $p_{m_i,k_x}(\omega)\exp{(ik_xx_i)}$ with $m_i=A, B$ according to the sublattice that the $i$-NP belongs to, should satisfy
\begin{equation}\label{coupled-dipole_bloch}
\begin{split}
&\frac{\omega^2}{c^2}\sum_{i=1,i\neq j}^{N}G_{0,xx}(\omega,\mathbf{r}_j,\mathbf{r}_i)p_{m_i,k_x}(\omega)\exp{(ik_xx_i)}\\&=\alpha^{-1}(\omega)p_{m_j,k_x}(\omega)\exp{(ik_xx_j)}.
\end{split}
\end{equation}
Here the $xx$-component of the Green's function is used:
\begin{equation}\label{Gxx}
G_{0,xx}(x)=-2\Big[\frac{i}{k|x|}-\frac{1}{(k|x|)^2}\Big]\frac{\exp{(ik|x|)}}{4\pi |x|},
\end{equation}
where $k=\omega/c$ is the free-space wavenumber. More specifically, by explicitly carrying out the summations, Eq.(\ref{coupled-dipole_bloch}) is rewritten in the following form
\begin{widetext}
\begin{equation}\label{coupled-emdipole_eigen4}
\frac{\omega^2}{c^2}\left(\begin{matrix}
\sum_{n\neq 0}G_{0,xx}(nd)\exp{(ik_xnd)} & \sum G_{0,xx}(nd+d_1)\exp{(ik_xnd)}\\
\sum G_{0,xx}(nd-d_1)\exp{(ik_xnd)} &
\sum_{n\neq 0}G_{0,xx}(nd)\exp{(ik_xnd)}
\end{matrix}\right)\left(\begin{matrix}p_{A,k_x}\\p_{B,k_x}\end{matrix}\right)=\alpha^{-1}(\omega)\left(\begin{matrix}p_{A,k_x}\\p_{B,k_x}\end{matrix}\right).
\end{equation}
\end{widetext}
For convenience, this equation is equivalently expressed as
\begin{equation}\label{eigenvalue_long}
\frac{\omega^3}{c^3}\left(\begin{matrix}
a_{11}^{L}(k_x) & a_{12}^{L}(k_x)\\
a_{21}^{L}(k_x) & a_{22}^{L}(k_x)
\end{matrix}\right)\left(\begin{matrix}p_{A,k_x}\\p_{B,k_x}\end{matrix}\right)=\alpha^{-1}(\omega)\left(\begin{matrix}p_{A,k_x}\\p_{B,k_x}\end{matrix}\right),
\end{equation}
where the superscript $L$ indicates the longitudinal modes. Hence we arrive at an eigenvalue problem whose solution corresponds to the dispersion relation (or band structure) of the longitudinal eigenmodes, and the matrix in the LHS can be regarded as the effective Hamiltonian $H(k_x)$ in the reciprocal space. For PhPs with ultranarrow linewidths, the lattice sum matrix in the LHS of above equation can be regarded as frequency-independent \cite{pocockArxiv2017}. Therefore, for a fixed $k_x$, $\alpha^{-1}(\omega)$ is the eigenvalue of that matrix. This fact allows us to straightforwardly calculate the eigenfrequency and thus the band structure.  Using the polylogrithm (or Jonqui\'ere's function) $\mathrm{Li}_s(z)=\sum_{n=1}^\infty z^n/n^s$ sum series \cite{NISThandbook}, the diagonal elements in the matrix are given by \cite{wang2018topological}
\begin{equation}
\begin{split}
&a_{11}^{L}(k_x)=a_{22}^{L}(k_x)=-i\frac{\mathrm{Li}_2(z^+)+\mathrm{Li}_2(z^-)}{2\pi k^2d^2}\\&+\frac{\mathrm{Li}_3(z^+)+\mathrm{Li}_3(z^-)}{2\pi k^3d^3},
\end{split}
\end{equation}
where $z^+=\exp{(i(k+k_x)d)}$ and $z^-=\exp{(i(k-k_x)d)}$. Using the Lerch transcendent $\Phi(z,s,a)=\sum_{n=0}^\infty z^n/(n+a)^s$ \cite{NISThandbook}, we can obtain the off-diagonal series sums as \cite{wang2018topological}
\begin{equation}\label{a12Leq}
\begin{split}
&a_{12}^{L}(k_x)=\Big[-i\frac{\Phi(z^+,2,\beta)}{2\pi k^2d^2}+\frac{\Phi(z^+,3,\beta)}{2\pi k^3d^3}\Big]\exp{(ik\beta d)}\\&+\Big[-i\frac{\Phi(z^-,2,1-\beta)}{2\pi k^2d^2}+\frac{\Phi(z^-,3,1-\beta)}{2\pi k^3d^3}\Big]z^-\exp{(-ik\beta d)},
\end{split}
\end{equation}
and
\begin{equation}\label{a21Leq}
\begin{split}
&a_{21}^{L}(k_x)=\Big[-i\frac{\Phi(z^+,2,1-\beta)}{2\pi k^2d^2}+\frac{\Phi(z^+,3,1-\beta)}{2\pi k^3d^3}\Big]\\&\times z^+\exp{(-ik\beta d)}+\Big[-i\frac{\Phi(z^-,2,\beta)}{2\pi k^2d^2}+\frac{\Phi(z^-,3,\beta)}{2\pi k^3d^3}\Big]\\&\times\exp{(ik\beta d)}.
\end{split}
\end{equation}
In the same way, for transverse eigenmodes whose dipole moments are perpendicular to the chain, by using the transverse (namely, $yy$ or $zz$) component of the Green's function as
\begin{equation}\label{Gyy}
G_{0,yy}(x)=[\frac{i}{k|x|}-\frac{1}{(k|x|)^2}+1]\frac{\exp{(ik|x|)}}{4\pi |x|},
\end{equation}
we have
\begin{equation}\label{eigenvalue_trans}
\frac{\omega^3}{c^3}\left(\begin{matrix}
a_{11}^{T}(k_x) & a_{12}^{T}(k_x)\\
a_{21}^{T}(k_x) &
a_{22}^{T}(k_x)
\end{matrix}\right)\left(\begin{matrix}p_{A,k_x}\\p_{B,k_x}\end{matrix}\right)=\alpha^{-1}(\omega)\left(\begin{matrix}p_{A,k_x}\\p_{B,k_x}\end{matrix}\right),
\end{equation}
where the superscript $T$ indicates the transverse modes. The diagonal elements read
\begin{equation}\label{a11T}
\begin{split}
&a_{11}^{T}(k_x)=a_{22}^{T}(k_x)=\frac{\mathrm{Li}_1(z^+)+\mathrm{Li}_1(z^-)}{4\pi kd}\\&+i\frac{\mathrm{Li}_2(z^+)+\mathrm{Li}_2(z^-)}{4\pi k^2d^2}-\frac{\mathrm{Li}_3(z^+)+\mathrm{Li}_3(z^-)}{4\pi k^3d^3}.
\end{split}
\end{equation}
The off-diagonal elements are given by
\begin{equation}\label{a12T}
\begin{split}
&a_{12}^{T}(k_x)=\Big[i\frac{\Phi(z^+,2,\beta)}{4\pi k^2d^2}-\frac{\Phi(z^+,3,\beta)}{4\pi k^3d^3}+\frac{\Phi(z^+,1,\beta)}{4\pi kd}\Big]\\&\times\exp{(ik\beta d)}+\Big[i\frac{\Phi(z^-,2,1-\beta)}{4\pi k^2d^2}-\frac{\Phi(z^-,3,1-\beta)}{4\pi k^3d^3}\\&+\frac{\Phi(z^-,1,1-\beta)}{4\pi kd}\Big]z^-\exp{(-ik\beta d)},
\end{split}
\end{equation}
and
\begin{equation}\label{a21T}
\begin{split}
&a_{21}^{T}(k_x)=\Big[i\frac{\Phi(z^+,2,1-\beta)}{4\pi k^2d^2}-\frac{\Phi(z^+,3,1-\beta)}{4\pi k^3d^3}\\&+\frac{\Phi(z^+,1,1-\beta)}{4\pi kd}\Big]z^+\exp{(-ik\beta d)}+\Big[i\frac{\Phi(z^-,2,\beta)}{4\pi k^2d^2}\\&-\frac{\Phi(z^-,3,\beta)}{4\pi k^3d^3}+\frac{\Phi(z^-,1,\beta)}{4\pi kd}\Big]\exp{(ik\beta d)}.
\end{split}
\end{equation}
Note above expressions are singular (divergent) at $k_x=\pm k$ due to the divergence of the summation involving long-range couplings which decay as $1/r$ (in the thermodynamic limit, i.e., the infinite chain), which can induce substantial non-Hermiticity and a significant influence on the topological invariant as well as bulk-boundary correspondence, as will be discussed later.

After the matrix elements are obtained, the eigenvalue problems of Eqs.(\ref{eigenvalue_long}) and (\ref{eigenvalue_trans}) result in two-band dispersion relations for both longitudinal and transverse eigenmodes as follows
\begin{equation}\label{dispersionrelation}
\begin{split}
&\frac{1}{4\pi (ka)^3}\frac{\varepsilon_p(\omega)+2}{\varepsilon_p(\omega)-1}-\frac{i}{6\pi }=a_{11}^{L(T)}(k_x)\\&\pm\sqrt{a_{12}^{L(T)}(k_x)}\sqrt{a_{21}^{L(T)}(k_x)}
\end{split}
\end{equation} 
from which the eigenfrequencies can be solved in the lower complex plane with respect to a fixed $k_x$. Sweeping $k_x$ across the entire Brillouin zone then gives the whole band structure. This equation shows that the radius $a$ only affects the detailed values of the eigenfrequencies, as long as the electric dipole approximation with the radiative correction and coupled-dipole model are valid, and thus the band structure in a quantitative manner (as shown in Appendix \ref{effect_of_radius}), but does not have any qualitative impact on the band topology. Therefore, without loss of generality, here we adopt a fixed radius of $a=0.1\mathrm{\mu m}$. The solved eigenfrequencies are generally expressed in the complex form of $\tilde{\omega}=\omega-i\Gamma/2$, where the real part $\omega$ amounts to the angular frequency of the eigenmode while the imaginary part $\Gamma$ corresponds to its linewidth (or decay rate of the eigenmode) \cite{caoRMP2015,pocockArxiv2017}. Due to their finite lifetimes, these eigenmodes are usually called quasi-normal modes for open systems \cite{caoRMP2015}. Actually, if we consider a lossless material with a real permittivity in the form of Eq.(\ref{permittivity}) with $\gamma=0$, implement the quasistatic approximation (i.e., no retardation effect) and take only the nearest-neighbor coupling into account, like the case studied by Ling \textit{et al.} \cite{lingOE2015}, we can immediately obtain that $a_{11}(k_x)=0$, and $a_{12}(k_x)$ and $a_{21}(k_x)$ are both real. This treatment gives rise to an Hermitian eigenvalue problem and thus a purely real band structure \cite{lingOE2015}. However, it should be noted that this is only a limit case valid when $kd\ll1$ and material dissipation is neglected, and hence is not a general situation encountered in our system, where $kd\sim1$ and dissipation should be included \cite{zhangPRB2018}. Therefore, we have to treat the present system as non-Hermitian and study the topological properties of the band structures consisting of complex eigenfrequencies. 

\subsection{Finite chains}
In fact, the calculation of the band structures of finite chains is rather straightforward and can be done by using Eq.(\ref{coupled_dipole_eq}) and setting the incident field to be zero \cite{weberPRB2004,pocockArxiv2017}. In this way, we obtain an eigenvalue equation in the form of
\begin{equation} \mathbf{G}|\mathbf{p}\rangle=\alpha^{-1}(\omega)|\mathbf{p}\rangle.
\end{equation} 
Here $\mathbf{G}$ stands for the interaction matrix whose elements are derived from the Green's function (Eqs.(\ref{Gxx}) and (\ref{Gyy}) according to the type of eigenmodes), and $|\mathbf{p}\rangle=[p_1p_2...p_j...p_N]$ is the right eigenvector, which stands for the dipole moment distribution of an eigenmode, where $p_j$ is the dipole moment of the $j$-th NP. Like the case of infinite chains, this equation also specifies a set of complex eigenfrequencies in the lower complex plane in the form of $\tilde{\omega}=\omega-i\Gamma/2$. The physical significance of these complex eigenfrequencies are the same as the ones appearing in Section \ref{model_infinite}. The corresponding wavenumber for an eigenmode is determined by \cite{weberPRB2004,pocockArxiv2017}
\begin{equation}\label{momentum}
k_x\Big(\frac{2\pi}{d}\Big)^{-1}=\frac{(N-2)n+1}{N(N-1)},
\end{equation}
where $n$ is the mode number of an eigenmode, which is 1 plus the number of times of sign changes of $\mathrm{Re}(p_j)$ along the chain for that eigenmode \cite{weberPRB2004,pocockArxiv2017}.

The topologically protected eigenmodes are highly localized over the boundary of the finite chain \cite{atalaNaturephys2013}. Therefore, to quantify the degree of eigenmode localization, we further calculate the inverse participation ratio (IPR) of an eigenmode from its eigenvector as \cite{wangOL2018,wang2018topological}. 
\begin{equation}
\mathrm{IPR}=\frac{\sum_{n=1}^{N}|p_j|^4}{(\sum_{n=1}^{N}|p_j|^2)^2}.
\end{equation}
For an IPR approaches $1/M$, where $M$ is an integer, the corresponding eigenmode involves the excitation of $M$ NPs \cite{wangOL2018}. Therefore, for a highly localized topological eigenmode, its IPR should be much larger compared to those of the bulk eigenmodes \cite{wangOL2018}. 

\section{Longitudinal modes}
In this section, we investigate the topological properties of longitudinal modes. For this non-Hermitian system, we first calculate the Bloch band structures, from which we derive the complex Zak phase by using the biorthogonality of the Bloch eigenvectors. Moreover, by computing the complex band structures of finite chains, we find highly localized edge modes which are topologically protected by the complex Zak phase if the dimerization parameter $\beta>0.5$, illustrating the validity of the bulk-boundary correspondence. The behavior of longitudinal modes is the same as that of the conventional SSH model, since the short-range near-field dipole-dipole interactions dominate in this situation.
\subsection{Bloch band structures}
\begin{figure}[htbp]
	\centering
	\subfloat{
		\includegraphics[width=0.46\linewidth]{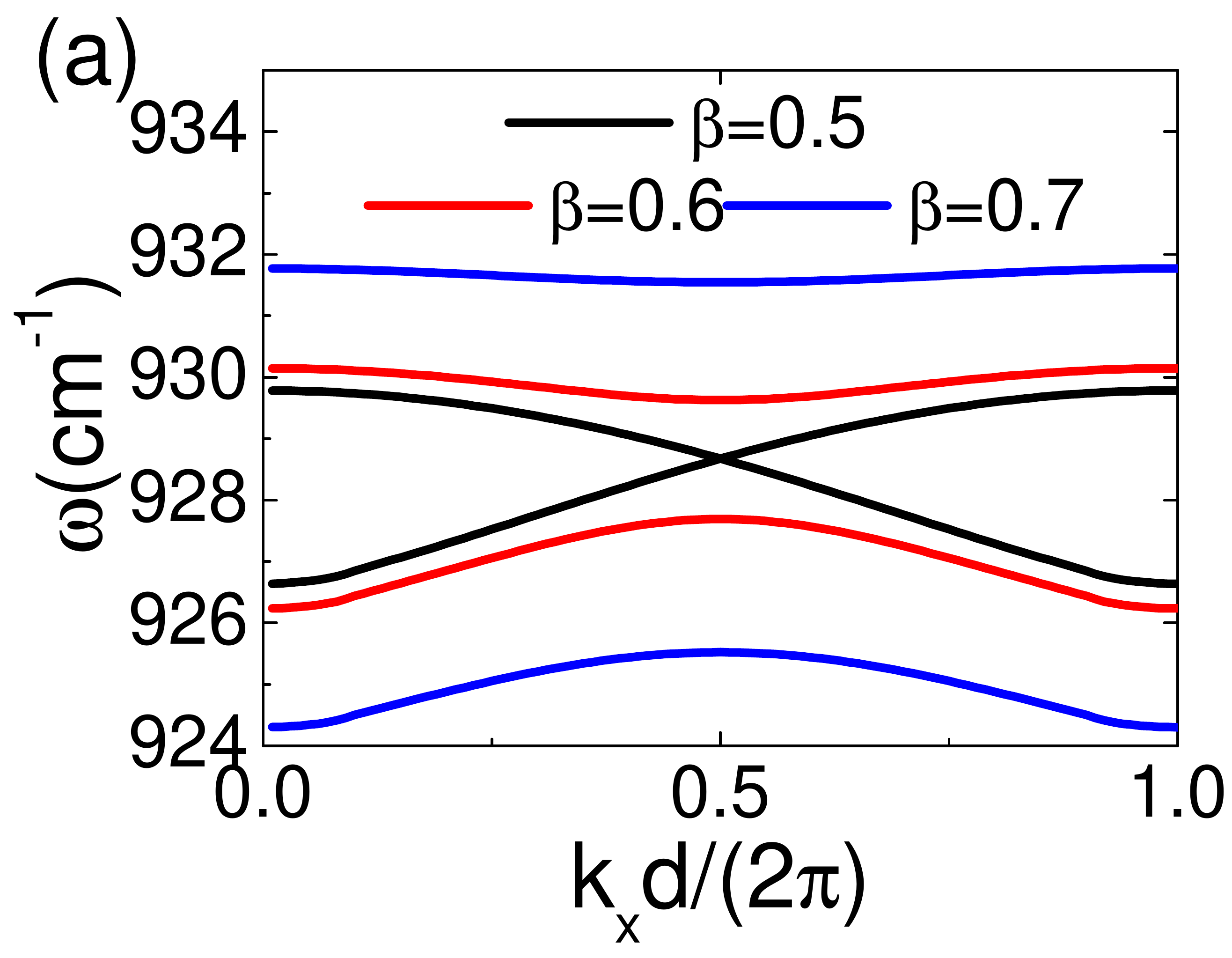}\label{bandstructurelongd1}
	}
	\hspace{0.01in}
	\subfloat{
		\includegraphics[width=0.46\linewidth]{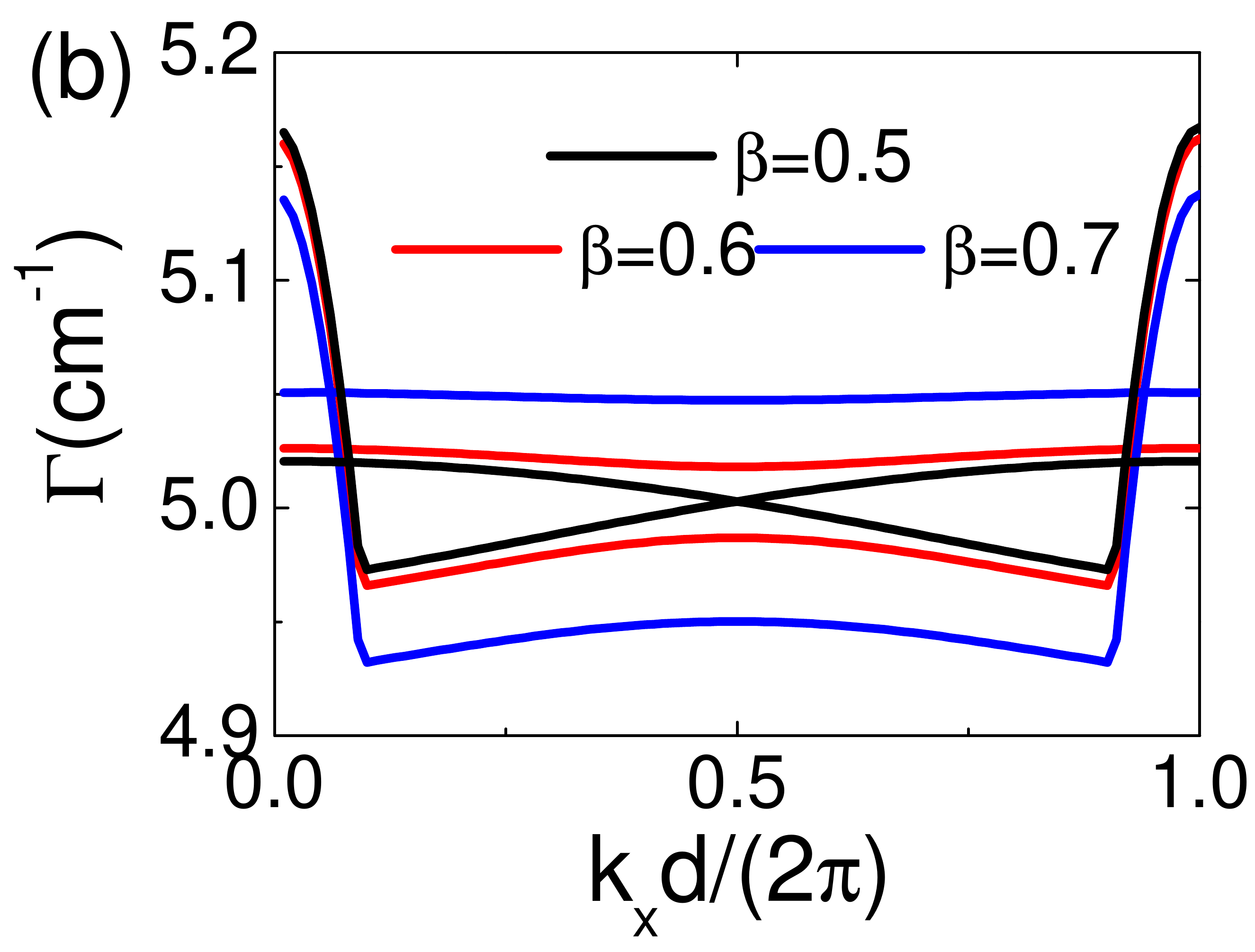}\label{bandstructurelongd1imag}
	}
	\caption{(a) Real and (b) imaginary parts of the longitudinal band structures of a dimerized chain with $d=d_1+d_2=1\mathrm{\mu m}$ for different dimerization parameters $\beta$. }\label{figcda}
	
\end{figure} 

We first investigate the band structure of a dimerized chain with an overall lattice constant of $d=1\mathrm{\mu m}$. In Figs.\ref{bandstructurelongd1} and \ref{bandstructurelongd1imag} we show the real and imaginary parts of the bulk band structures for different dimerization parameters $\beta=0.5, 0.6, 0.7$ for both longitudinal eigenmodes respectively. Note the bulk band structures for $\beta=0.3$ and $\beta=0.4$ are the same as those for $\beta=0.7$ and $\beta=0.6$ correspondingly. It is found that for $\beta\neq0.5$, bandgaps in the real frequency space are opened in both cases and a larger $|\beta-0.5|$ (i.e., the deviation from a non-bipartite chain) gives rise to a wider bandgap. This behavior is consistent with the conventional SSH model \cite{atalaNaturephys2013}. On the other hand, the imaginary spectrum is ungapped for all dimerization parameters.

\subsection{The complex Zak phase}\label{zakphasesec}
In the last several years, especially recent two years, there were fierce discussions on the topological properties of non-Hermitian systems \cite{huPRB2011,esakiPRB2011,liangPRA2013,schomerusOL2013,leePRL2016,lingSR2016,leykamPRL2017,jinPRA2017,weimannNaturemat2017,lieuPRB2018,yucePRA2018,xiongJPC2018,shenPRL2018,yao2018edge,yinPRA2018,alvarez2018topologicalreview,dangel2018topological,kunst2018biorthogonal,gong2018topological,kawabata2018nonhermitian,wang2018topological}. Most of these discussions were devoted to the topologically nontrivial properties related to exceptional points (EPs) \cite{fengNaturephton2017,elganainyNaturephys2018}, which are singularities in the energy spectra of non-Hermitian Hamiltonians where the eigenvalues and eigenwavefunctions coalesce \cite{heissJPA2012}. On the other hand, for non-Hermitian systems without EPs, it was recently shown that the complex Zak phase, which is defined by simultaneously using the left-eigenvectors and right-eigenvectors of the non-Hermitian Hamiltonian \cite{garrisonPLA1988,liangPRA2013,wagnerAP2017,lieuPRB2018,partoPRL2018,dangel2018topological}, can be exploited to characterize the band topology of 1D systems. In our recent paper \cite{wang2018topological}, we studied the non-Hermitian topological optical states in 1D dimerized ultracold atomic chains, and showed that as long as the band structure is separable in the complex plane, the complex Zak phase is always quantized and the bulk-boundary correspondence applies. In fact, Shen \textit{et al.} \cite{shenPRL2018} presented a rigorous definition for the non-Hermitian band structures. In a non-Hermitian system, that a band with a band number of $n$ is separable means that for any $m\neq n$ in the entire band structure, the energies (complex frequencies) also satisfy $\tilde{\omega}_{m,\mathbf{k}}\neq \tilde{\omega}_{n,\mathbf{k}}$ in the complex plane for all possible $\mathbf{k}$ \cite{shenPRL2018} . Moreover, if their energies (complex frequencies) further fulfill $\tilde{\omega}_{m,\mathbf{k}'}\neq \tilde{\omega}_{n,\mathbf{k}}$ in the complex plane for all $\mathbf{k}$ and $\mathbf{k}'$, the band $n$ is then regarded as isolated or gapped \cite{shenPRL2018}. According to Eq.(\ref{dispersionrelation}), since $a_{12}(k_x)a_{21}(k_x)$ is always not exactly zero when $\beta\neq0.5$, the bulk band structures are always separable in the complex frequency plane (which can be seen in the examples exhibited below). Hence we anticipate that for the present non-Hermitian system the complex Zak phase is quantized and can describe the topological phase transition, where the transition point is the gap closing point, i.e., $\beta=0.5$ \cite{lieuPRB2018,wang2018topological}.

Another issue that is necessary to be addressed before calculating the complex Zak phase is that the present non-Hermitian Hamiltonian exhibits a breaking of chiral symmetry (or sublattice symmetry) due to the existence of diagonal elements $a_{11}(k_x)$ and $a_{22}(k_x)$. According to the conventional AZ classification, this system belongs to the AI class and is topologically trivial \cite{schnyderPRB2008,asboth2016short}. Nevertheless, since $a_{11}(k_x)$ is always equal to $a_{22}(k_x)$, the eigenvectors of the matrix are not affected by these diagonal terms at all \cite{wang2018topological,pocockArxiv2017}. Actually, the normalized (i.e., $\langle p_{k_x}^L|p_{k_x}^R\rangle=1$) left and right eigenvectors for longitudinal eigenmodes are solved as follows:
\begin{equation}\label{lefteigenvector}
|p_{k_x}^{L}\rangle=\left(\begin{matrix}p_{A,k_x}^{L}\\p_{B,k_x}^{L}\end{matrix}\right)=\frac{1}{\sqrt{2}}\left(\begin{matrix}\mp\frac{\sqrt{a_{21}^{L,*}(k_x)}}{\sqrt{a_{12}^{L,*}(k_x)}}\\1\end{matrix}\right),
\end{equation}
\begin{equation}\label{righteigenvector}
|p_{k_x}^{R}\rangle=\left(\begin{matrix}p_{A,k_x}^{R}\\p_{B,k_x}^{R}\end{matrix}\right)=\frac{1}{\sqrt{2}}\left(\begin{matrix}\mp\frac{\sqrt{a_{12}^L(k_x)}}{\sqrt{a_{21}^L(k_x)}}\\1\end{matrix}\right).
\end{equation}
The left eigenvector is solved through the relation of  $H^\dag(k_x)|p_{k_x}^{L}\rangle=E_{k_x}^*|p_{k_x}^{L}\rangle$. Therefore, the eigenvectors of the present Hamiltonian is the same as those of its chirally-symmetric counterpart with zero diagonal elements. This chiral symmetry breaking can be viewed as trivial, as pointed out by Pocock \textit{et al.} \cite{pocockArxiv2017}. Hence the complex Zak phase is still quantized in a similar way as in a chirally symmetric system. This complex Zak phase can be conveniently used to determine the topology of bulk band structure, as indicated by Lieu \cite{lieuPRB2018}. Moreover, this quantization does not require the inversion symmetry unlike the real Zak phase. The real Zak phase is defined solely based on right eigenvectors, and in that circumstance, the inversion symmetry is necessary to make it quantized \cite{downing2018topological,lieuPRB2018}. However, it is not appropriate to use the real Zak phase in a non-Hermitian system \cite{lieuPRB2018}.

\begin{figure}[htbp]
	\centering
	\flushleft
	\subfloat{
		\includegraphics[width=0.46\linewidth]{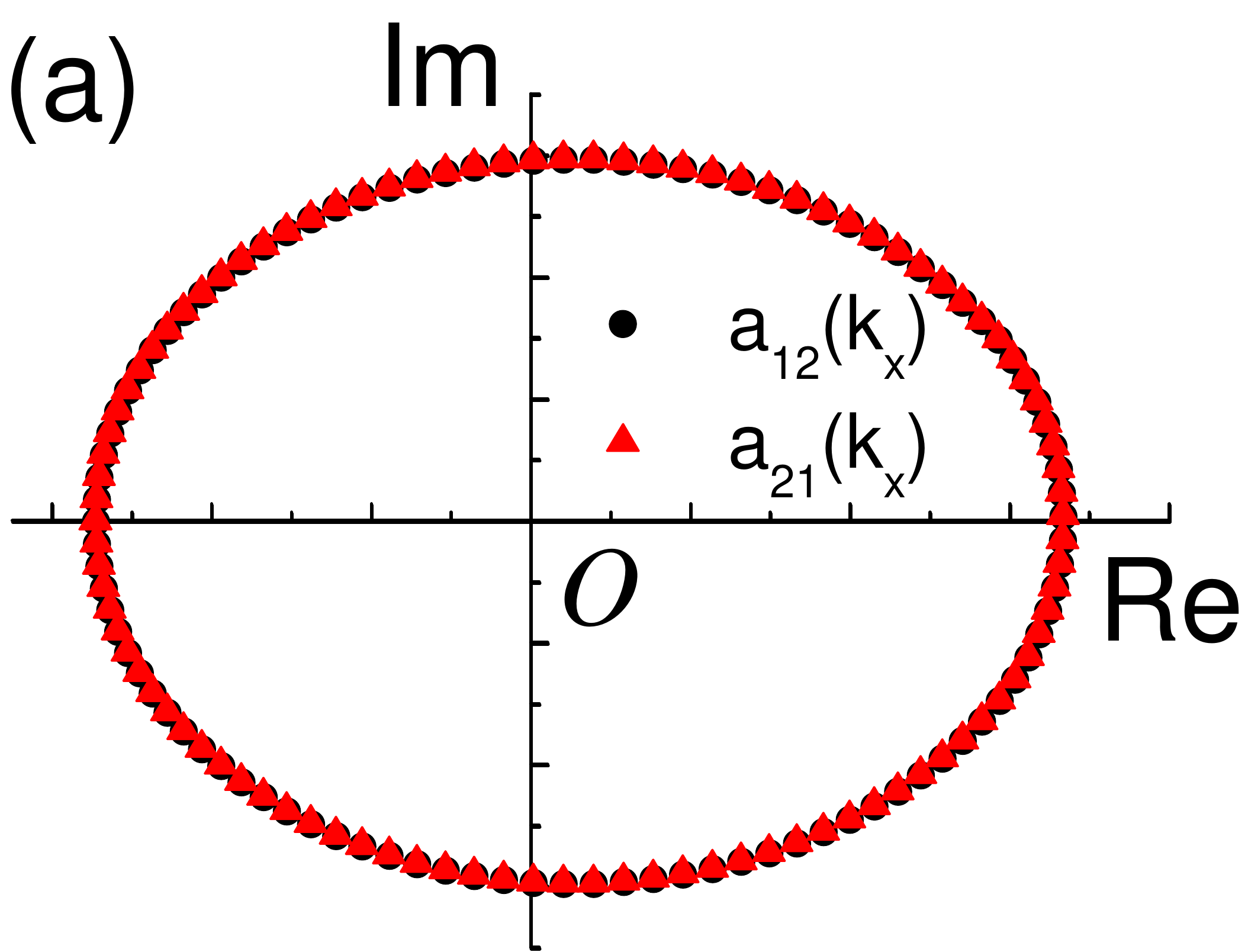}\label{vorticitybeta07d1}
	}
	\hspace{0.01in}
	\subfloat{
		\includegraphics[width=0.46\linewidth]{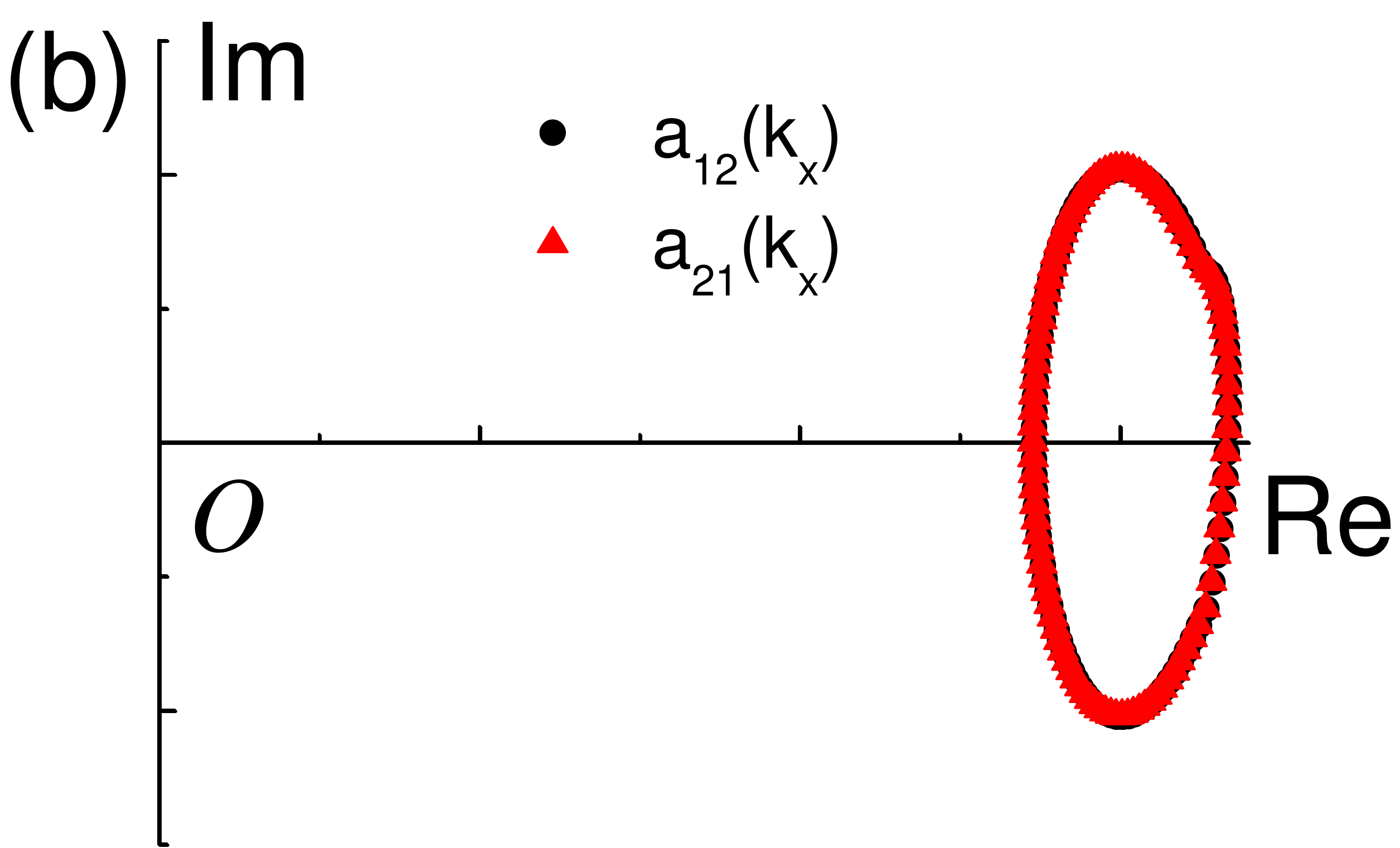}\label{vorticitybeta03d1}
	}
	\hspace{0.01in}
	\subfloat{
		\includegraphics[width=0.46\linewidth]{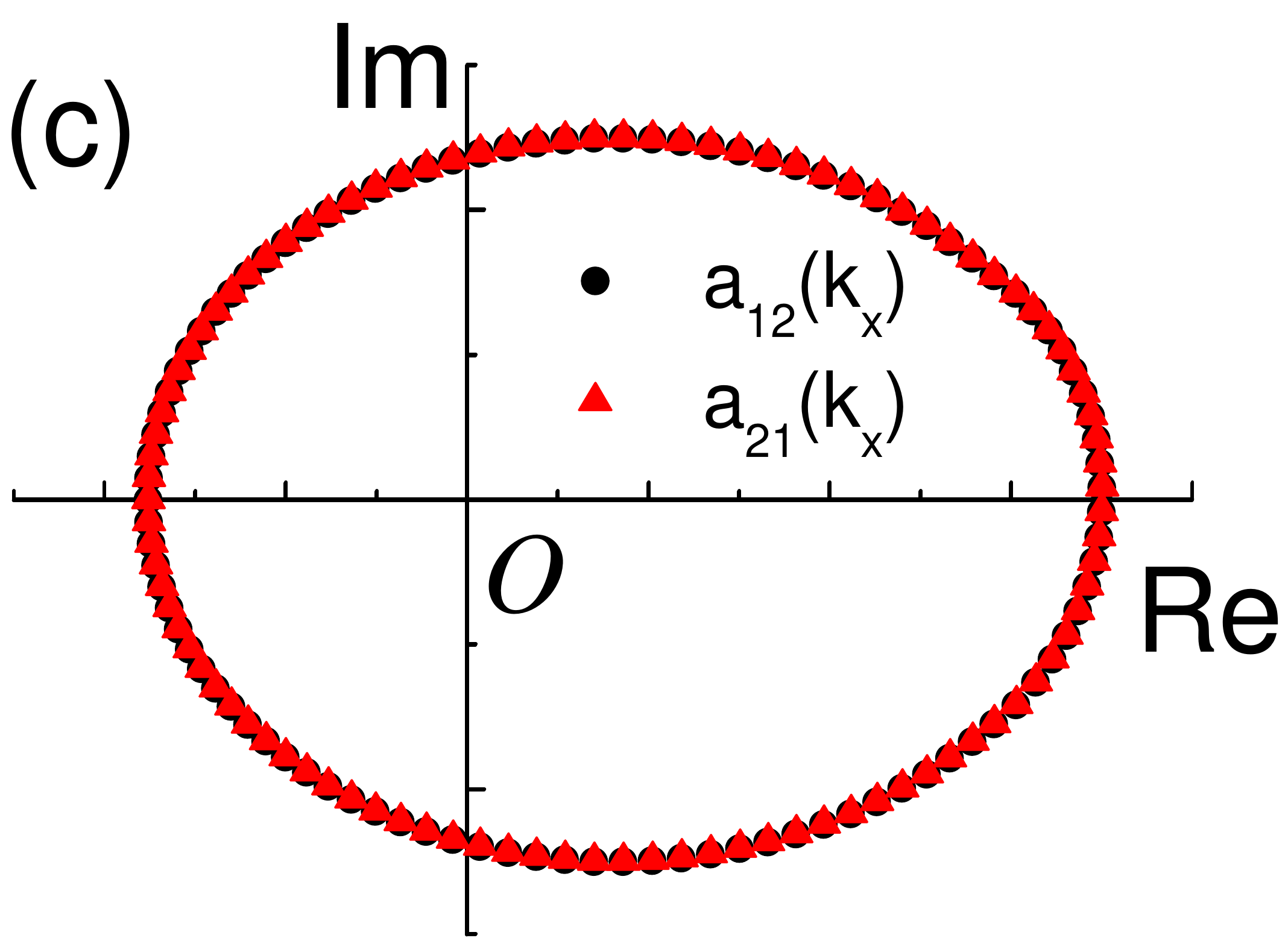}\label{vorticitybeta06d1}
	}
	\hspace{0.01in}
	\subfloat{
		\includegraphics[width=0.46\linewidth]{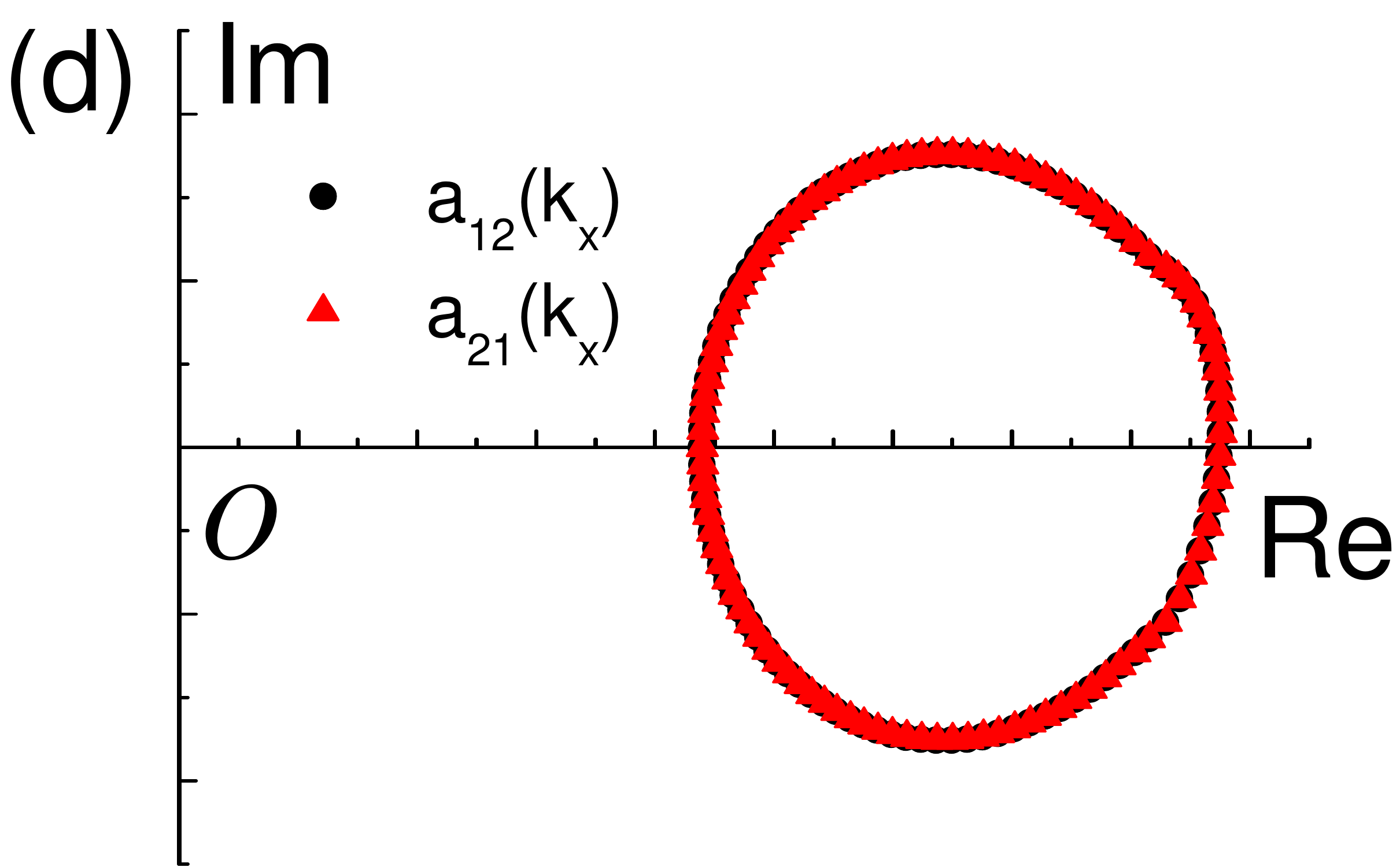}\label{vorticitybeta04d1}
	}
	\caption{The winding of $a_{12}(k_x)$ and $a_{21}(k_x)$ over the origin in the complex plane. The lattice constant is $d=1\mathrm{\mu m}$ with different dimerization parameters (a) $\beta=0.7$. (b) $\beta=0.3$. (c) $\beta=0.6$. (d) $\beta=0.4$.}\label{windingnumber}
	
\end{figure}

Based on the orthogonality of left and right eigenmodes (namely, biorthogonality) \cite{esakiPRB2011,schomerusOL2013,lingSR2016,weimannNaturemat2017,dingPRB2015,jinPRA2017,lieuPRB2018,shenPRL2018,partoPRL2018,yucePLA2015,wagnerAP2017,yucePRA2018,lieuPRB2018,alvarezPRB2018} in 1D non-Hermitian systems, the complex Zak phase, as the geometric phase picked up by an eigenmode when it adiabatically evolves across the first Brillouin zone, is expressed as
\begin{equation}\label{cberryphase}
\begin{split}
\theta_\mathrm{Z}&=\int_\mathrm{BZ}dk_x\mathcal{A}(k_x)\\&=i\int_{-\pi/d}^{\pi/d}\Big[p_{A,k_x}^{L,*}\frac{\partial p_{A,k_x}^R}{\partial k_x}+p_{B,k_x}^{L,*}\frac{\partial p_{B,k_x}^R}{\partial k_x}\Big]dk_x\\&=\frac{\arg[a_{21}(k_x)]-\arg[a_{12}(k_x)]}{4}\Big|_{-\pi/d}^{\pi/d},
\end{split}
\end{equation}
where $\mathcal{A}(k_x)$ is the Berry connection. According to Eq.(\ref{cberryphase}), the complex Zak phase is actually a real quantity \cite{wang2018topological}, and it is simply half the difference of the winding numbers of $a_{21}(k_x)$ and $a_{12}(k_x)$ encircling the origin multiplied by $\pi$.  To calculate $\theta_\mathrm{Z}$, the winding of $a_{12}(k_x)$ and $a_{21}(k_x)$ around the origin of the complex plane by sweeping $k_x$ for $0$ to $2\pi/d$ is illustrated in Fig.\ref{windingnumber}, for the cases of $\beta=0.7$, $\beta=0.3$, $\beta=0.6$ and $\beta=0.4$ with the lattice constant $d=1\mathrm{\mu m}$. Note the directions of the encircling of $a_{12}(k_x)$ and $a_{21}(k_x)$ are always opposite because $a_{12}(k_x)=a_{21}(-k_x)$. Therefore the winding numbers of $a_{12}(k_x)$ and $a_{21}(k_x)$ are +1 and -1 respectively when the dimerization parameter is $\beta=0.7$ and $\beta=0.6$, and are both zero when $\beta=0.3$ and $\beta=0.4$. As a consequence, the complex Zak phase for $\beta=0.7$ and $\beta=0.6$ is $\pi$ and is $0$ for $\beta=0.3$ and $\beta=0.4$. 

\subsection{Bulk-boundary correspondence and midgap modes}
In Hermitian systems, the principle of bulk-boundary correspondence indicates that the topological invariant (for 1D, i.e., the Zak phase) determines the existence of edge modes, and the total winding number $\mathcal{W}=\theta_\mathrm{Z}/\pi$ is equivalent to the number of edge modes localized over the boundary of the systems \cite{luNPhoton2014,khanikaevNPhoton2017,rhimPRB2017,ozawa2018topological}. However, it was surprisingly found that the conventional bulk-boundary correspondence becomes invalid for some specific 1D non-Hermitian Hamiltonians \cite{yao2018edge,kunst2018biorthogonal,xiongJPC2018}. This is because in those systems, the Bloch band structures under the PBC are much different from the band structures calculated from the OBC \cite{yao2018edge,kunst2018biorthogonal}. Therefore, to appropriately investigate the topological properties of the present non-Hermitian system, we must first directly compute the band structures of finite chains with open boundaries.

\begin{figure}[htbp]
	\flushleft
	\subfloat{
	\includegraphics[width=0.46\linewidth]{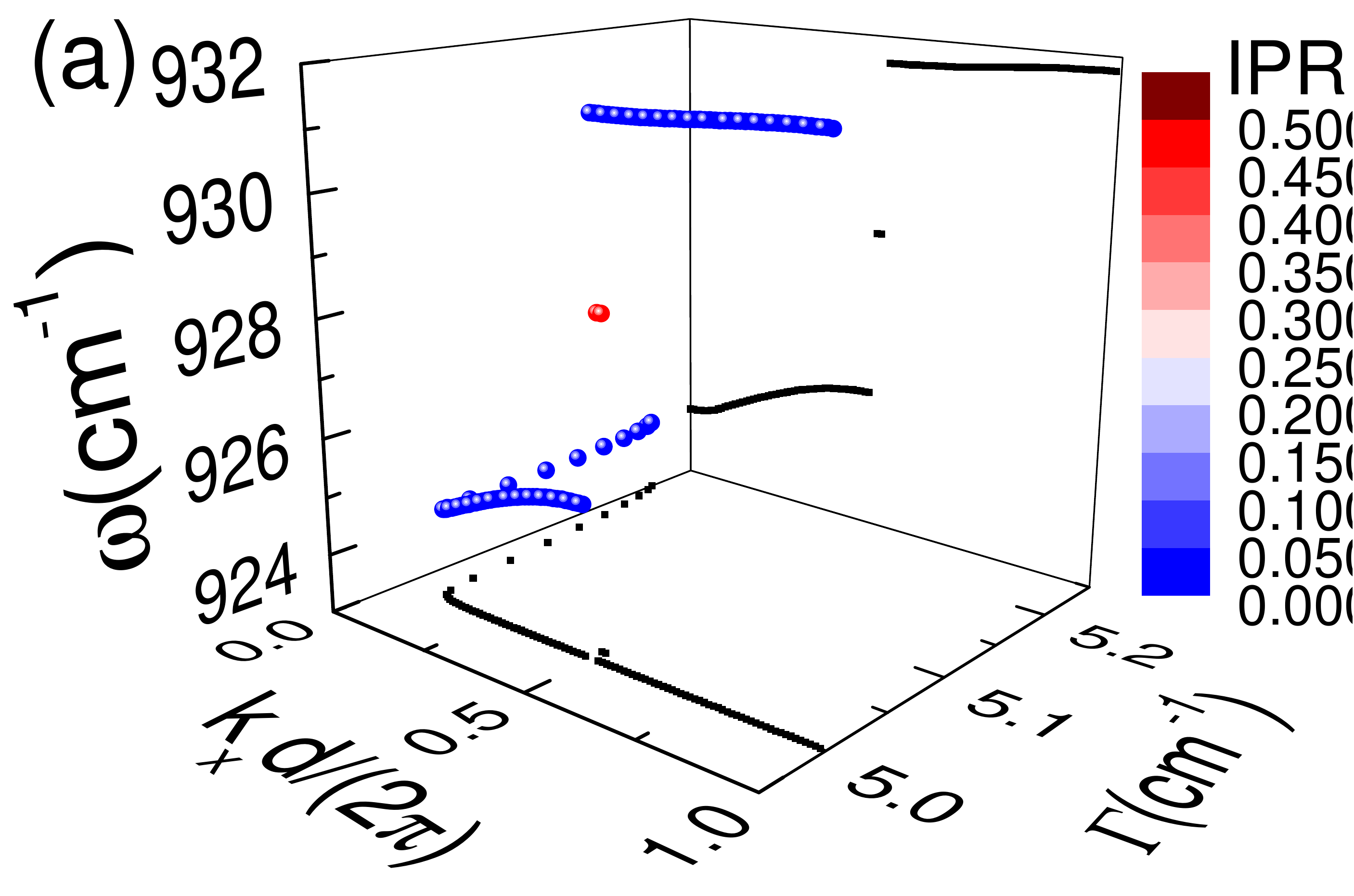}\label{beta07longband}
}
	\subfloat{
	\includegraphics[width=0.46\linewidth]{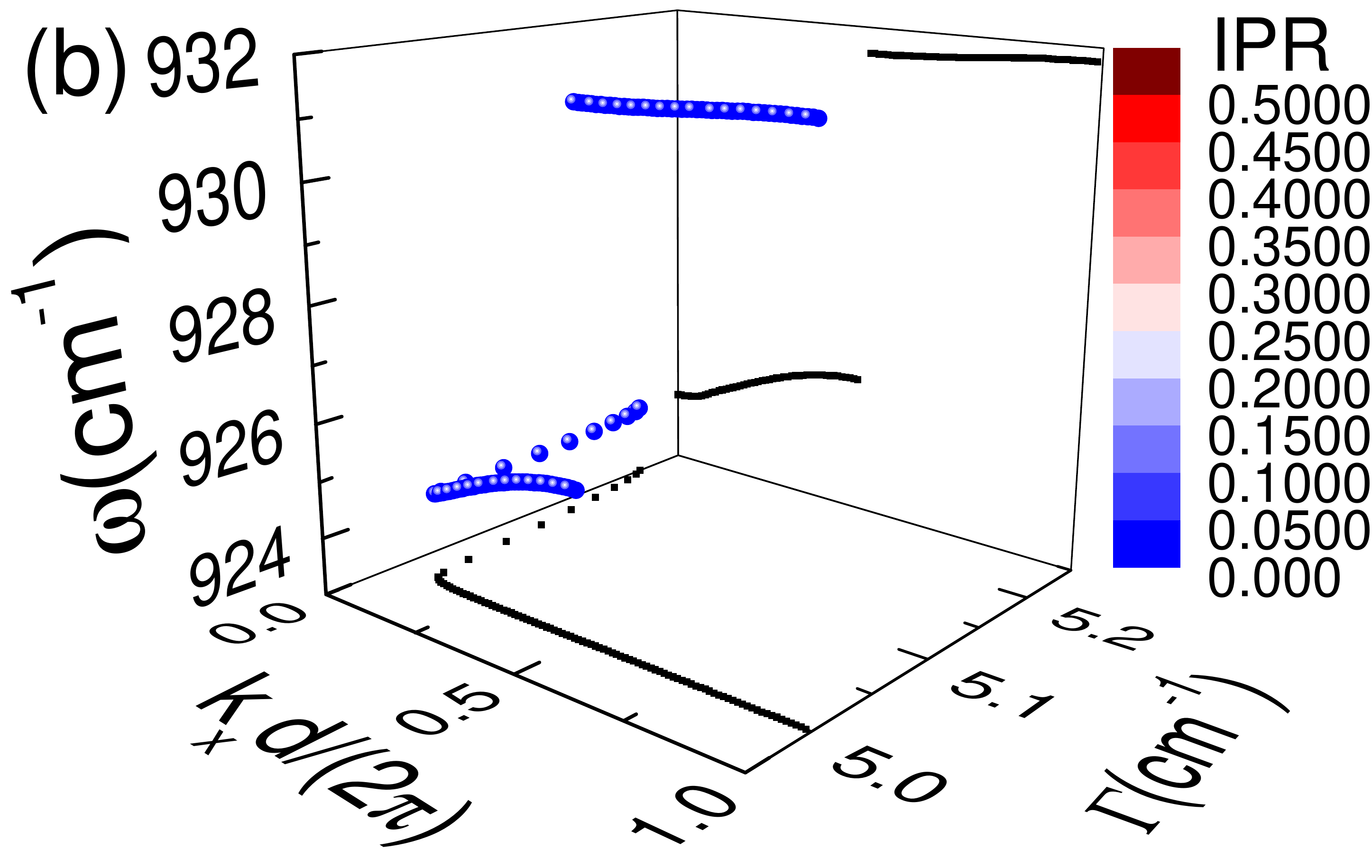}\label{beta03longband}
}
	\hspace{0.01in}
	\subfloat{
		\includegraphics[width=0.46\linewidth]{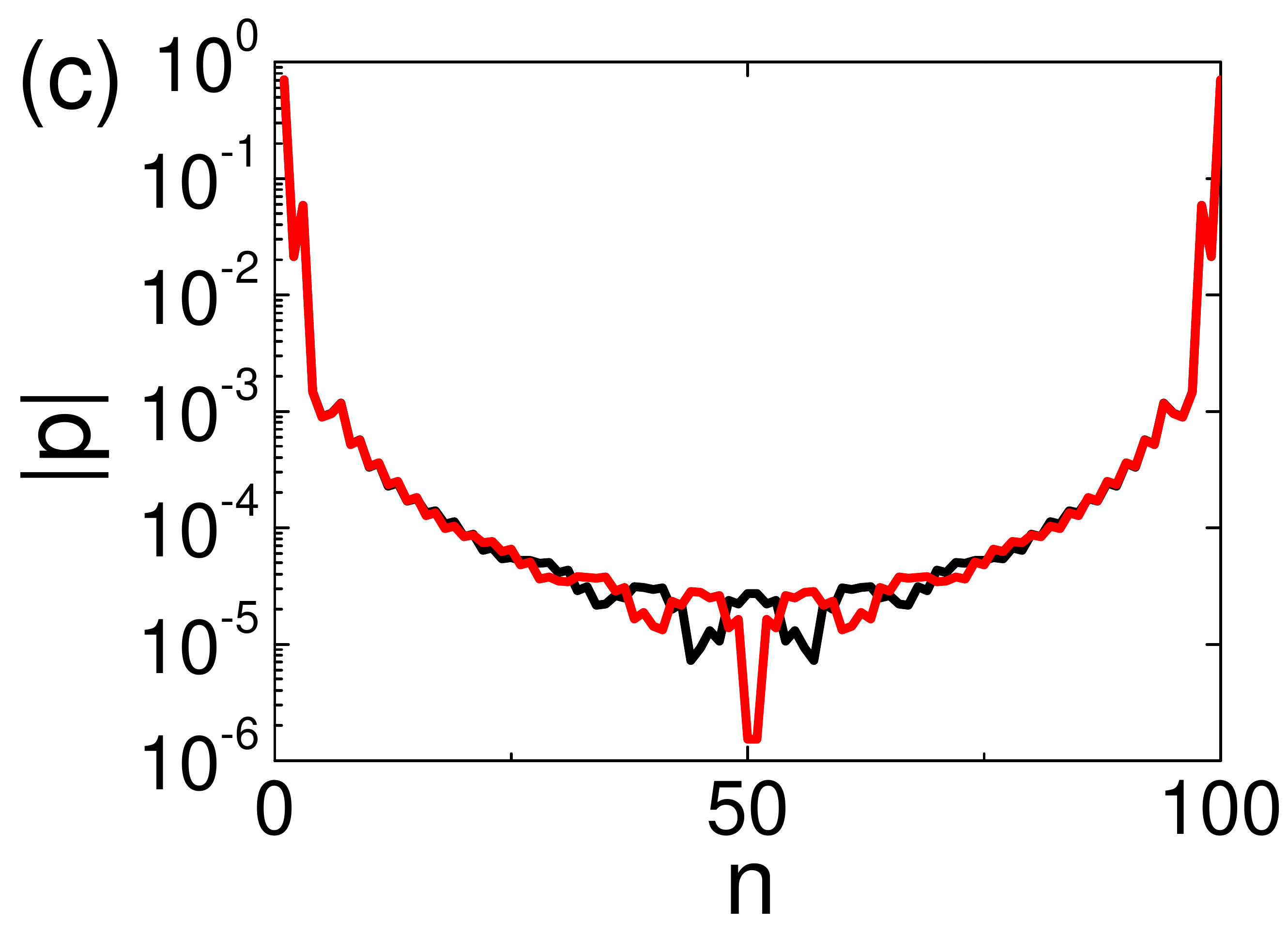}\label{midgapmode}
	}
	\subfloat{
	\includegraphics[width=0.49\linewidth]{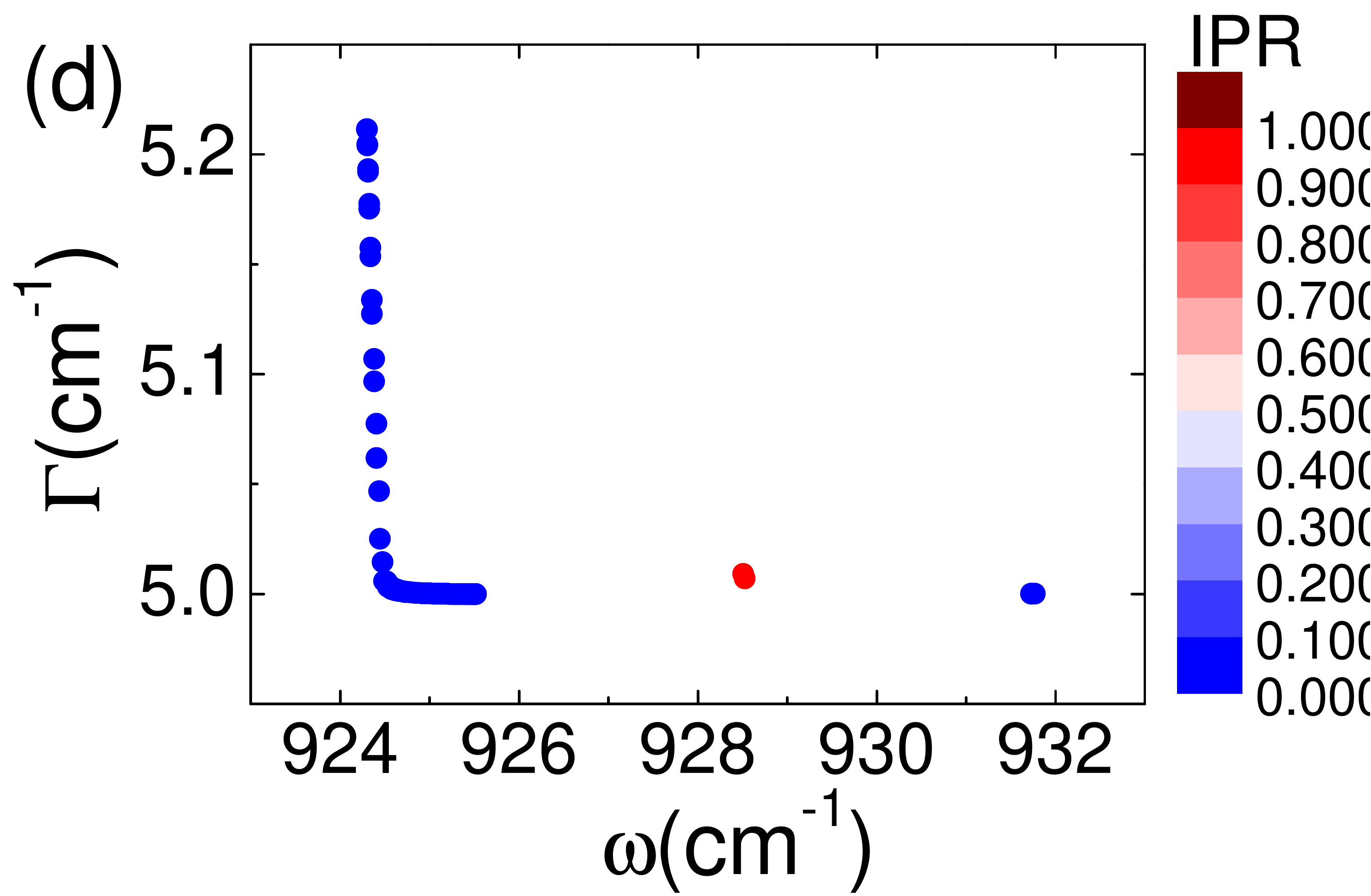}\label{interfacemodebeta07}
	}
	\hspace{0.01in}
	\subfloat{
	\includegraphics[width=0.46\linewidth]{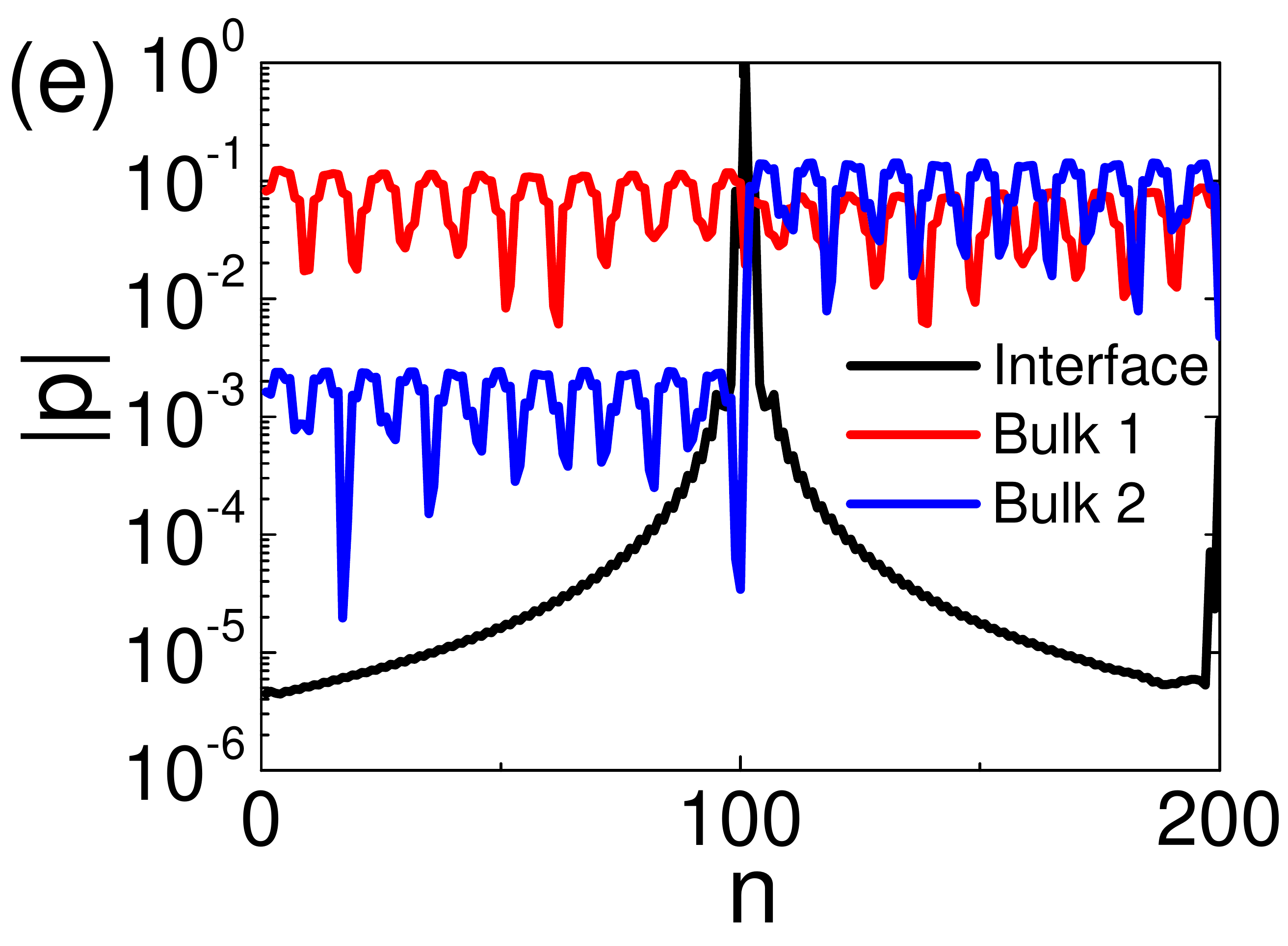}\label{interfacemodedipole}
	}
	\subfloat{
	\includegraphics[width=0.46\linewidth]{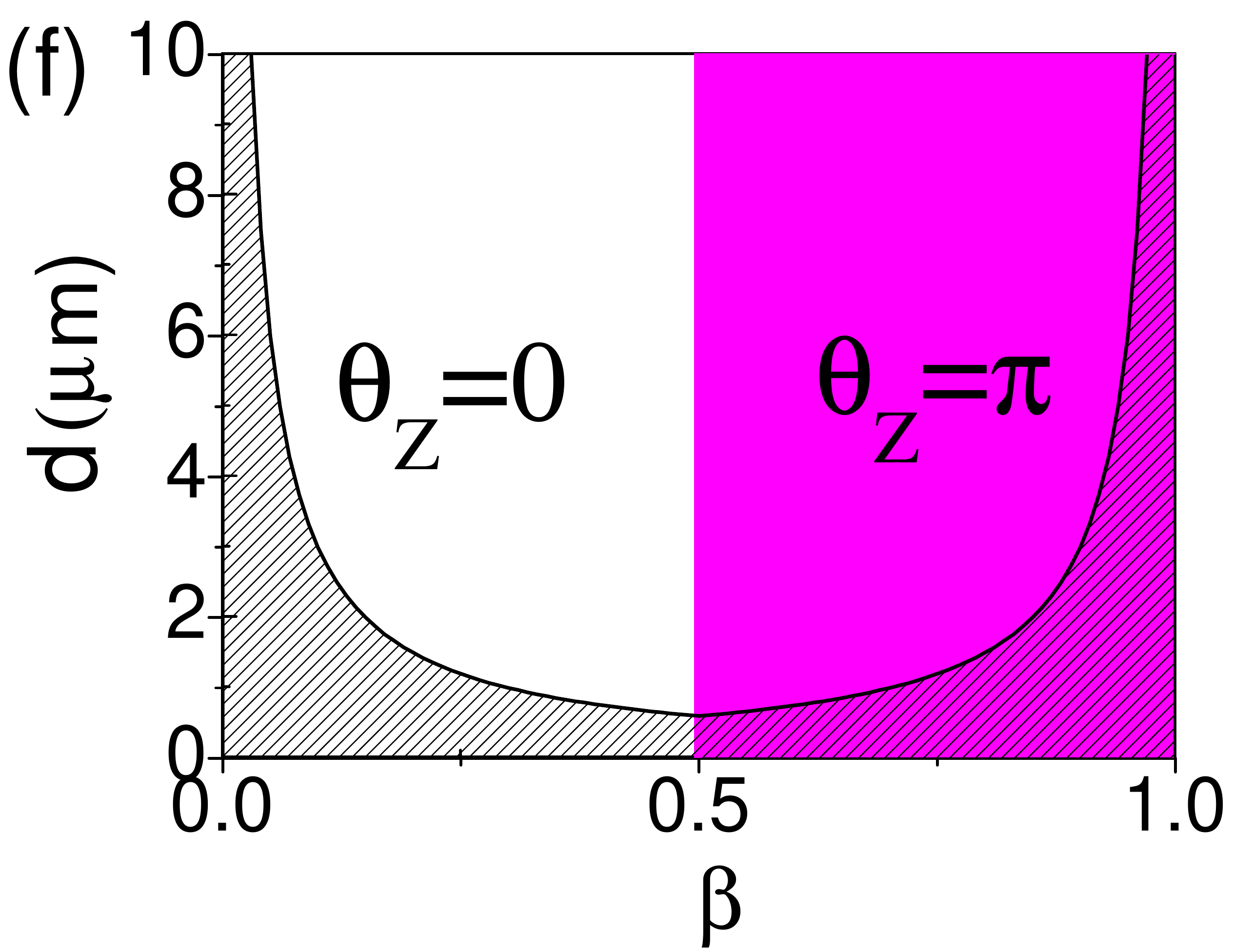}\label{phasediagram}
	}

	\caption{(a) Complex band structure of longitudinal eigenmodes of a dimerized chain with $N=100$ NPs under $\beta=0.7$ and $d=1\mathrm{\mu m}$. Note there are two midgap modes. The black dots in $dO\omega$ and $dO\Gamma$ planes are projections of the spectrum. (b) The same as (a) but here $\beta=0.3$. (c) Dipole moment distribution of the midgap edge modes. (d) Longitudinal eigenmode distribution for a connected chain. (e) Dipole moment distribution of the interface  mode at ($\omega/\gamma=928.5277\mathrm{cm}^{-1}$, $\Gamma=5.0070\mathrm{cm}^{-1}$) in (d), compared with those of bulk eigenmodes at ($\omega=924.4771\mathrm{cm}^{-1}$, $\Gamma=5.0145\mathrm{cm}^{-1}$) and ($\omega=931.5735\mathrm{cm}^{-1}$, $\Gamma=5.0001\mathrm{cm}^{-1}$).(f) Phase diagram with respect to the complex Zak phase calculated from bulk band structures with different lattice constants and dimerization parameters. The shaded area indicates the invalidity of dipole approximation.} \label{eigenmodelong}
\end{figure}

In Fig.\ref{beta07longband}, we show the complex band structures of the longitudinal eigenmodes for $\beta=0.7$ and $\beta=0.3$ with a lattice constant of $d=1\mathrm{\mu m}$. Here the number of NPs in the chain is fixed as 100 (50 dimers). In both cases we observe that bandgaps are opened in the complex frequency plane. The corresponding range of the bandgaps also agrees well with the Bloch band structure. The difference between the $\beta=0.7$ and $\beta=0.3$ cases is that two midgap modes with high IPRs emerge in the complex bandgap in the former case. The dipole moment distributions of the two midgap modes are shown in Fig.\ref{midgapmode}, which exhibit an exponential localization behavior from the boundary. Hence, by combining the nontrivial complex Zak phase of the $\beta=0.7$ case, we can conclude that these midgap modes are topologically protected edge modes.

To further examine the bulk-boundary correspondence, we can check whether the topologically protected interface  mode can emerge at the boundary involving two topologically different media. Fig.\ref{interfacemodebeta07} presents the eigenmode distribution of a 1D connected chain, which consists of a topologically trivial chain with $\beta=0.3$ in the left and a topologically nontrivial chain with $\beta=0.7$ in the right. The distance between the two chains is set to be $1\mathrm{\mu m}$ (more specifically, the distance between the centers of the rightmost NP in the left chain and the leftmost NP in the right chain.). Two midgap modes with high IPRs are also observed, in which one is the interface  mode while the other is the edge mode localized at the right boundary of the right chain. Their IPRs are much larger than the midgap modes in the single chain case, both reaching unity. In Fig.\ref{interfacemodedipole} we show the dipole moment distribution for the interface  mode, which is indeed highly localized over the interface, compared with those of two typical bulk eigenmodes in the upper band and lower band. Therefore, so far, we can conclude that the previously derived complex Zak phase is able to characterize these edge states, and the principle of bulk-boundary correspondence is still valid for the longitudinal modes.

We then briefly discuss the effect of lattice constant on the topological properties of longitudinal modes. A phase diagram of the complex Zak phase by varying the lattice constants and dimerization parameters is presented in Fig.\ref{phasediagram}. It is observed that the gap closing point $\beta=0.5$ indeed amounts to a topological transition from the topologically trivial phase to the topologically nontrivial phase as expected. Note there is a shaded area where the dipole approximation does not hold, and thus the parameters in this area are not investigated. In a brief equation, the complex Zak phase of longitudinal modes ($\theta_\mathrm{Z}^L$) is:
\begin{equation}
\theta_\mathrm{Z}^L=\begin{cases}
\pi &{\beta>0.5, (1-\beta)\geq 3a},\\
0 & {\beta<0.5,\beta d\geq 3a}.
\end{cases}
\end{equation}

\begin{figure}[htbp]
	\flushleft
	\subfloat{
		\includegraphics[width=0.46\linewidth]{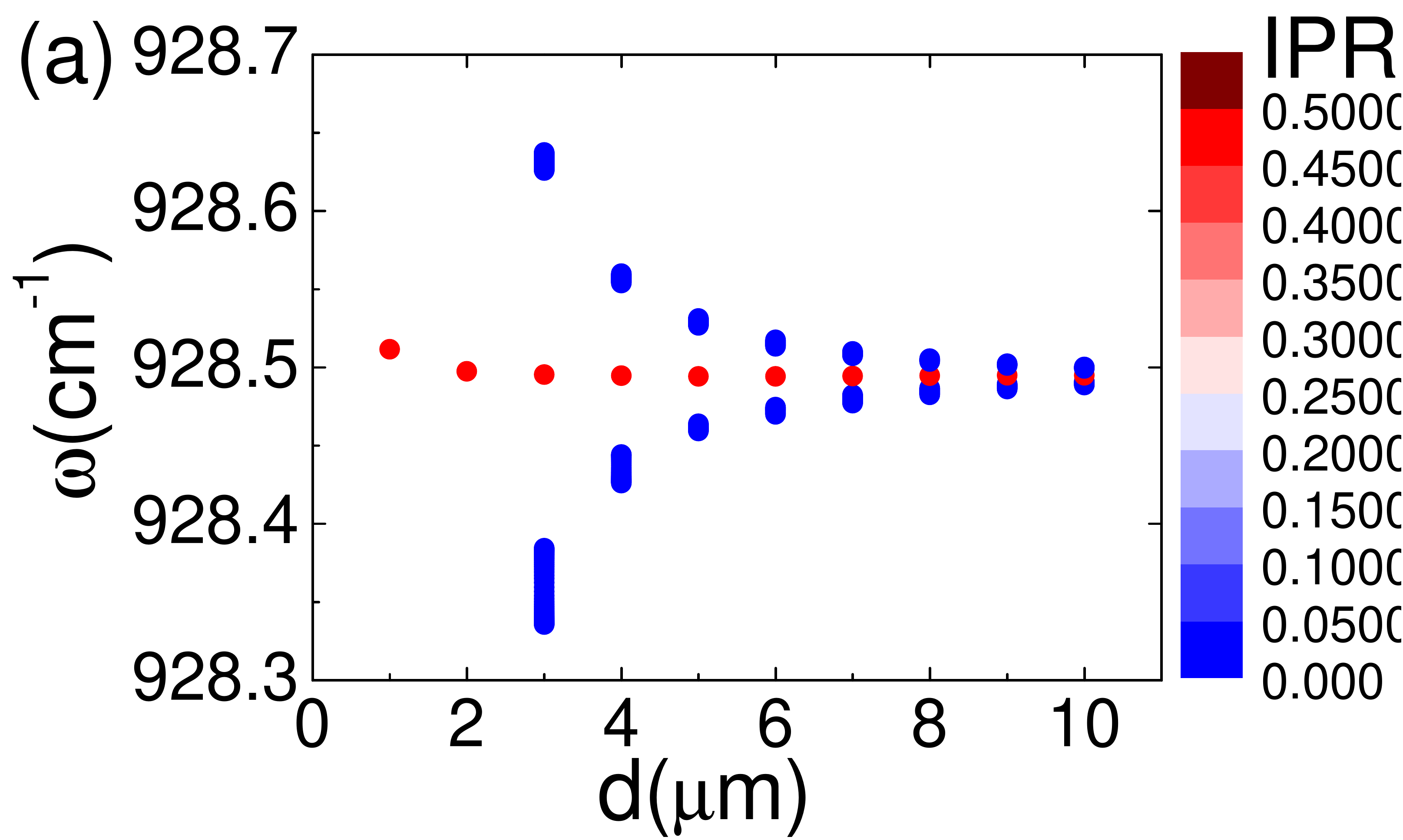}\label{bandevolutionlongreal}
	}
	\subfloat{
	\includegraphics[width=0.46\linewidth]{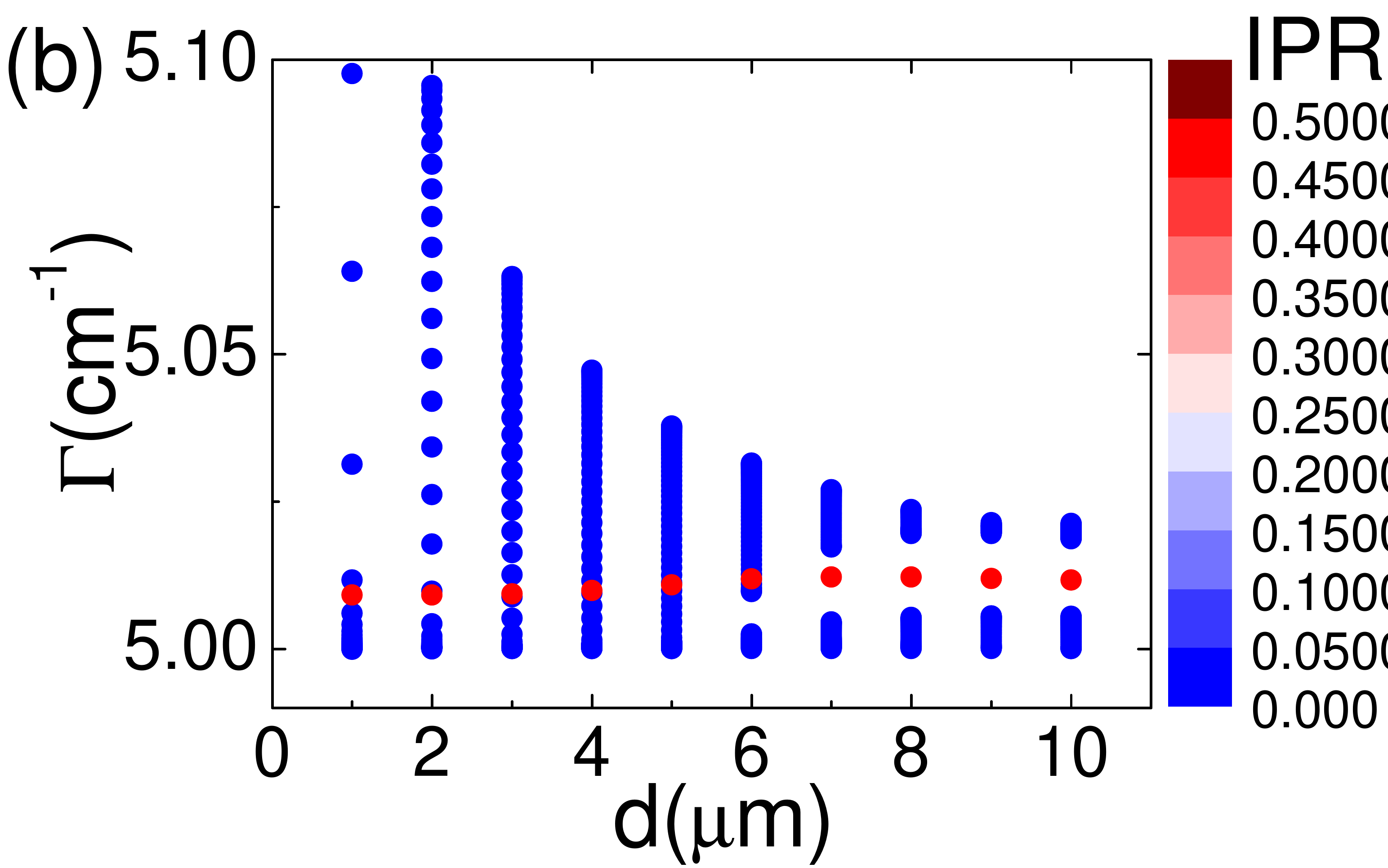}\label{bandevolutionlongimag}
	}
	\caption{Real (a) and imaginary (b) parts of the complex eigenfrequency spectrum of longitudinal eigenmodes as a function of the lattice constant $d$ for $\beta=0.7$.}\label{bandedgeevolutionlong}
\end{figure}
	
Along with the phase diagram, we further plot the eigenfrequency spectrum of a finite chain with 100 NPs, as a function of the lattice constant, shown in Fig.\ref{bandedgeevolutionlong}. The dimerization parameter is set to be $\beta=0.7$. It is obviously seen that the complex bandgaps are always open and eigenmodes with high IPRs are well situated in the complex bandgaps, despite that the bandgaps in the real-frequency plane are almost closed at large lattice constants. Therefore taking the entire complex band structure into account in a non-Hermitian system is necessary and critical to correctly study its topological properties \cite{shenPRL2018,pocockArxiv2017,wang2018topological}. We also confirm that for the cases of $\beta<0.5$, no localized edge eigenmodes are observed. Therefore, the bulk-boundary correspondence for longitudinal modes is verified as predicted by the phase diagram with respect to the quantized complex Zak phase.

\section{Transverse modes}\label{transmode_sec}
Now we turn our attention to the transverse modes. As mentioned in Section \ref{model}, the transverse modes are strongly affected the very slowly decaying $1/r$ long-range far-field dipole-dipole interactions, and therefore the effect of long-range interactions are much more significant than the case of longitudinal modes. More specifically, the discontinuities in the band structures derived from the long-range, far-field (transverse) interactions are critical to the band topology \cite{leporiNJP2017,bettlesPRA2017}. Since conventional bulk-boundary correspondence is established for Hermitian systems with only short-range interactions, this section is mainly dedicated to further investigations on the role of long-range interactions on the topological physics of this system \cite{bettlesPRA2017,perezArxiv2018,gongPRB2016}. Recently, there are a great deal of studies on the topological properties of certain long-range interacting quantum mechanical models, like the 1D Kitaev model with long-range (usually in power law as $1/r^{\alpha}$, $\alpha>0$) pairing \cite{vodolaPRL2014}, 1D Kitaev model with both long-range hopping and pairing \cite{viyuelaPRB2016,leporiNJP2017} as well as 2D $p$-wave superconductor models with long-range hopping or (and) pairing amplitudes \cite{viyuelaPRL2018,leporiPRB2018}. Novel topological phases with massive, nonlocal Dirac edge modes are found for certain 1D lattice model consisting of long-range interactions with a small decay exponent $\alpha\lesssim1$ \cite{viyuelaPRB2016}. For sufficiently strong long-range interactions, correlation functions and edge modes were predicted to decay purely algebraically. The weakening of bulk-boundary correspondence and fractional topological invariant (e.g., winding number \cite{leporiNJP2017} and Chern number \cite{leporiPRB2018}) were also discovered due to the presence of the second type of singularities in energy spectrum introduced by long-range interactions \cite{leporiNJP2017}.  Recently, in a dense two-dimensional atomic lattice gas interacting with light, Bettles \textit{et al}. \cite{bettlesPRA2017} found a significant overlap between edge and bulk modes, and showed that the bulk-boundary correspondence does not hold for the cases with strong long-range interactions.  

Hence, in this section, we focus on the role of long-range interactions in the topological properties of the system. According to the relative strength between long-range far-field interactions and short-range near-field ones, we study the case of a small lattice constant $d=1\mathrm{\mu m}$ and cases of large lattice constants including $d=2\mathrm{\mu m}$ and $d=5\mathrm{\mu m}$. Note here interactions decaying as $1/r^2$ or faster are called as ``short-range" for the convenience of discussion. The more generalized meaning of short-range interactions usually indicated those exponentially decayed ones whose decay length is comparable or even smaller than the lattice constant \cite{viyuelaPRB2016}. By calculating the Bloch band structure for an infinite lattice and corresponding complex Zak phase, we predict a topological phase transition brought by the increase of lattice constant, when the long-range interactions are sufficiently strong, as also recently observed by Pocock \textit{et al.} for an extended SSH model for plasmonic NPs \cite{pocockArxiv2017}. By carefully tackling with the eigenmodes near the divergences of the spectrum, we further find the bulk-boundary correspondence breaks down in this situation, which is attributed to the emergence of localized bulk eigenmodes in a finite lattice due to the long-range interactions induced non-Hermiticity, as a manifestation of the non-Hermitian skin effect \cite{yao2018edge}. We then show that in this situation the bulk-boundary correspondence can still be recovered by qualitatively introducing a modified, or non-Bloch, complex Zak phase. This part of our investigation has implications for the role of long-range interactions in non-Hermitian topological systems.
\subsection{The case of a small lattice constant}
\begin{figure}[htbp]
	\centering
	\subfloat{
		\includegraphics[width=0.46\linewidth]{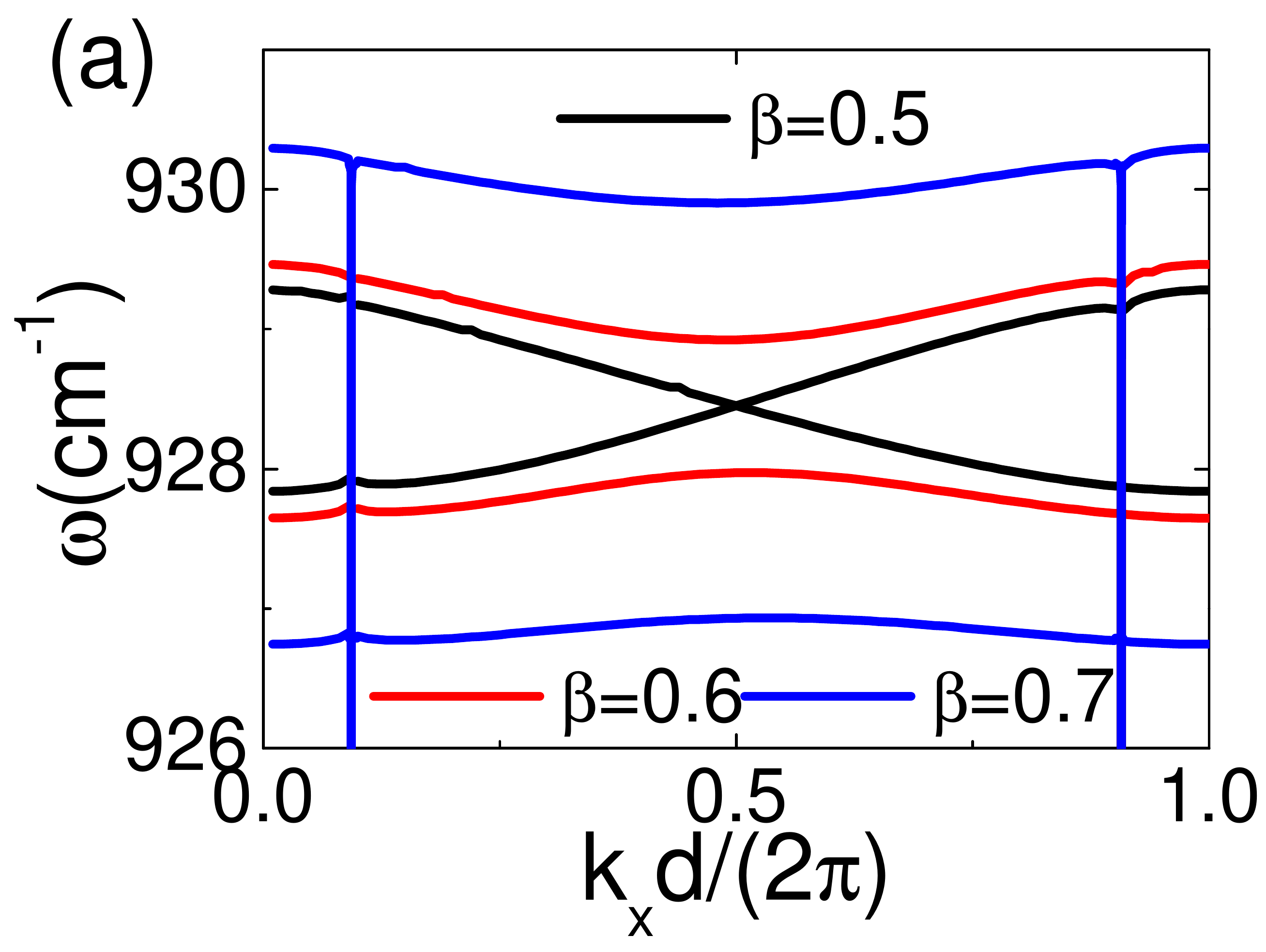}\label{realbandstructuretransd1}
	}
	\hspace{0.01in}
	\subfloat{
	\includegraphics[width=0.46\linewidth]{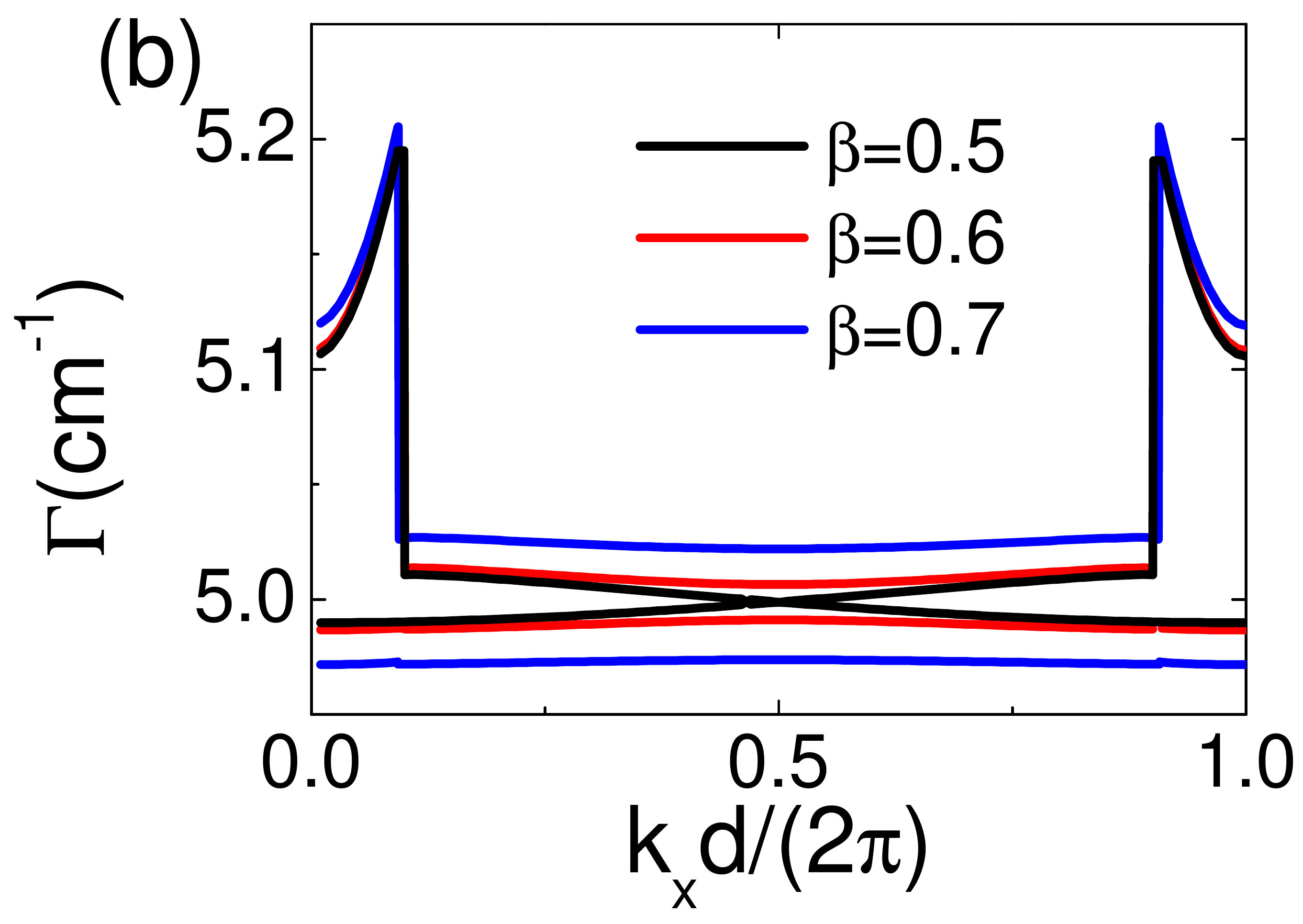}\label{imagbandstructuretransd1}
}
		\hspace{0.01in}
	\subfloat{
		\includegraphics[width=0.46\linewidth]{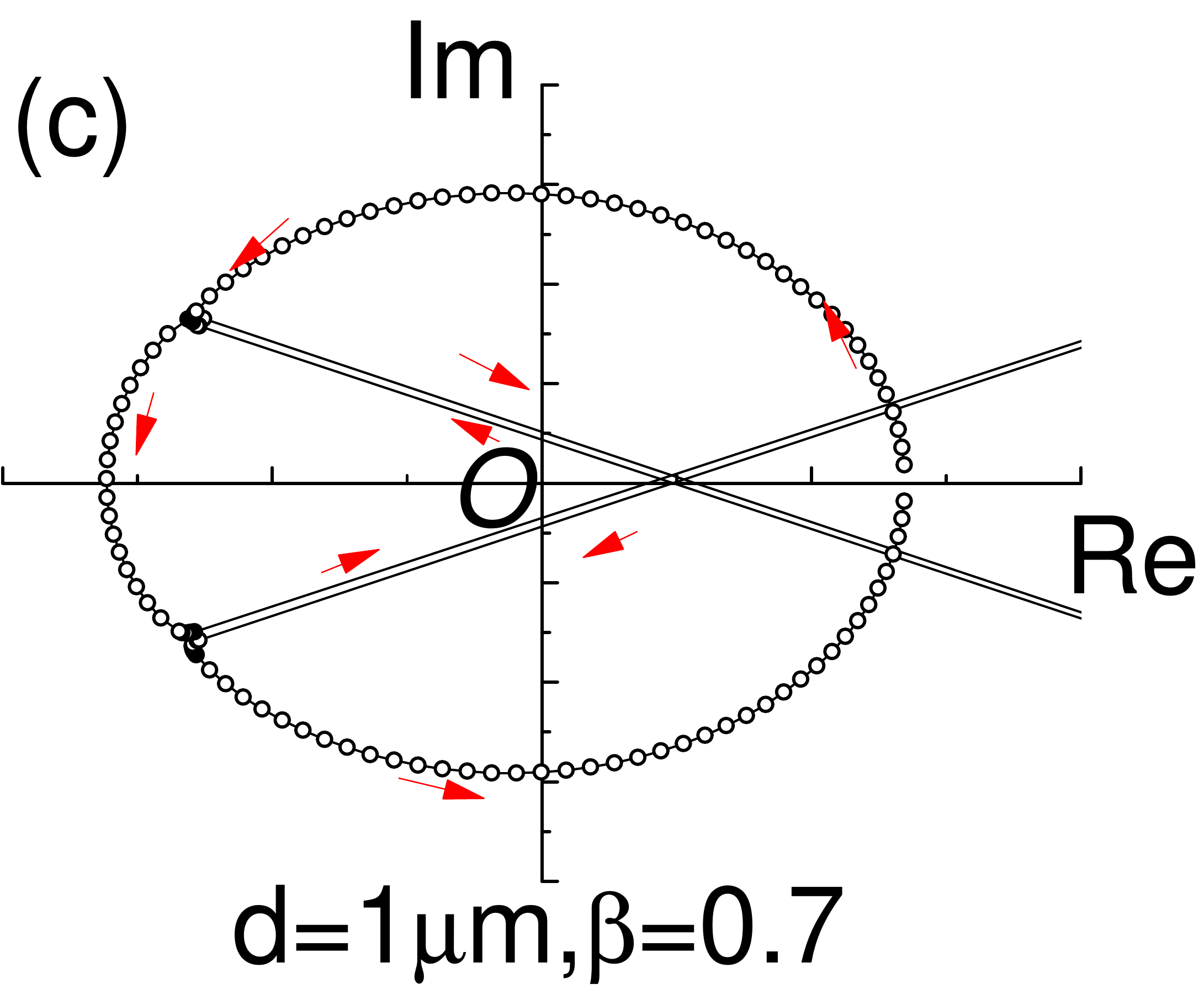}
		\label{windingbeta07d1trans}
}
	\hspace{0.01in}
\subfloat{
	\includegraphics[width=0.46\linewidth]{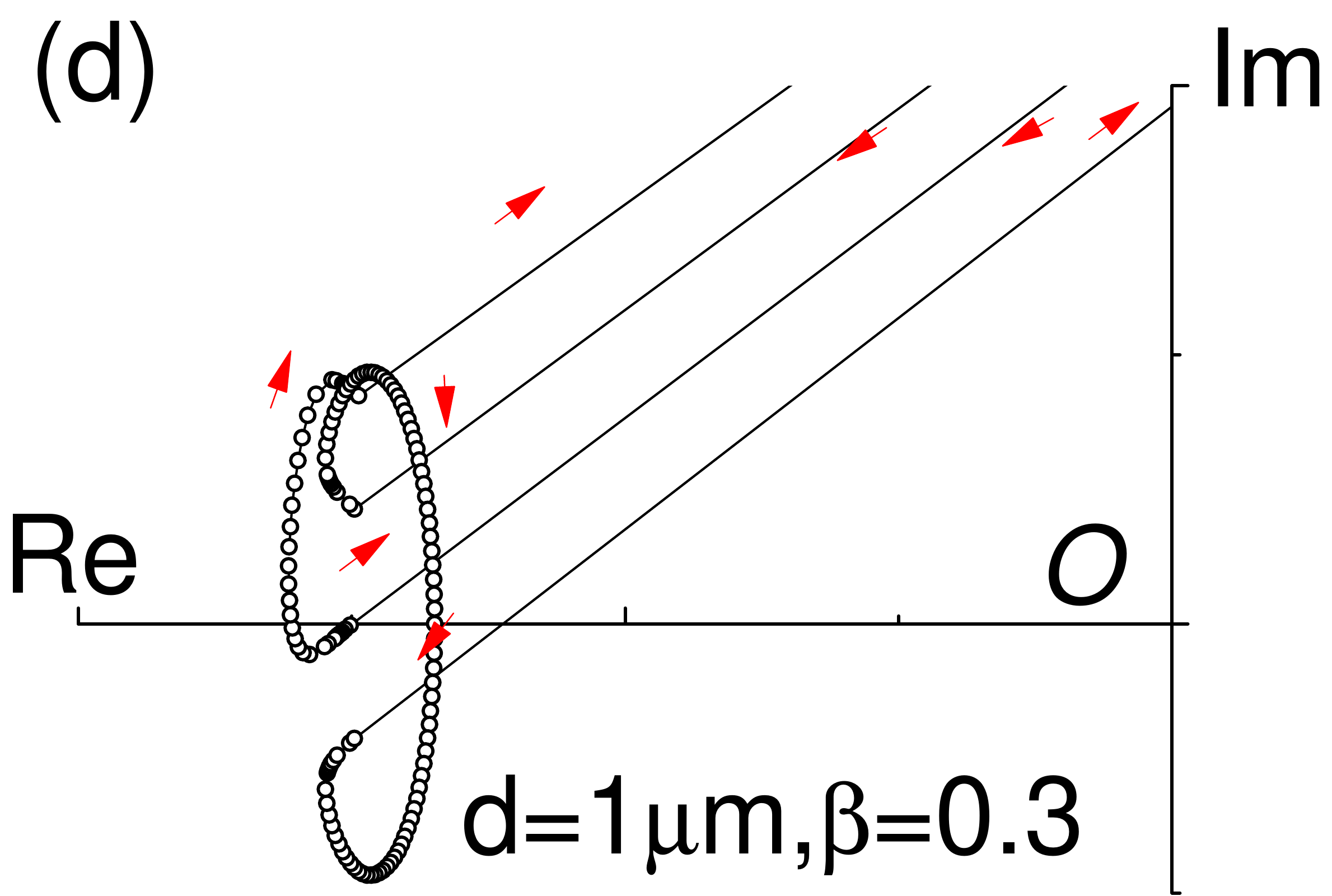}\label{windingbeta03d1trans}
}
\caption{(a) Real parts of band structures of a dimerized chain with $d=d_1+d_2=1\mathrm{\mu m}$ for different dimerization parameters $\beta$. (b) Imaginary parts of the transverse band structures of a dimerized chain with $d=1\mathrm{\mu m}$ for different dimerization parameters $\beta$. Note here we cut the infinitely diverging singularity in the spectra off at some limited values. (c) The winding of $a_{12}(k_x)$ around the origin for $d=1\mathrm{\mu m}$ and $\beta=0.7$. (d) The same as that of (c) but for $\beta=0.3$. The arrows indicate the direction of winding.}\label{transbandstructure}
\end{figure}

We first start with a dimerized NP chain with a relatively small lattice constant $d=1\mathrm{\mu m}$. In Fig.\ref{transbandstructure}, we show the real and imaginary parts of the complex band structures under different dimerization parameters. The most prominent feature is a strong divergence of the upper band near the light cone for all dimerization parameters, due to the coupling of Bloch modes with free-space radiation \cite{pocockArxiv2017,bettlesPRA2017,downing2018topological}. Mathematically, as manifested in Eqs.(\ref{a11T})-(\ref{a21T}), this is because the functions $\mathrm{Li}_1(z)$ and $\Phi(z,1,a)$ are logarithmically divergent near $z=1$ \cite{NISThandbook}, which physically arise from the long-range dipole-dipole interactions slowly decaying as $1/r$ with the distance $r$. Another feature is that, away from the singularities, the band gaps of transverse eigenmodes are substantially smaller than those of longitudinal eigenmodes under the same system parameters. This phenomenon is not only due to the long-range far-field interactions but also the factor of two difference in the near-field dipole-dipole interactions. 

Due to the existence of divergences brought by the infinite summation of $1/r$ terms, the winding of $a_{12}(k_x)$ and $a_{21}(k_x)$ also encounters discontinuities when $k_x=\pm k$ as shown in Figs.\ref{windingbeta07d1trans} and \ref{windingbeta03d1trans}. Here only the data of $a_{12}(k_x)$ is shown since the winding procedure of $a_{21}(k_x)$ is exactly the same with $a_{12}(k_x)$ but in the opposite direction. If we follows the winding of $a_{12}(k_x)$ around the origin, we can find that the divergences introduce an abrupt phase shift. Take the limit of $k_x\rightarrow k$ as an example, i.e., $z^-\rightarrow1$. In this circumstance, $a_{12}^{T}$ and $a_{21}^{T}$ both diverge. We can analyze the asymptotic behavior near the singularity at $z^-=\exp{[i(k-k_x)d]}=1$. By invoking the logarithmic divergence properties of Lerch transcendent and polylogarithm function, namely,
$\lim_{z\rightarrow1}\frac{\Phi(z,1,a)}{-\ln(1-z)}=1$ and
$\lim_{z\rightarrow1}\frac{\mathrm{Li}_1(z)}{-\ln(1-z)}=1$,
for a real infinitesimal quantity $\eta\rightarrow0$ and $>0$ \cite{NISThandbook}, we have
\begin{equation}\label{a12kplus}
\begin{split}
a_{12}^{T}(k+\eta)\rightarrow\frac{-\ln(i\eta d)}{4\pi kd}\exp{(-ik\beta d)},
\end{split}
\end{equation}
and
\begin{equation}\label{a12kminus}
\begin{split}
a_{12}^{T}(k-\eta)\rightarrow\frac{-\ln(-i\eta d)}{4\pi kd}\exp{(-ik\beta d)}.
\end{split}
\end{equation}
Similarly, for $a_{21}^T(k_x\rightarrow k)$, we have
\begin{equation}\label{a21kplus}
\begin{split}
a_{21}^{T}(k+\eta)\rightarrow\frac{-\ln(i\eta d)}{4\pi kd}\exp{(ik\beta d)}
\end{split}
\end{equation}
and
\begin{equation}\label{a21kminus}
\begin{split}
a_{21}^{T}(k-\eta)\rightarrow\frac{-\ln(-i\eta d)}{4\pi kd}\exp{(ik\beta d)}
\end{split}
\end{equation}
As a result, the winding angle will undergo an abrupt change near the singularities, as readily seen in Figs.\ref{windingbeta07d1trans} and \ref{windingbeta03d1trans}. However, the difference between $a_{12}(k_x)$ and $a_{21}(k_x)$ is finite and can be evaluated as $a_{12}^{T}(k+\eta)-a_{12}^{T}(k-\eta)\rightarrow i\exp{(-ik\beta d)}/(4\pi kd)$, which can be neglected compared to the infinite divergences. Hence the encircling procedure of $a_{12}(k_x)$ and $a_{21}(k_x)$ can be regarded as continuous if we appropriately track the abrupt phase shift near the singularity. In that sense, the complex Zak phase for transverse modes $\theta_\mathrm{Z}^T$ can be still calculated. Therefore, for the $d=1\mathrm{\mu m}$ case, when $\beta<0.5$, $\theta_\mathrm{Z}^T=0$ and when $\beta<0.5$, $\theta_\mathrm{Z}^T=\pi$. 

\begin{figure}[htbp]
	\flushleft
	\subfloat{
		\includegraphics[width=0.46\linewidth]{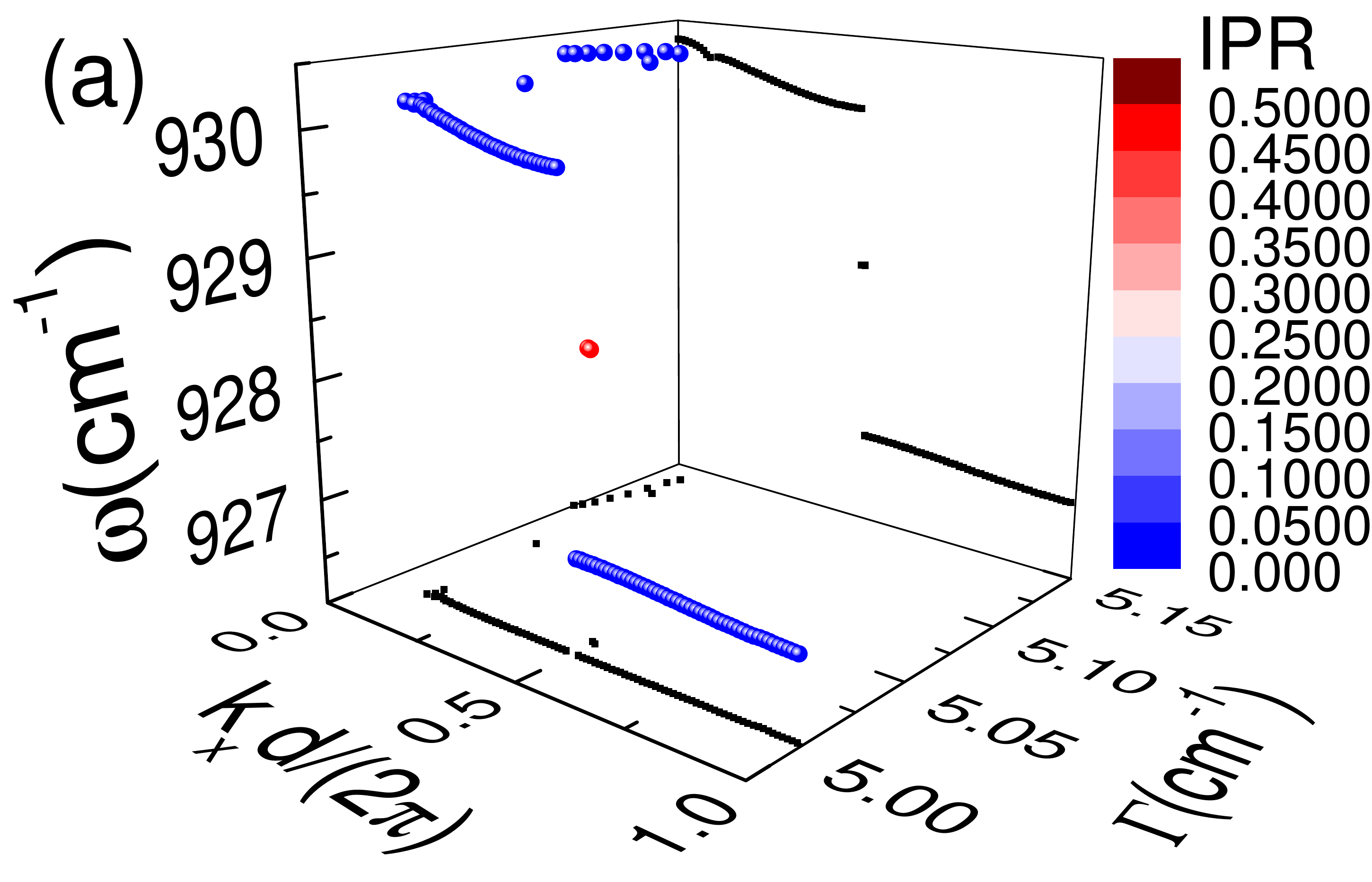}\label{beta07transband}
	}
	\subfloat{
		\includegraphics[width=0.46\linewidth]{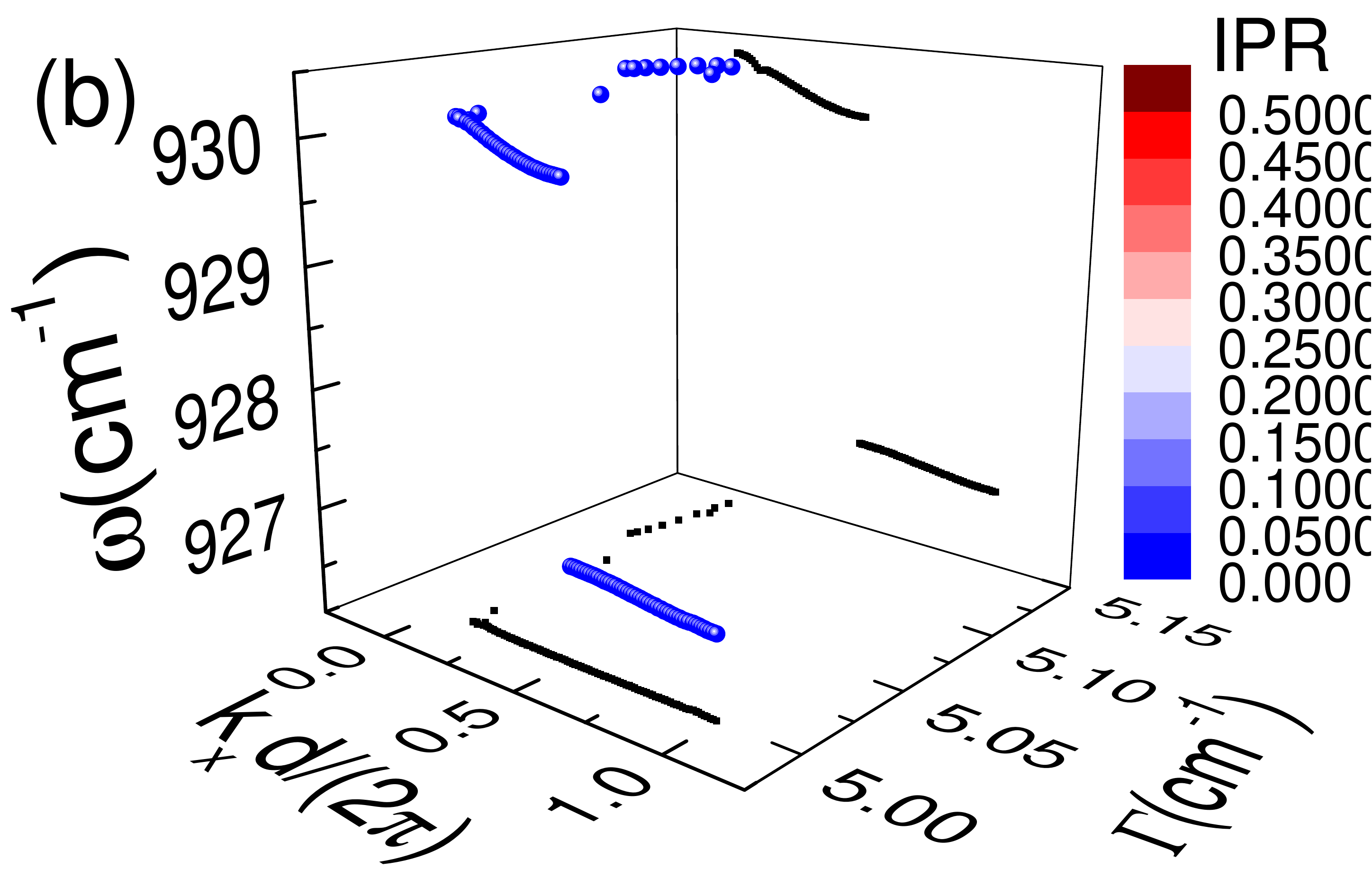}\label{beta03transband}
	}
	\hspace{0.01in}
	\subfloat{
		\includegraphics[width=0.46\linewidth]{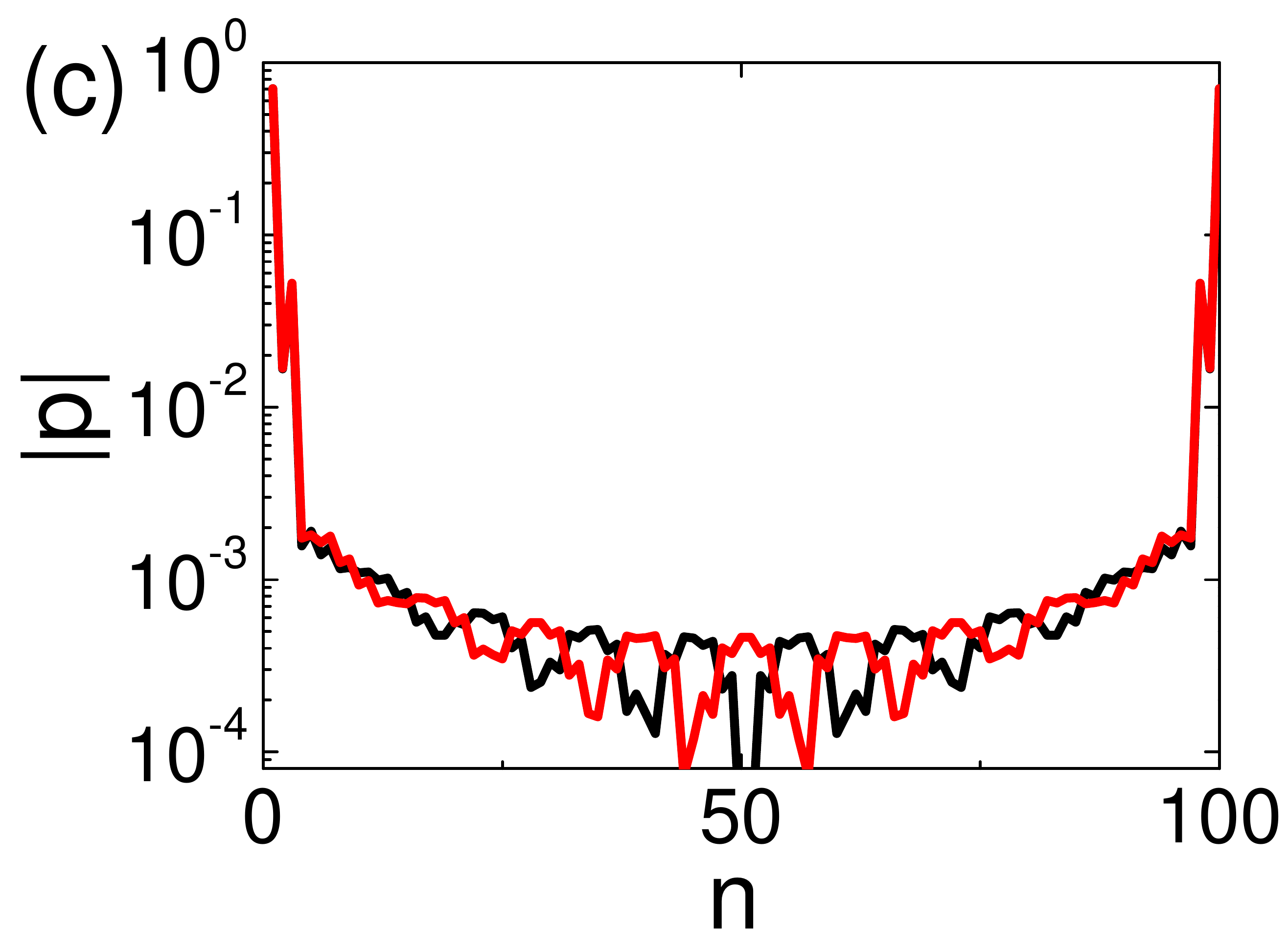}\label{midgapmodetrans}
	}
	\subfloat{
		\includegraphics[width=0.46\linewidth]{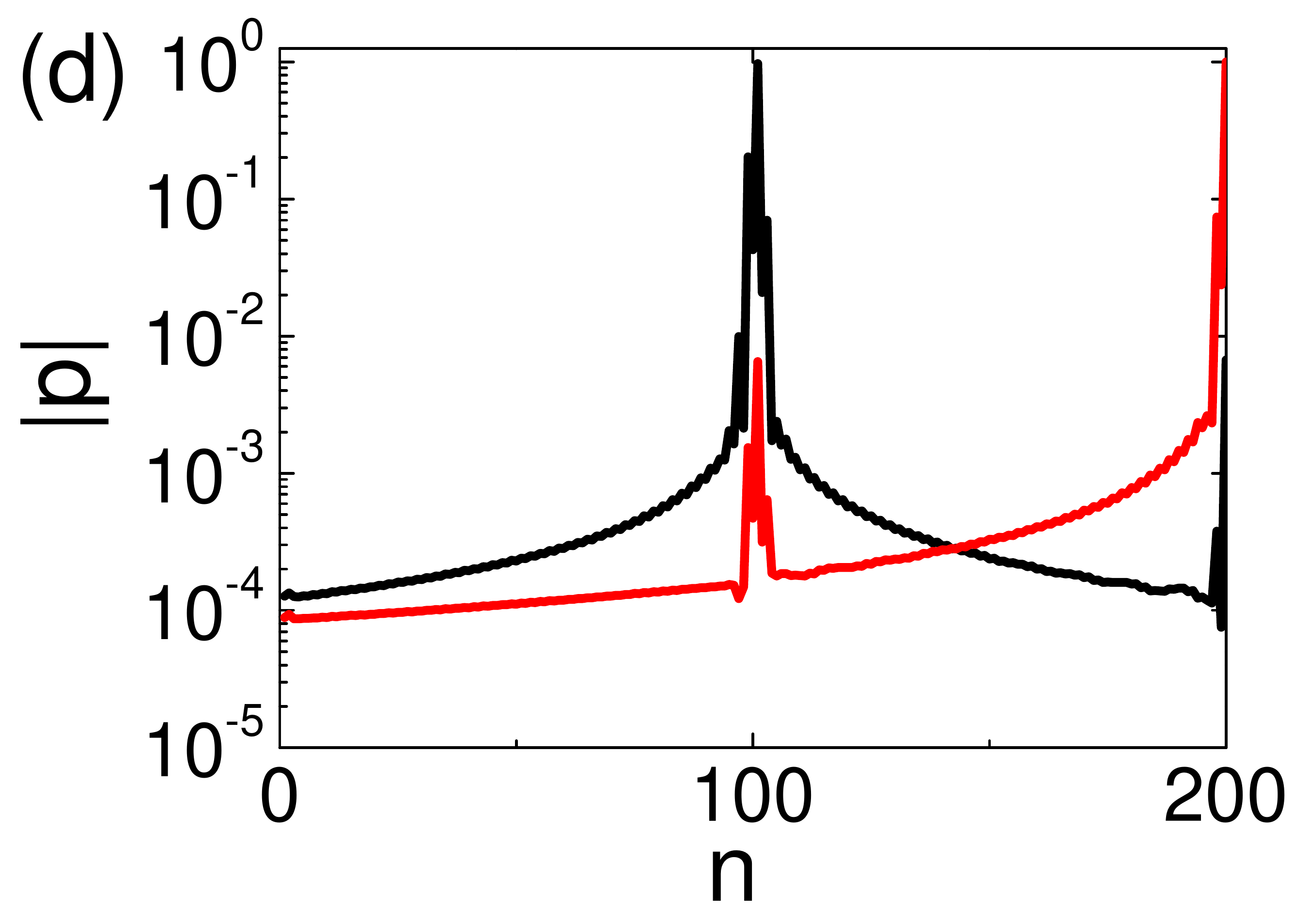}\label{interfacemodetrans}
	}
	\caption{(a) Complex band structure of transverse eigenmodes of a dimerized chain with $N=100$ NPs under $\beta=0.7$ and $d=1\mathrm{\mu m}$. Note there are two midgap modes. (b) The same as (a) but here $\beta=0.3$. (c) Dipole moment distribution of the midgap edge states. No midgap modes with high IPRs are observed. (d) Interface and right edge midgap modes for a connected chain. The IPRs of the two eigenmodes are 0.9063 and 0.9878. The complex frequencies are ($\omega=928.458190\mathrm{cm}^{-1}$ and $\Gamma=5.005121\mathrm{cm}^{-1}$) and ($\omega=928.489008\mathrm{cm}^{-1}$ and $\Gamma=5.009574\mathrm{cm}^{-1}$).}\label{edgemodetrans}
\end{figure}

To examine the bulk-boundary correspondence in this case, we calculate the eigenmode distribution for $\beta=0.7$ and $\beta=0.3$, as shown in Figs.\ref{beta07transband} and \ref{beta03transband} respectively. It is observed that there are two midgap modes with high IPRs ($\sim0.5$) when $\beta=0.7$, while no midgap modes exist in the $\beta=0.3$ case. Fig.\ref{midgapmodetrans} shows the dipole moment distribution of the two midgap edge modes. To further illustrate the bulk-boundary correspondence, we combine two NP chains with different complex Zak phases together. The distance between the two chains (more specifically, the distance between the centers of the rightmost NP in the left chain and the leftmost NP in the right chain) is set to be $0.5\mathrm{\mu m}$.  We show the dipole moment distributions of interface and right edge midgap modes for the connected chain in Fig.\ref{interfacemodetrans}, which both exhibit very high IPRs approaching 1. Moreover, a hybridization between the interface and right edge modes is observed, which is also due to the existence of long-range interactions. This phenomenon was also seen by Bettles \textit{et al.} in a dense two-dimensional atomic lattice gas interacting with light \cite{bettlesPRA2017}.

\subsection{The cases of larger lattice constants and the role of long-range interactions}
\begin{figure}[htbp]
	\flushleft
	\subfloat{
		\includegraphics[width=0.46\linewidth]{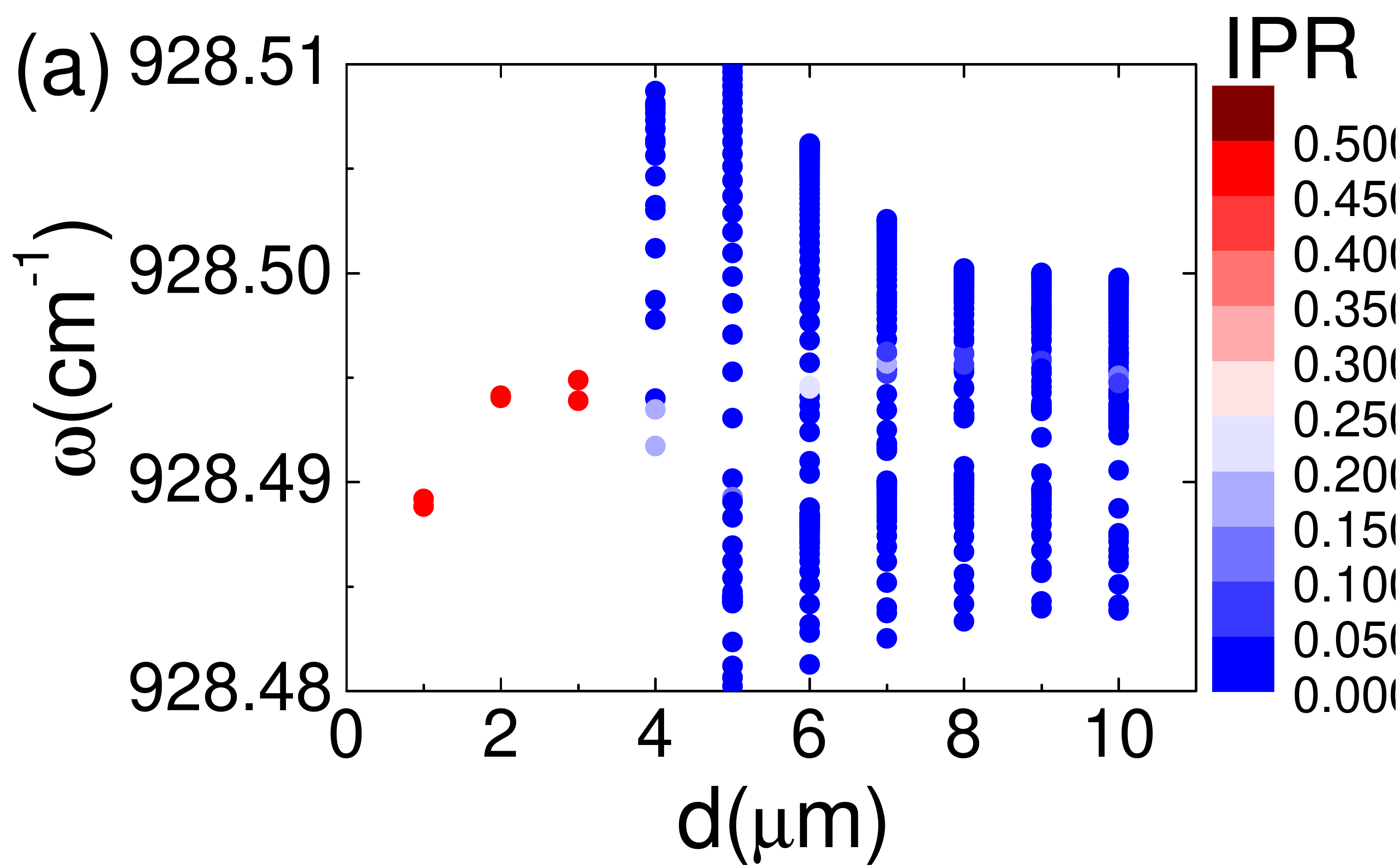}\label{bandevolutiontransreal}
	}
	\hspace{0.01in}
	\subfloat{
		\includegraphics[width=0.46\linewidth]{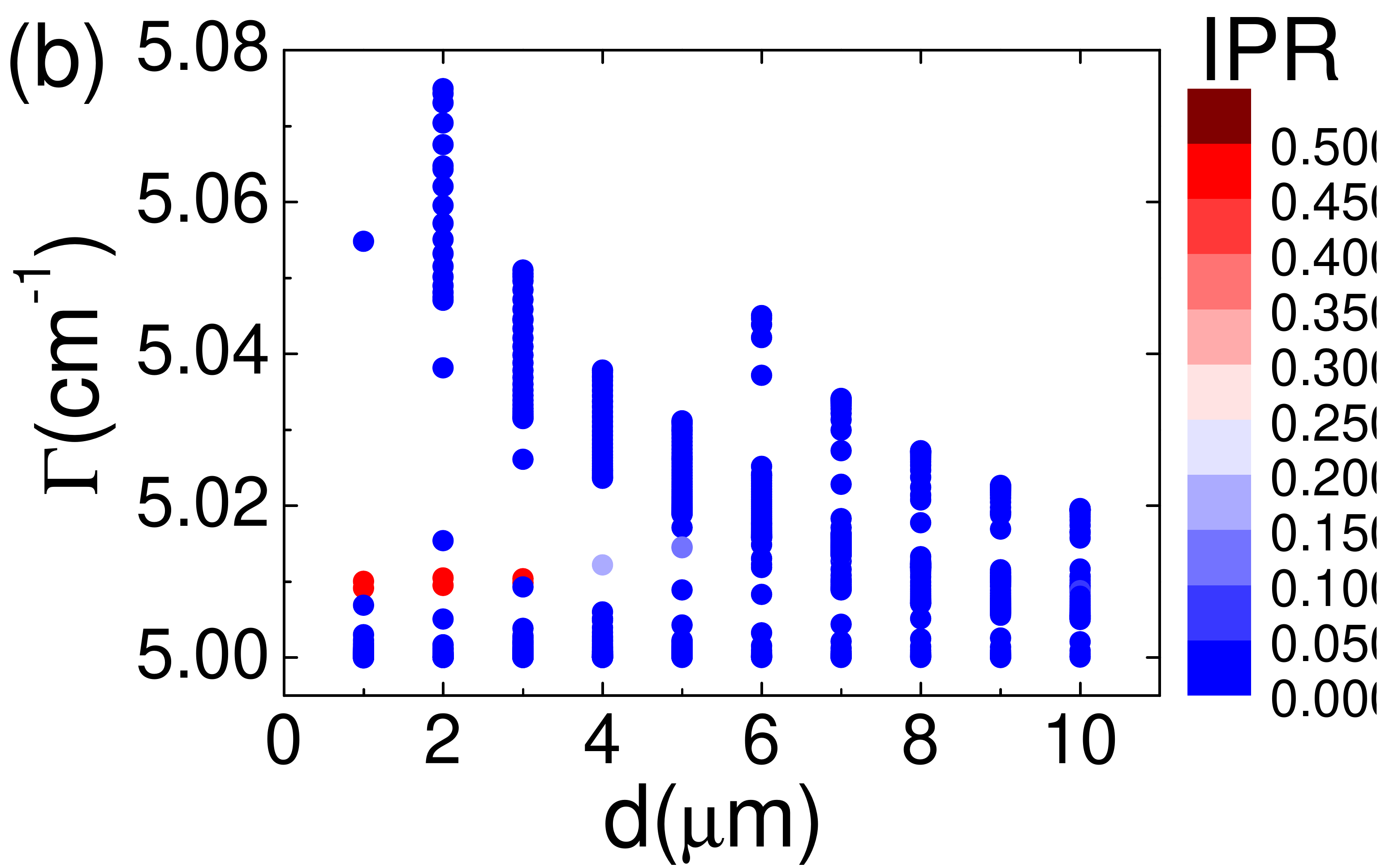}\label{bandevolutiontransimag}
	}
	\caption{Real (a) and imaginary (b) parts of the complex eigenfrequency spectrum of transverse eigenmodes as a function of the lattice constant $d$ for $\beta=0.7$. The chain contains 100 NPs (50 dimers).}\label{bandedgeevolutiontrans}
\end{figure}

In this subsection, we investigate how the increase of lattice constant influences the topological properties of the system, including the complex Zak phase and edge modes. We first show the evolution of eigenmode distributions with the lattice constant $d$, as shown in Fig.\ref{bandedgeevolutiontrans}. It is found that when $d\lesssim4\mathrm{\mu m}$, eigenmodes with high IPRs are apparently observed, while for $d\geq5\mathrm{\mu m}$, all the eigenmodes show substantially lower IPRs. Since a sufficiently large lattice constant that makes $kd>1$, can lead to a situation in which the strength of long-range interactions can exceed that of short-range ones, we may expect that some new behaviors can emerge in this circumstance \cite{pocockArxiv2017}. 

\begin{figure}[htbp]
\centering
\subfloat{
	\includegraphics[width=0.48\linewidth]{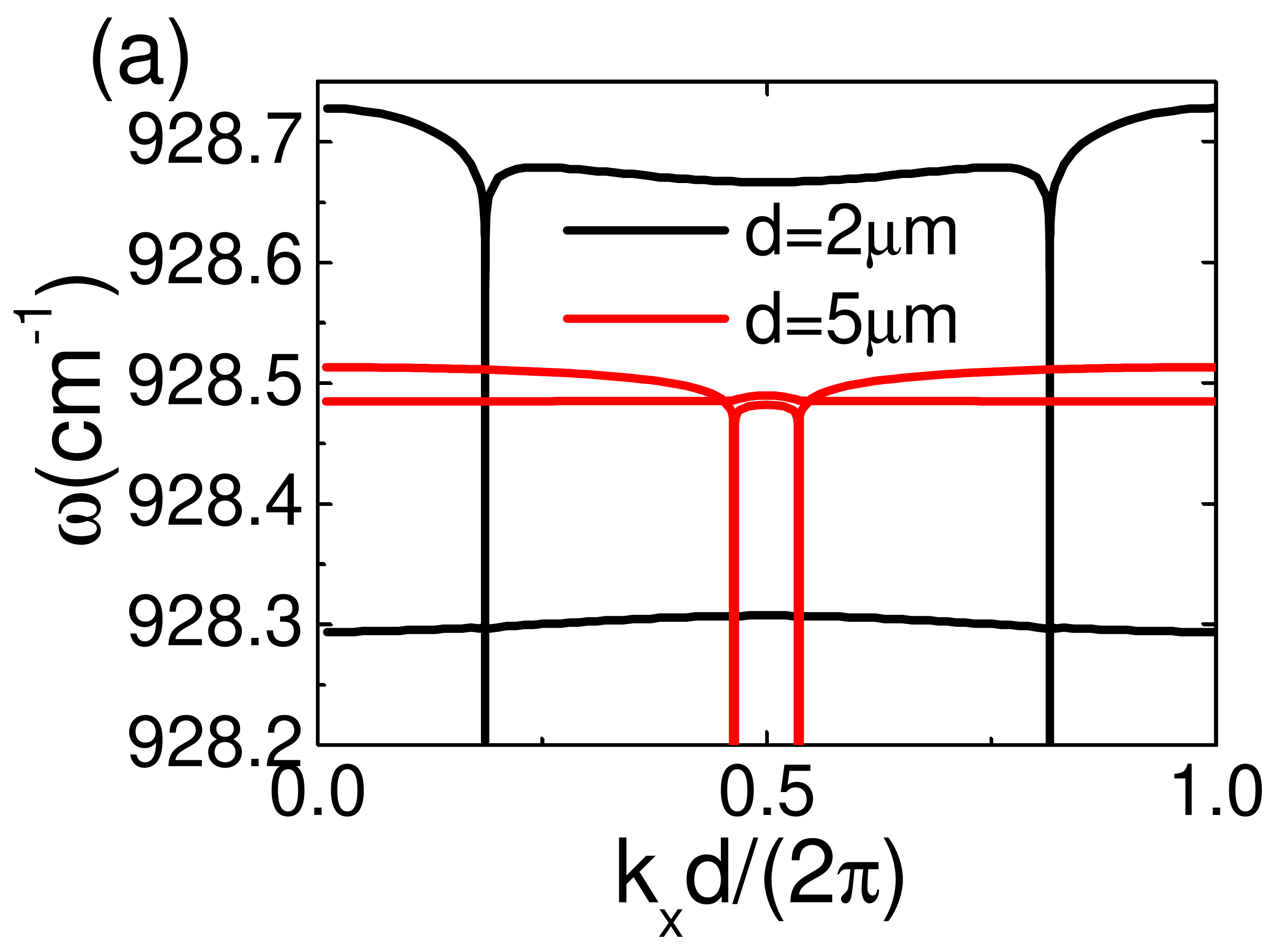}\label{bandstructuretransd5}
}
\subfloat{
		\includegraphics[width=0.44\linewidth]{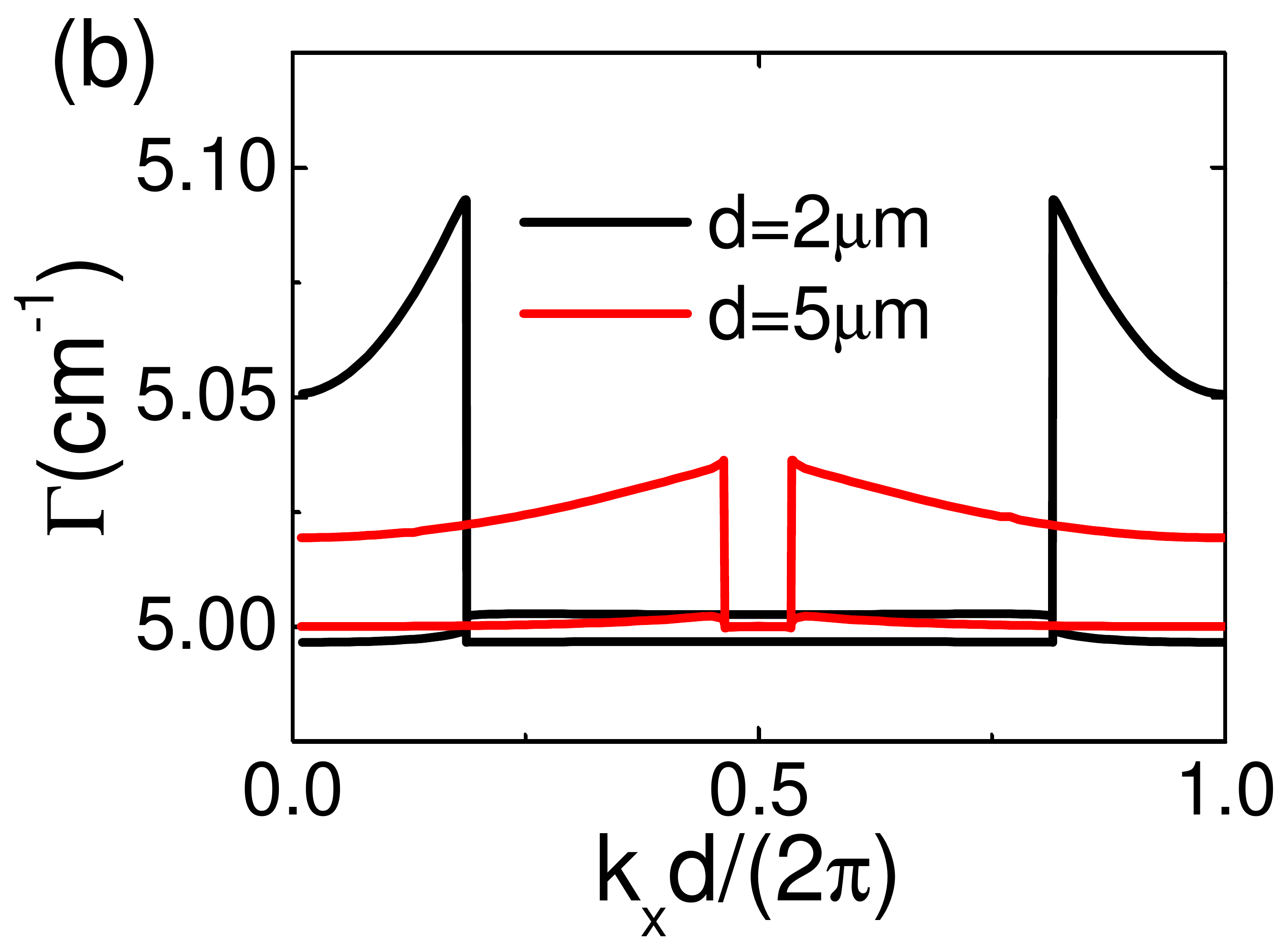}\label{bandstructuretransd5imag}
	}
		\hspace{0.01in}
		\subfloat{
			\includegraphics[width=0.46\linewidth]{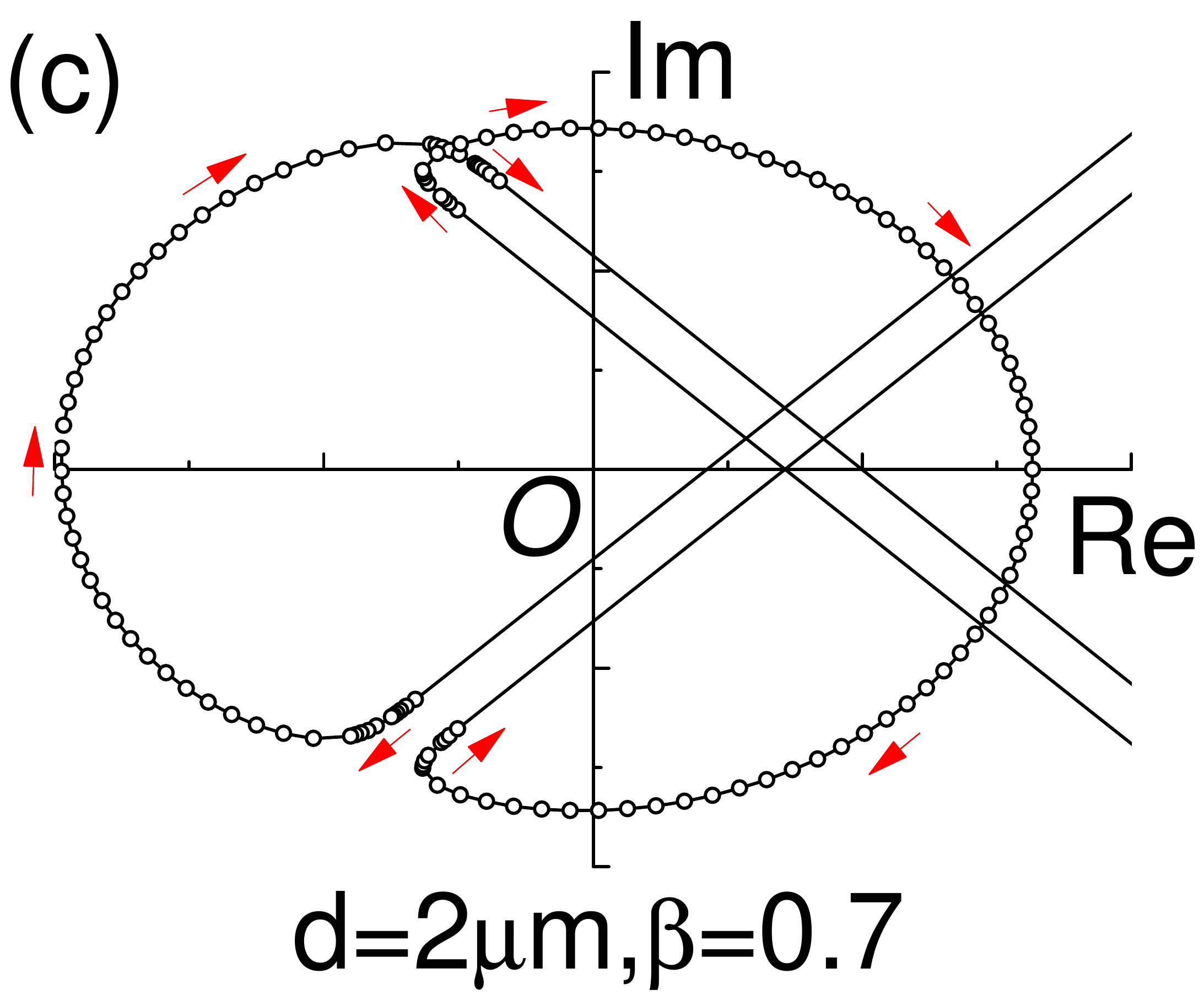}\label{windingbeta07d2trans}
		}
		\hspace{0.01in}
		\subfloat{
			\includegraphics[width=0.46\linewidth]{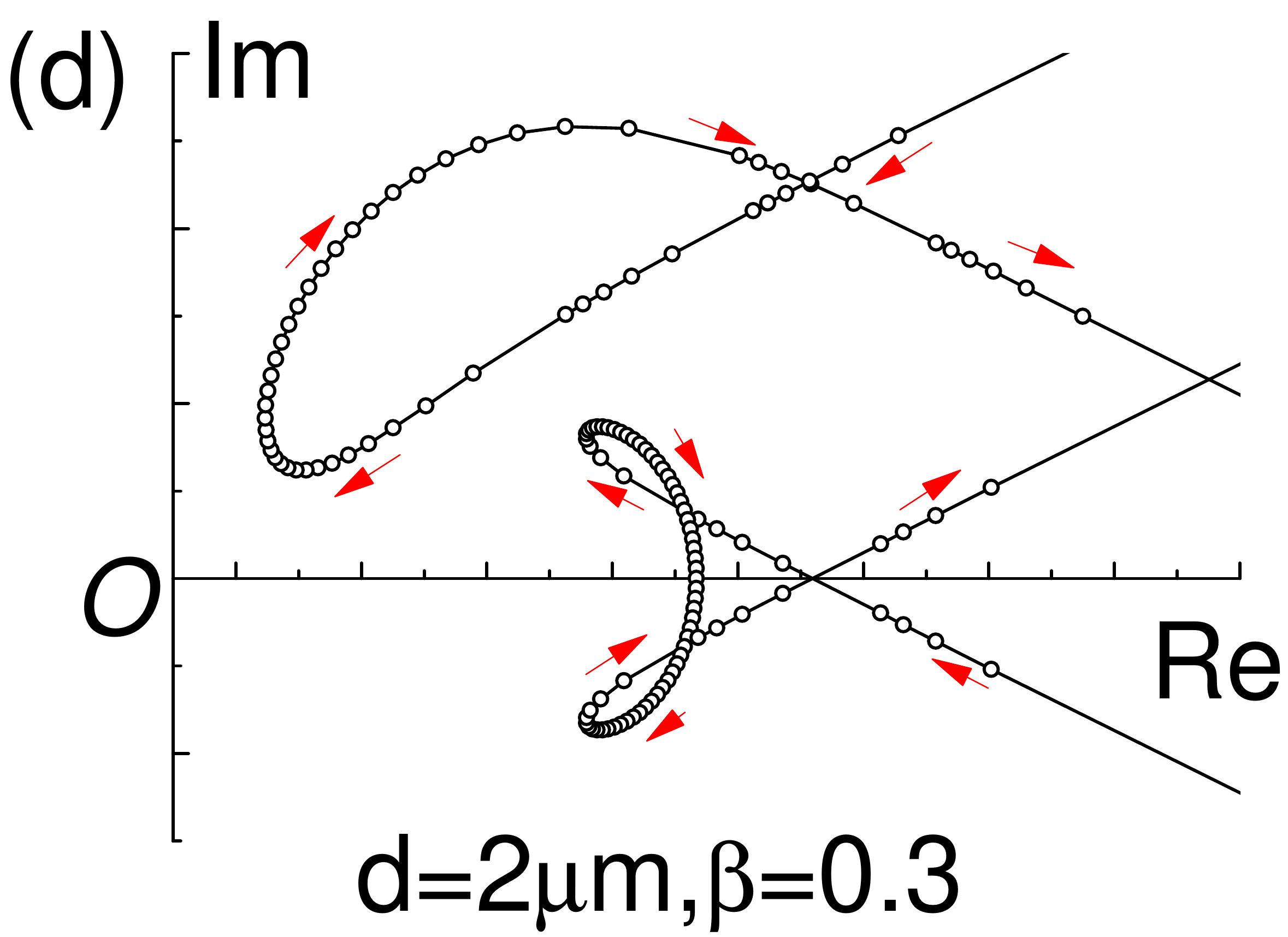}\label{windingbeta03d2trans}
		}
		\hspace{0.01in}
		\subfloat{
			\includegraphics[width=0.46\linewidth]{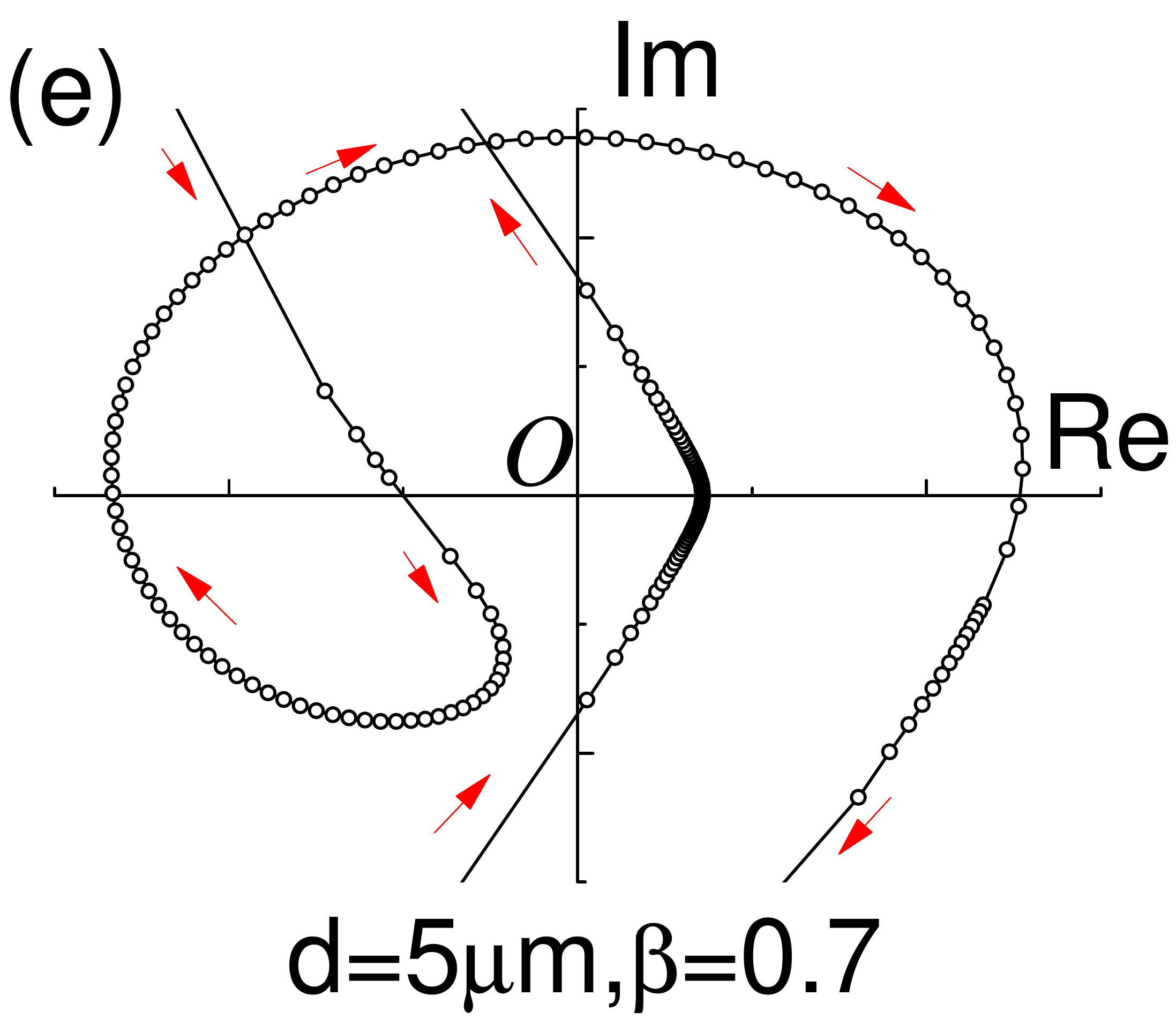}\label{windingbeta07d5trans}
		}
		\hspace{0.01in}
		\subfloat{
			\includegraphics[width=0.46\linewidth]{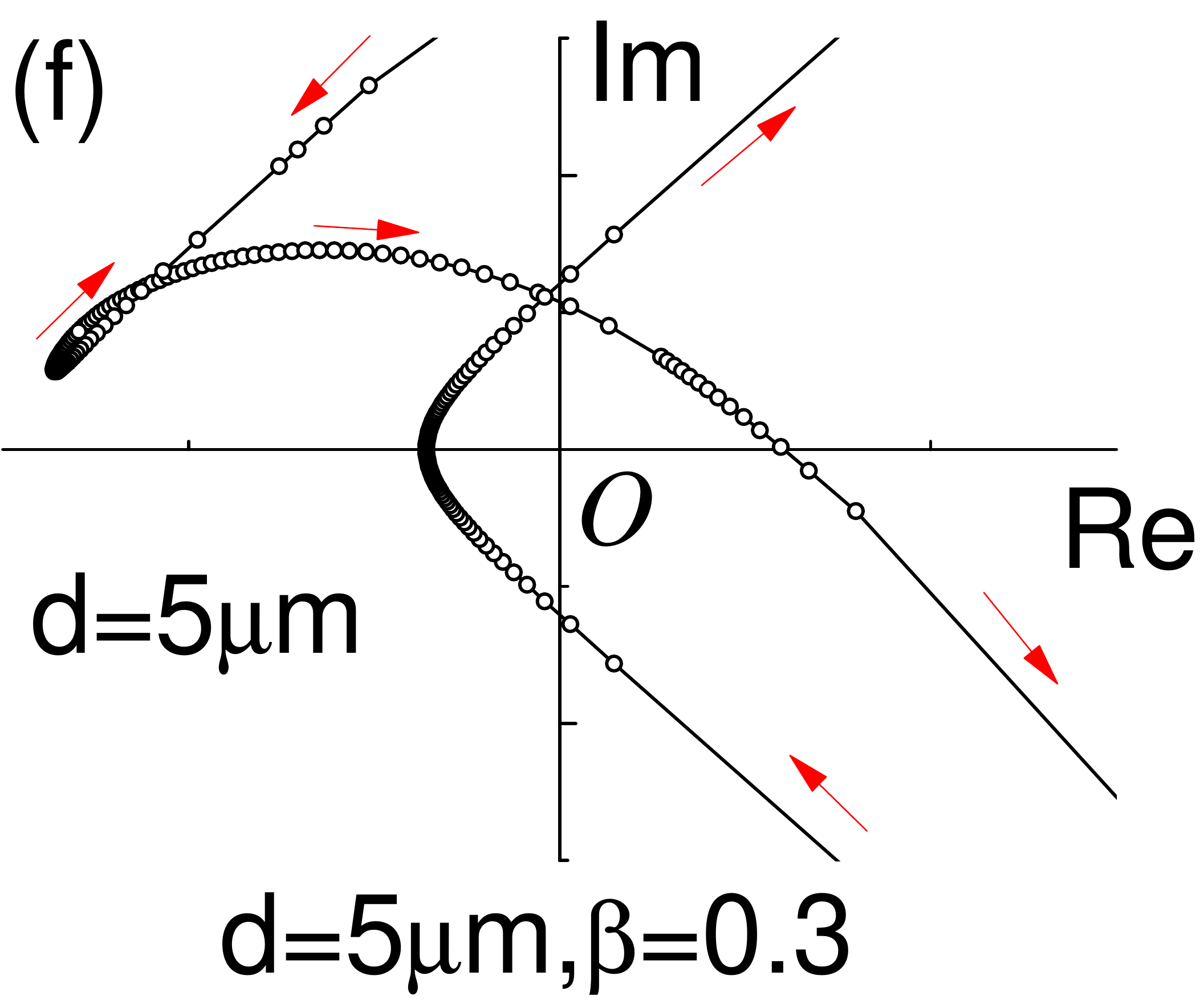}\label{windingbeta03d5trans}
		}
\caption{Real (a) and imaginary (b) parts of the transverse band structures of dimerized chain with $d=2\mathrm{\mu m}$ and $d=5\mathrm{\mu m}$ for a fixed dimerization parameter $\beta=0.7$. (c-f) The winding of $a_{12}(k_x)$ with $k_x$ around the origin in the complex plane. System parameters are (c) $\beta=0.7$, $d=2\mathrm{\mu m}$. (d) $\beta=0.3$, $d=2\mathrm{\mu m}$. (e) $\beta=0.7$, $d=5\mathrm{\mu m}$. (f) $\beta=0.3$, $d=5\mathrm{\mu m}$. }\label{bandstructuretransmore}
\end{figure}

According to the results presented in Fig.\ref{bandedgeevolutiontrans}, we then choose the cases of $d=2\mathrm{\mu m}$ and $d=5\mathrm{\mu m}$, and first calculate the complex Bloch band structures of them as shown in Fig.\ref{bandstructuretransmore} with a fixed dimerization parameter $\beta=0.7$. From the real parts of the band structures, the $d=2\mathrm{\mu m}$ case is shown to be gapped while the $d=5\mathrm{\mu m}$ seems to display a ``metallic" behavior, where a substantial part of the upper-in-frequency band overlaps with the lower-in-frequency band near the band edges ($k=\pm\pi/d$). Note the divergences are not necessarily considered when determining the real band gap \cite{pocockArxiv2017}. However, if the imaginary band structure shown in Fig.\ref{bandstructuretransd5imag} and the definition of a complex (or non-Hermitian) band gap mentioned in Section \ref{zakphasesec} are taken into account, the complex band structures are still gapped for the $d=5\mathrm{\mu m}$. In other words, while the Bloch band gaps may close in the real or imaginary plane when $\beta\neq0.5$, the only time they close simultaneously is at the transition point, i.e., $\beta=0.5$. This is a key difference of non-Hermitian systems with Hermitian ones, e.g., the band gap of an extended Hermitian SSH model with a certain type of long-range hoppings may close and then do not display any topological edge modes, as pointed out by P\'erez-Gonz\'alez \textit{et al.} \cite{perezArxiv2018}.

By using the Bloch band eigenvectors for an infinite system we then turn our attention to the complex Zak phase for the transverse band. Specifically, we show the winding processes of $a_{12}(k_x)$ for the $d=2\mathrm{\mu m}$ and $d=5\mathrm{\mu m}$ in Figs.\ref{windingbeta07d2trans}-\ref{windingbeta03d5trans} (the winding of $a_{21}(k_x)$ is the reversal of $a_{12}(k_x)$ and is not given here). Apparently, the winding phase angle of $a_{12}(k_x)$ in the $d=2\mathrm{\mu m}$, $\beta=0.7$ case is $2\pi$, while that of $d=2\mathrm{\mu m}$, $\beta=0.3$ case is 0. On the other hand, the winding phase angle of $a_{12}(k_x)$ in the $d=5\mathrm{\mu m}$, $\beta=0.7$ case is $0$, and that of $d=5\mathrm{\mu m}$, $\beta=0.3$ case is $2\pi$. According to Eq.(\ref{cberryphase}), we may find a topological phase transition, that is, the Zak phase becomes zero for $\beta>0.5$ and $\pi$ for $\beta<0.5$ when the lattice constant becomes larger than a certain value between $2\mathrm{\mu m}$ and $5\mathrm{\mu m}$, as also observed by Pocock \textit{et al}. \cite{pocockArxiv2017} for dimerized plasmonic NP chains. In brief equations, this conclusion is expressed as
\begin{equation}
\theta_\mathrm{Z}^T=\begin{cases}
\pi &{\beta>0.5, d=2\mathrm{\mu m}},\\
0 & {\beta<0.5, d=2\mathrm{\mu m}},
\end{cases}
\end{equation}
and
\begin{equation}
\theta_\mathrm{Z}^T=\begin{cases}
0 &{\beta>0.5, d=5\mathrm{\mu m}},\\
\pi & {\beta<0.5, d=5\mathrm{\mu m}}.
\end{cases}
\end{equation}

\begin{figure}[htbp]
	\flushleft
	\subfloat{
	\includegraphics[width=0.46\linewidth]{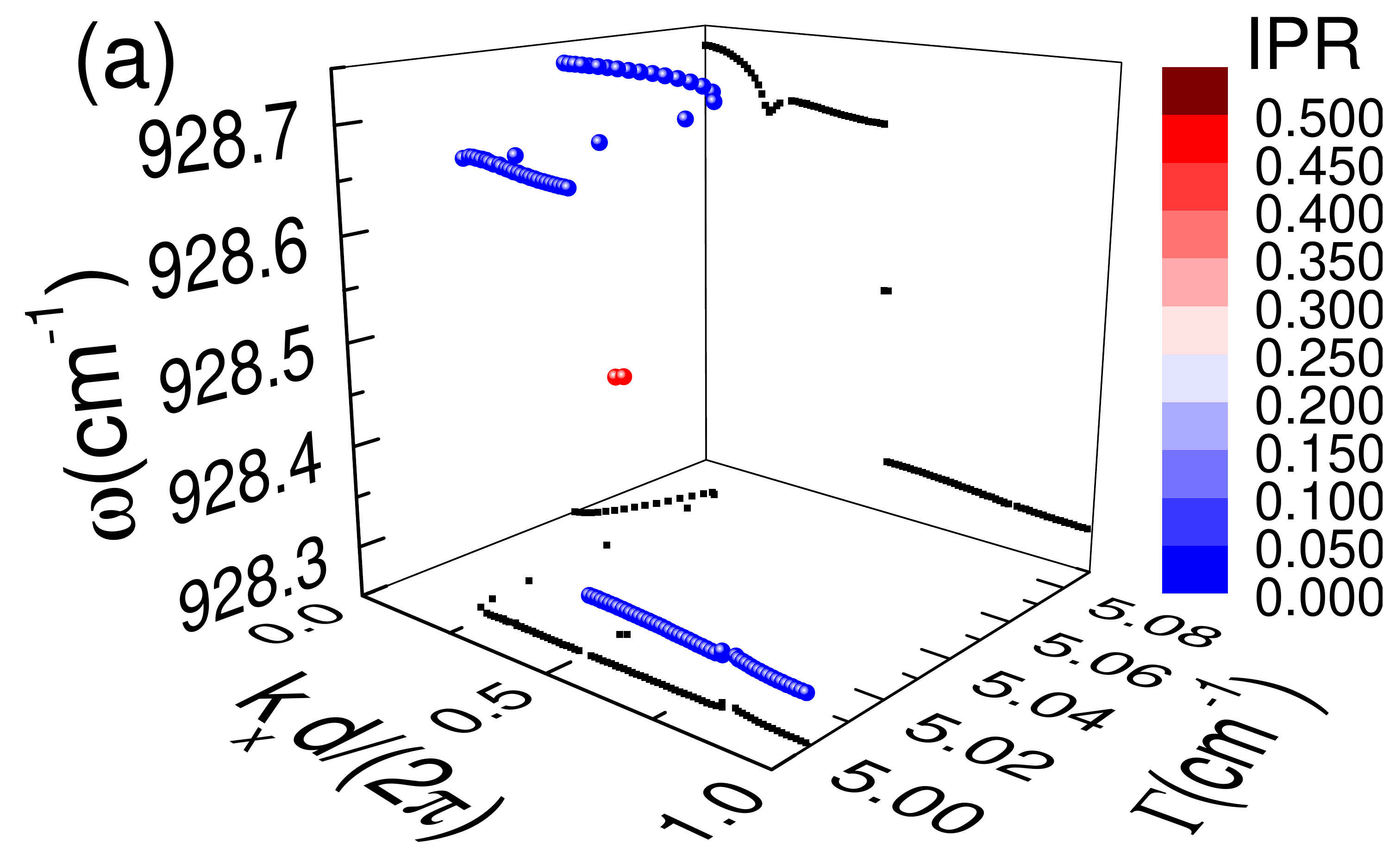}\label{d2transband}
	}
	\hspace{0.01in}
	\subfloat{
		\includegraphics[width=0.46\linewidth]{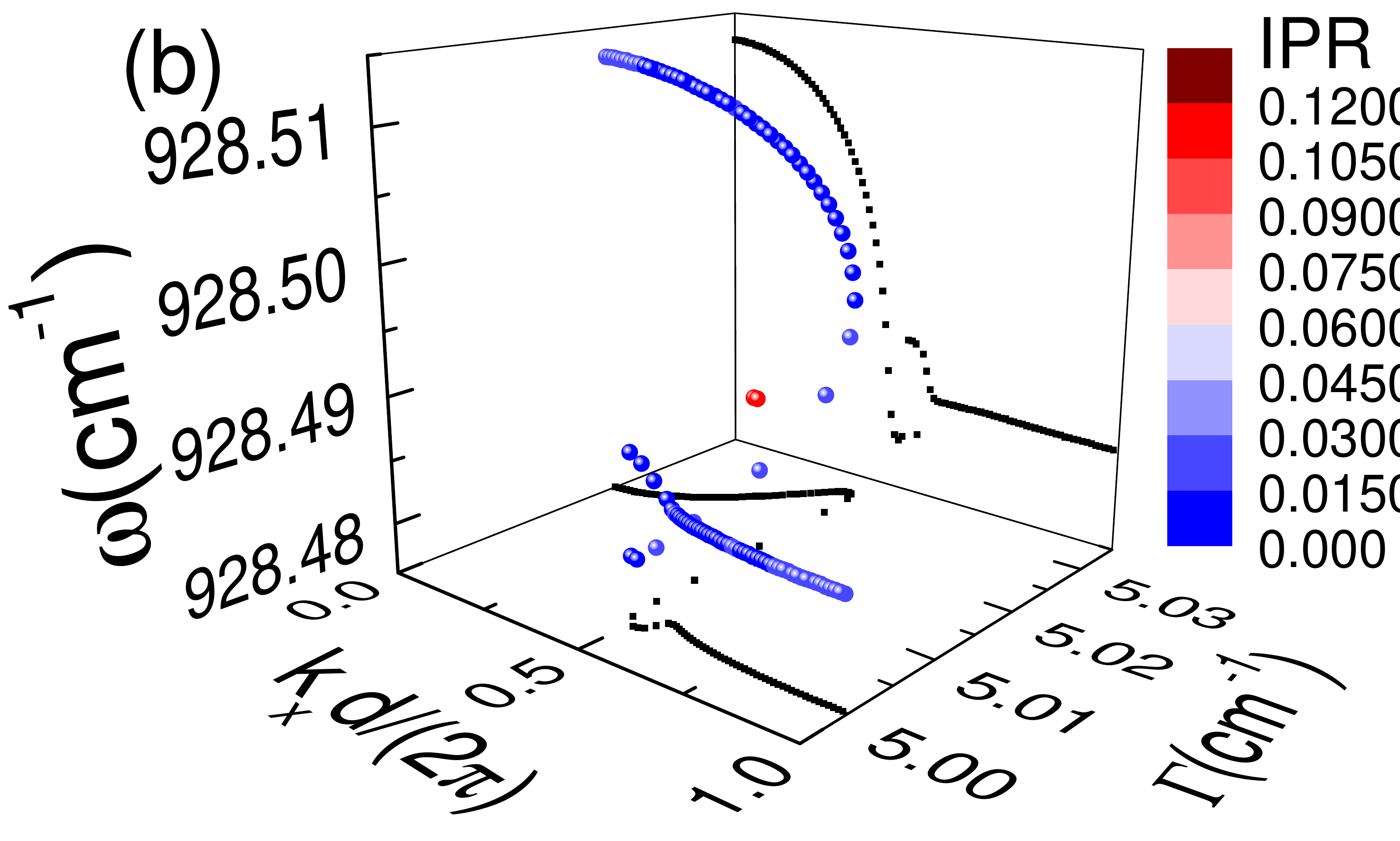}\label{d5transband}
	}
	\hspace{0.01in}
	\subfloat{
		\includegraphics[width=0.46\linewidth]{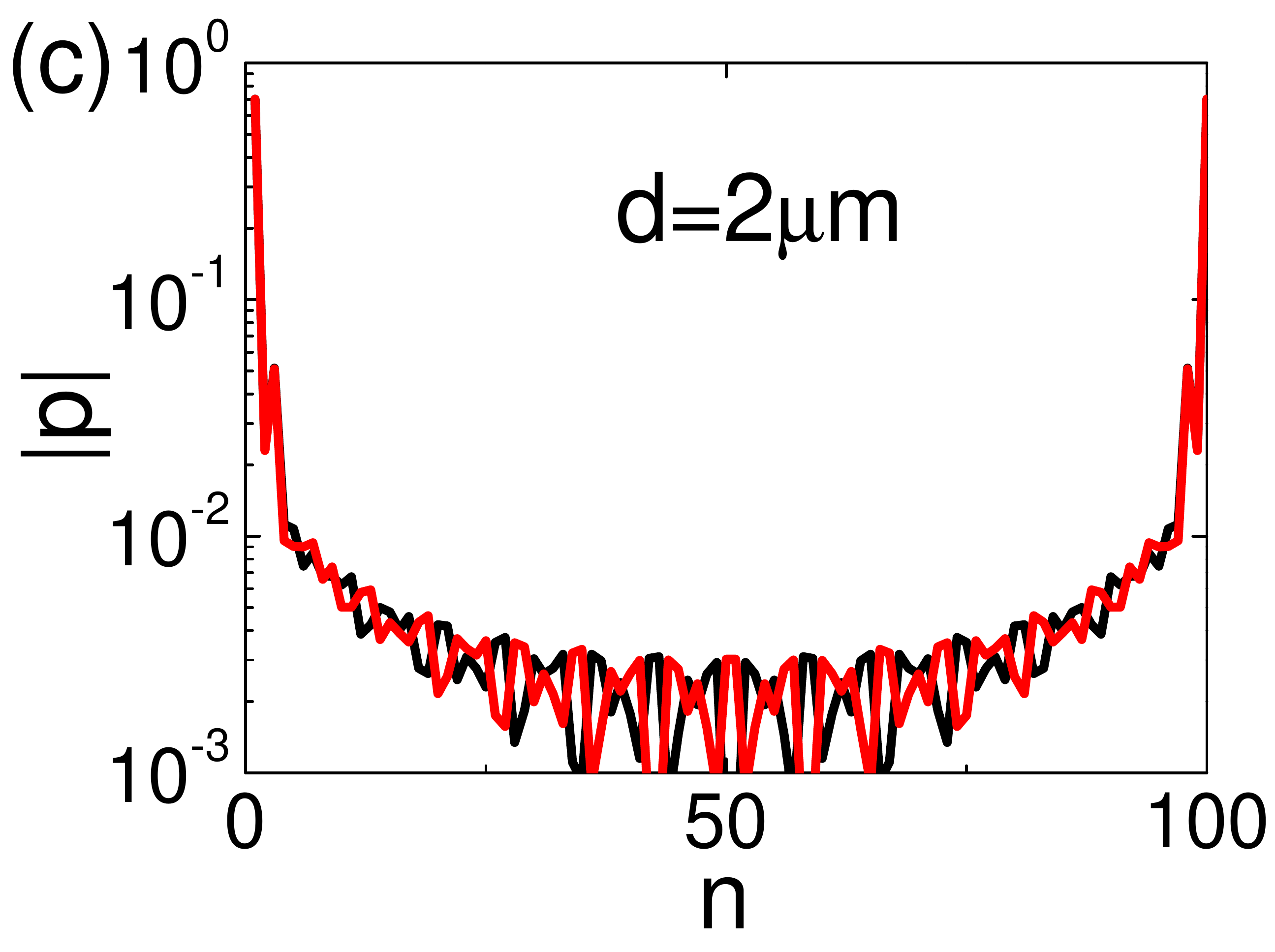}\label{d2transedgemode}
	}
	\hspace{0.01in}
	\subfloat{
		\includegraphics[width=0.46\linewidth]{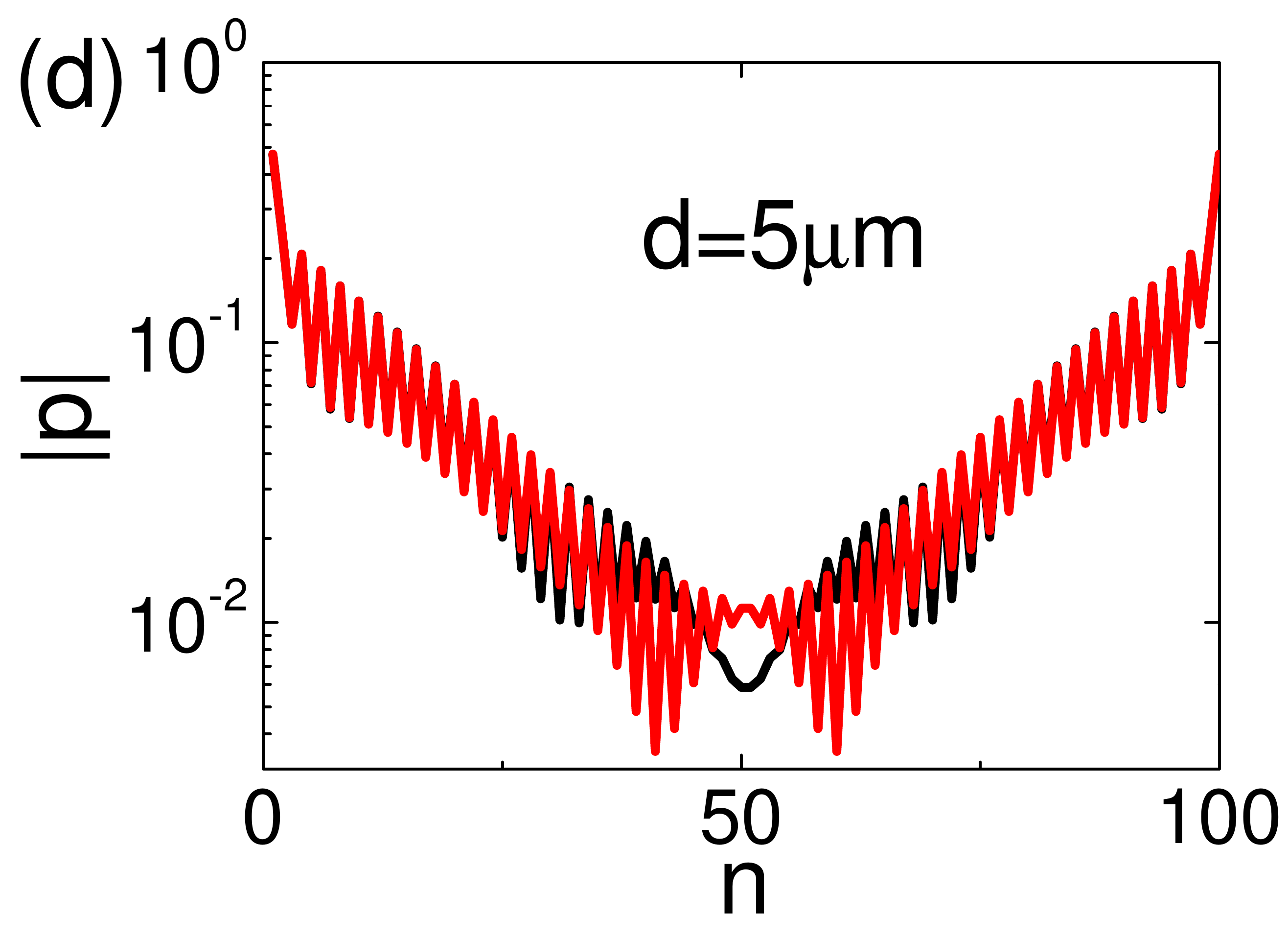}\label{d5transedgemode}
	}
	\hspace{0.01in}
	\subfloat{
	\includegraphics[width=0.46\linewidth]{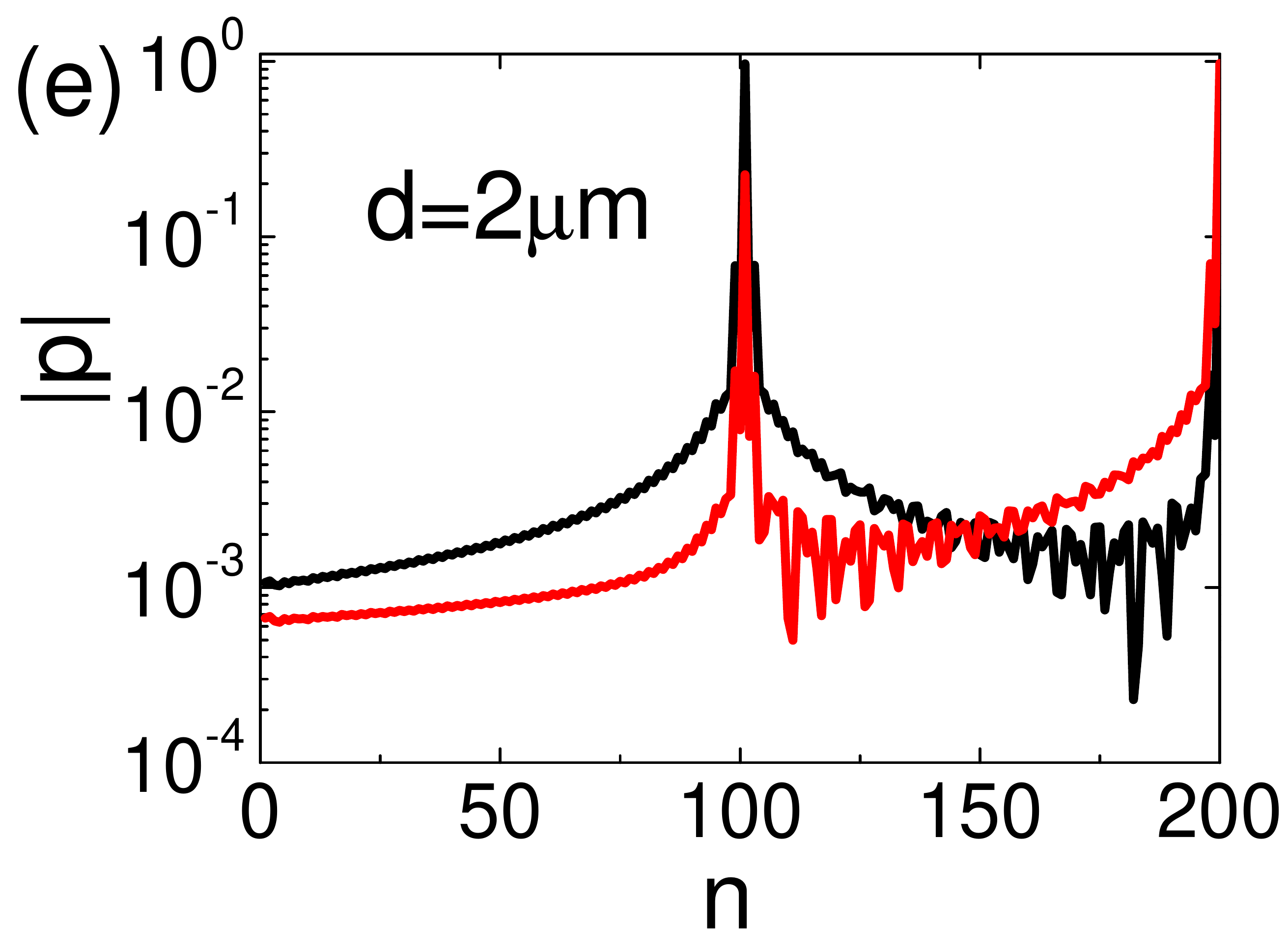}\label{interfacemodebeta07d2trans}
	}
	\hspace{0.01in}
	\subfloat{
	\includegraphics[width=0.46\linewidth]{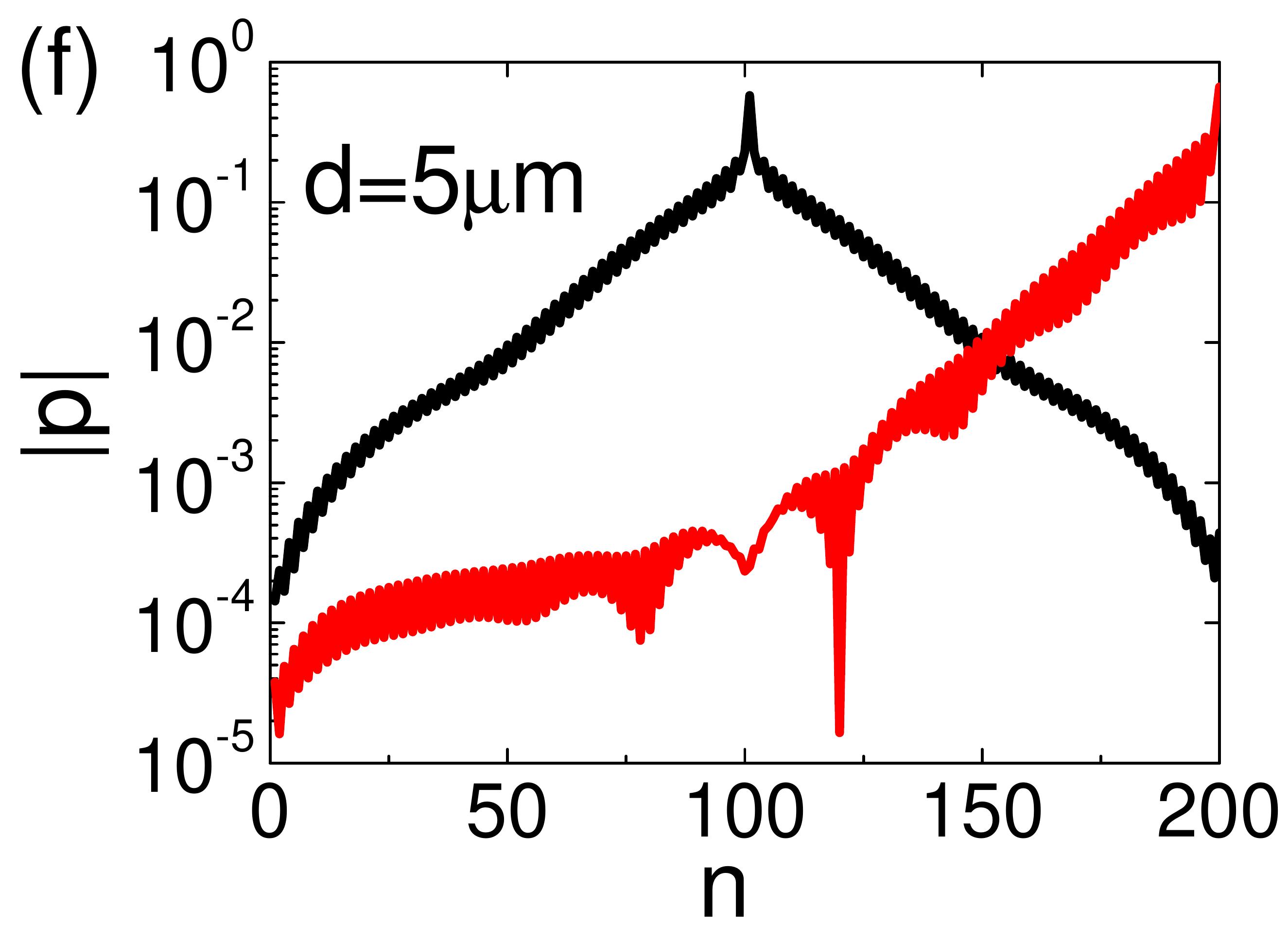}\label{interfacemodebeta07d5trans}
	}

	\caption{(a) Complex band structure of transverse eigenmodes of a dimerized chain with $N=100$ NPs under $\beta=0.7$ and $d=2\mathrm{\mu m}$. Note there are two midgap modes. (b) The same as (a) but here $d=5\mathrm{\mu m}$. (c-d) Dipole moment distributions of the midgap edge modes for the cases of (c) $d=2\mathrm{\mu m}$ and (d) $d=5\mathrm{\mu m}$. (e-f) Dipole moment distributions of the interface  modes and right edge modes for the cases of (e) $d=2\mathrm{\mu m}$ and (f) $d=5\mathrm{\mu m}$. }\label{moreedgemodetrans}
\end{figure}

To examine the bulk-boundary correspondence for this exotic observation, we compute the transverse eigenmode distributions for a finite chain with 100 NPs (50 dimers) as plotted in Fig.\ref{moreedgemodetrans}. There are two midgap modes with high IPRs near 0.5 in the chain with $d=2\mathrm{\mu m}$, $\beta=0.7$ as shown in Fig.\ref{d2transband}. The dipole moment distributions shown in Fig.\ref{d2transedgemode} confirm that these midgap modes are topologically protected edge modes. Surprisingly, we find the chain with $d=5\mathrm{\mu m}$, $\beta=0.7$ still supports midgap edge modes (whose dipole moment distributions are also given in Fig.\ref{d5transedgemode}), as shown in Fig.\ref{d5transband}, while for a chain with $\beta<0.5$ no localized edge modes can be found (not shown here). Therefore, in this circumstance, this flip of Zak phase of the Bloch band structure does not induce a topological phase transition. Moreover, for the $d=5\mathrm{\mu m}$ case, the chain with $\beta=0.7$ and that with $\beta=0.3$ seem to be topologically different, which can be verified by the localized interface modes shown in Fig.\ref{interfacemodebeta07d5trans}, where the $\beta=0.3$ chain is positioned on the left and the $\beta=0.7$ one is position on the right. It is noted that when compared to the right edge and interface modes of the $d=2\mathrm{\mu m}$ case in Fig.\ref{interfacemodebeta07d2trans}, the localization lengths of both modes in the $d=5\mathrm{\mu m}$ case are substantially longer (in units of particle numbers).

Above phenomenon appears to be a paradox, and leads us to consider whether the bulk-boundary correspondence still holds in the case of relatively strong long-range far-field interactions (when compared with short-range near-field interactions) at $d=5\mathrm{\mu m}$, as suggested by Bettles \textit{et al.} \cite{bettlesPRA2017}. To this end, we may anticipate this breakdown of bulk-boundary correspondence arises from the substantial difference in eigenmode properties between an infinitely large system and a finite system due to extremely long-range dipole-dipole interactions \cite{pocockArxiv2017}. This is because for conventional nontrivial topological systems with short-range interactions, the delocalized bulk eigenmodes are not influenced by the system boundaries, and therefore the knowledge of the Bloch band structure of an infinite system under PBC can tell us the properties of edge modes in a finite system under OBC. Hence the so-called bulk-boundary correspondence hold in that circumstance. However, the long-range interactions can result in a situation where the bulk eigenmodes are strongly affected by the system edges \cite{vodolaPRL2014,bettlesPRA2017}, and then the bulk-boundary correspondence might be violated.

\begin{figure}[htbp]
	\flushleft
	\subfloat{
		\includegraphics[width=0.47\linewidth]{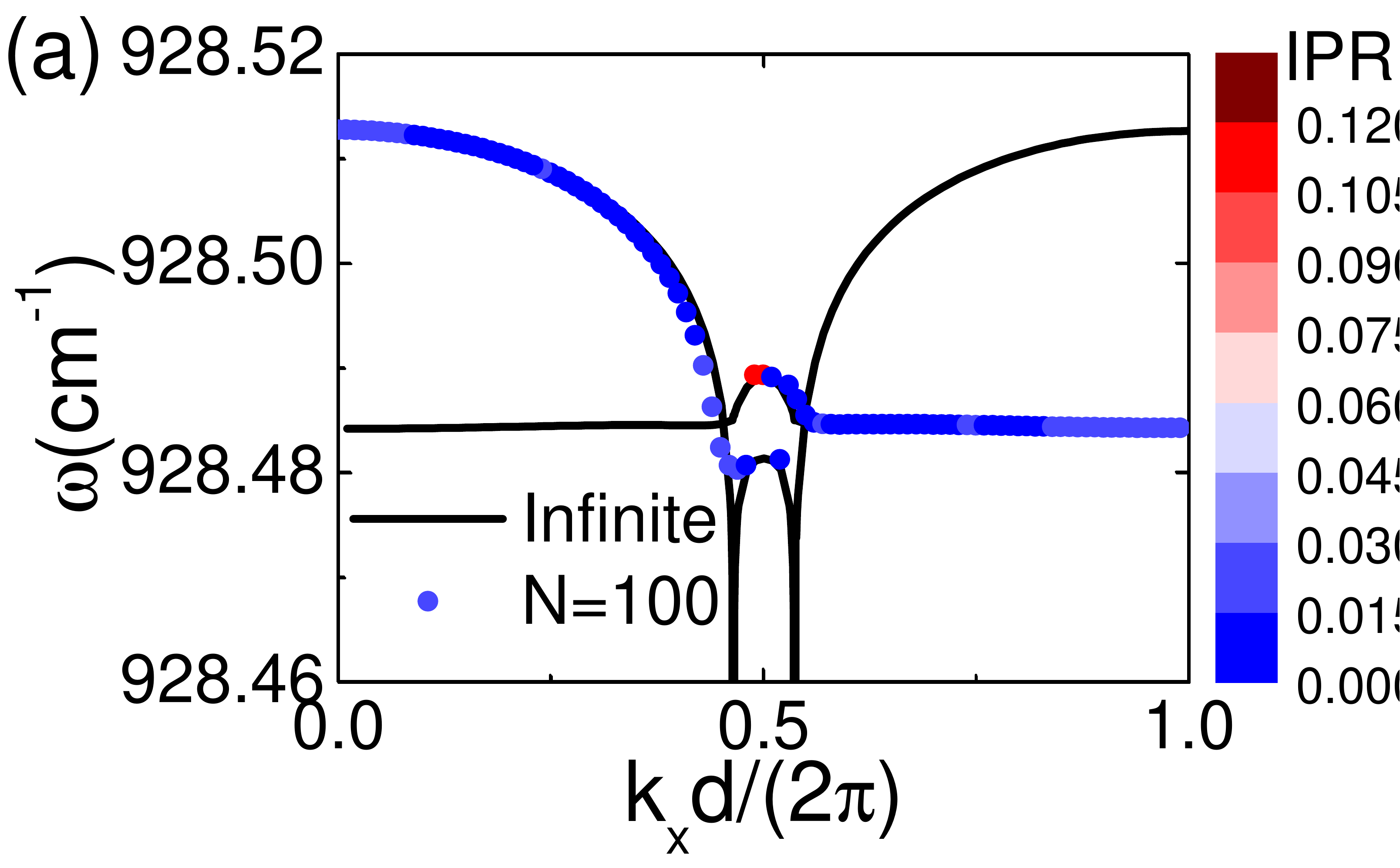}\label{comparisonreal}
	}
	\hspace{0.01in}
	\subfloat{
		\includegraphics[width=0.46\linewidth]{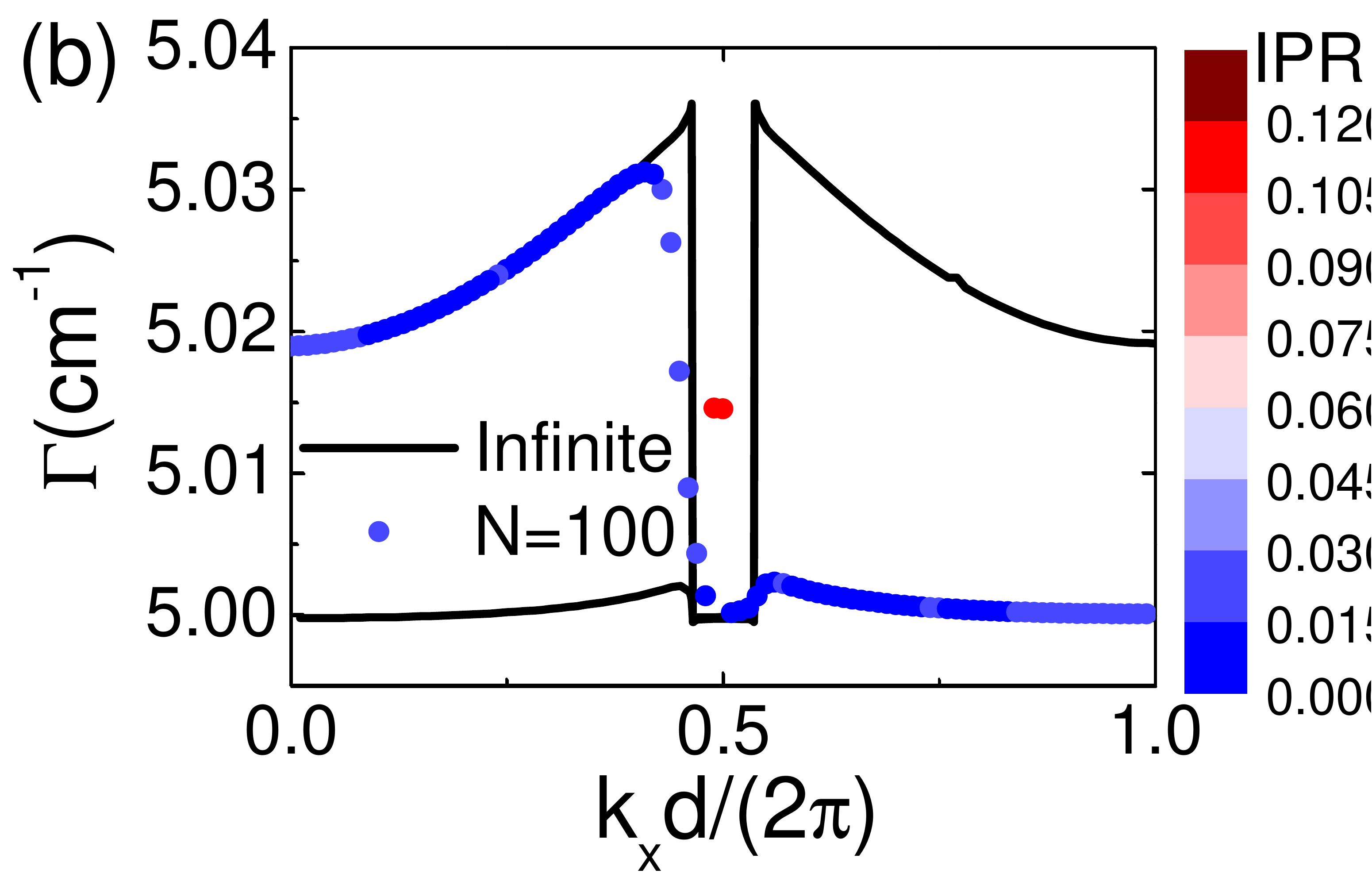}\label{comparisonimag}
	}	
	\hspace{0.01in}
	\subfloat{
		\includegraphics[width=0.46\linewidth]{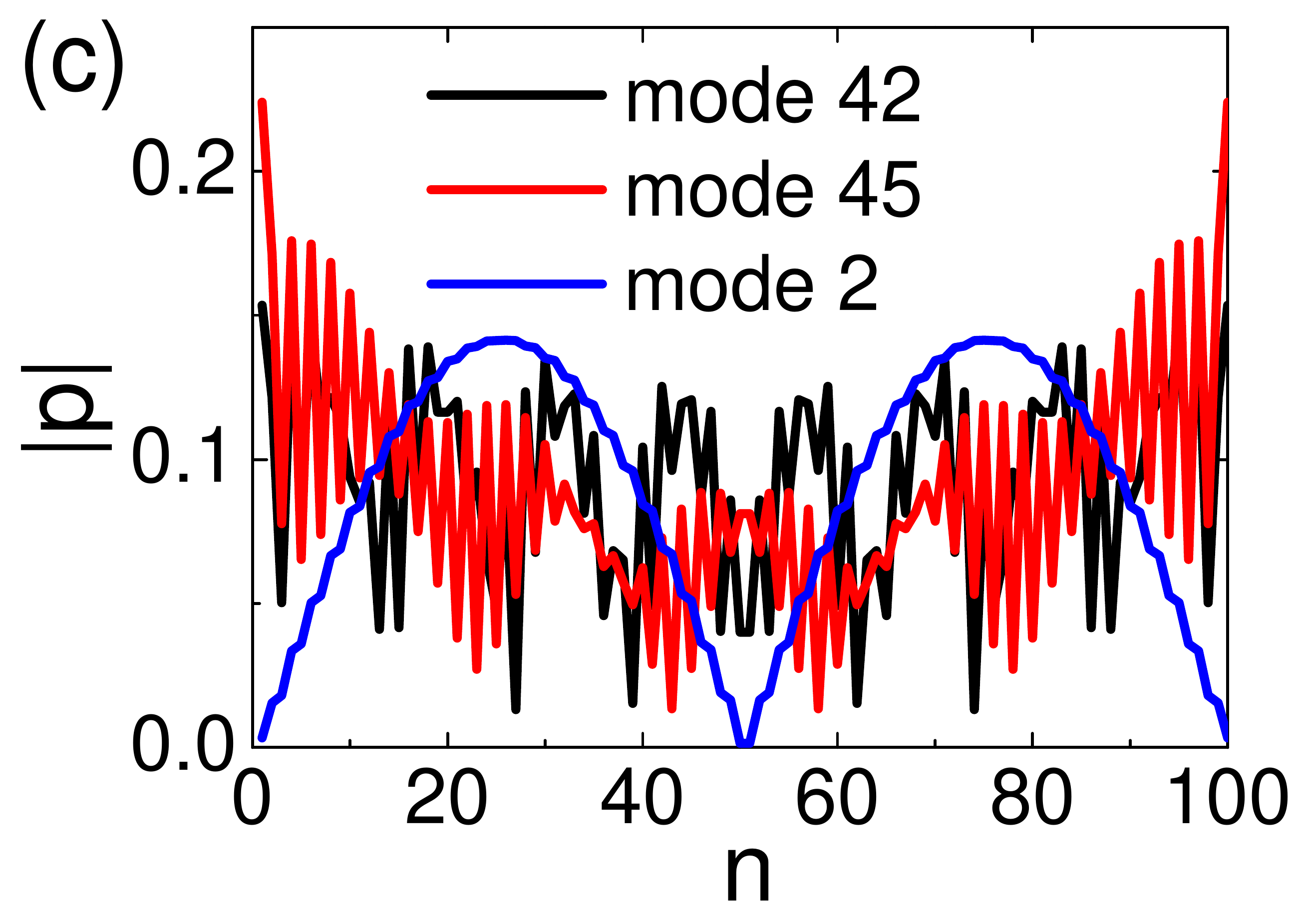}\label{edgebulkmode}
	}
	\hspace{0.01in}
	\subfloat{
		\includegraphics[width=0.46\linewidth]{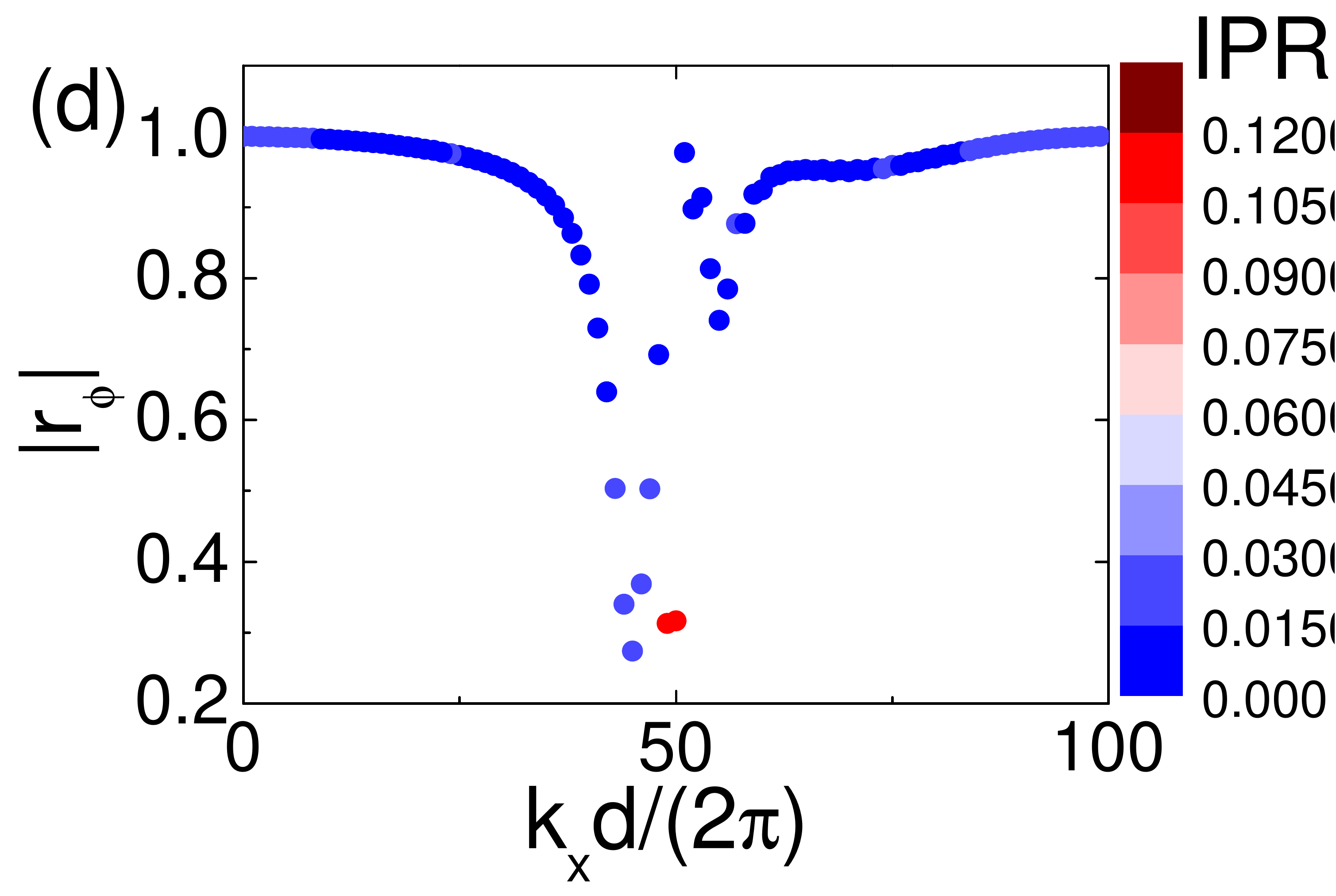}\label{phase_rigidity}
	}
	\hspace{0.01in}
	\subfloat{
		\includegraphics[width=0.46\linewidth]{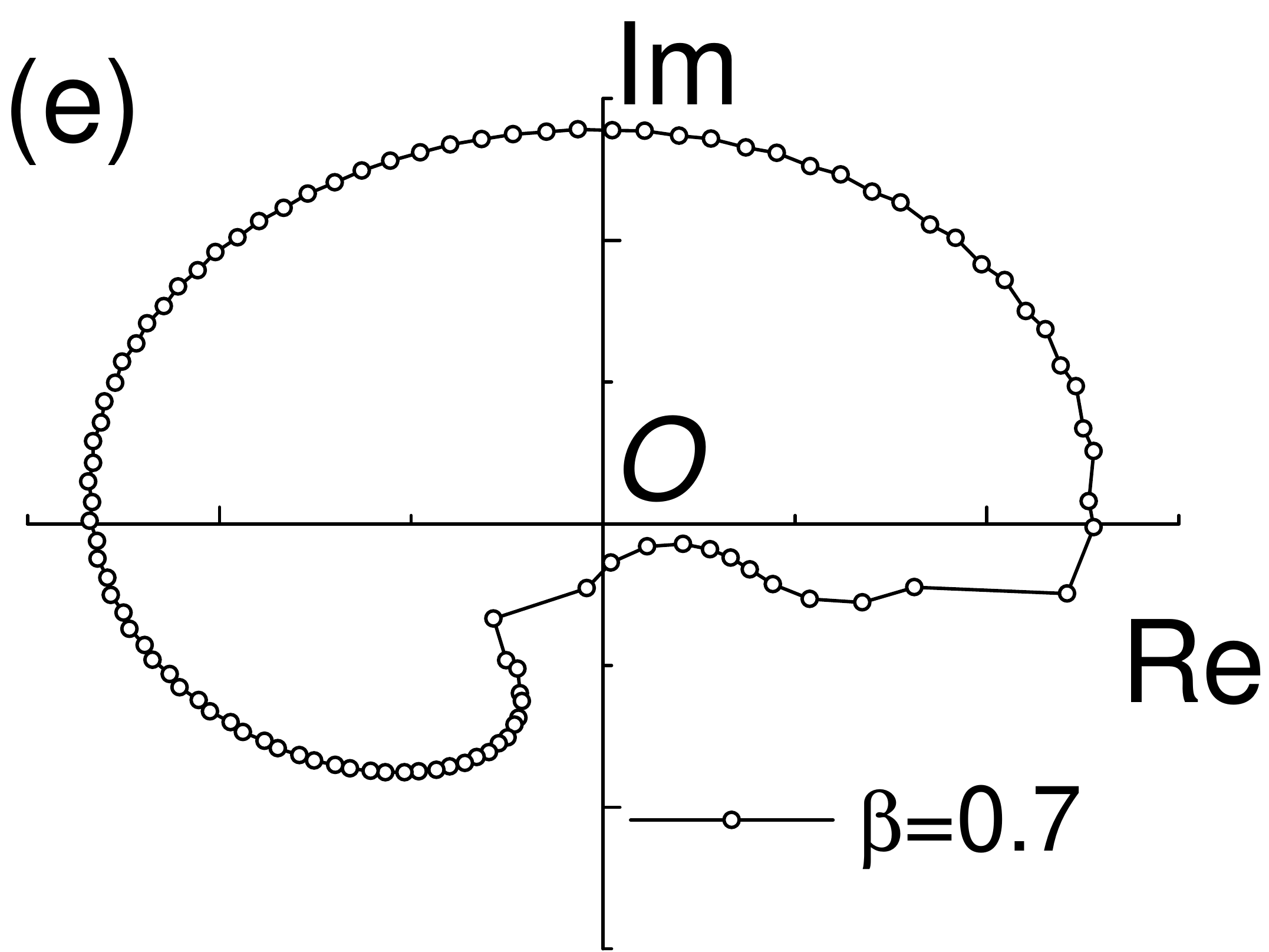}\label{modifiedwindingtransd5beta07}
	}
	\hspace{0.01in}
	\subfloat{
		\includegraphics[width=0.46\linewidth]{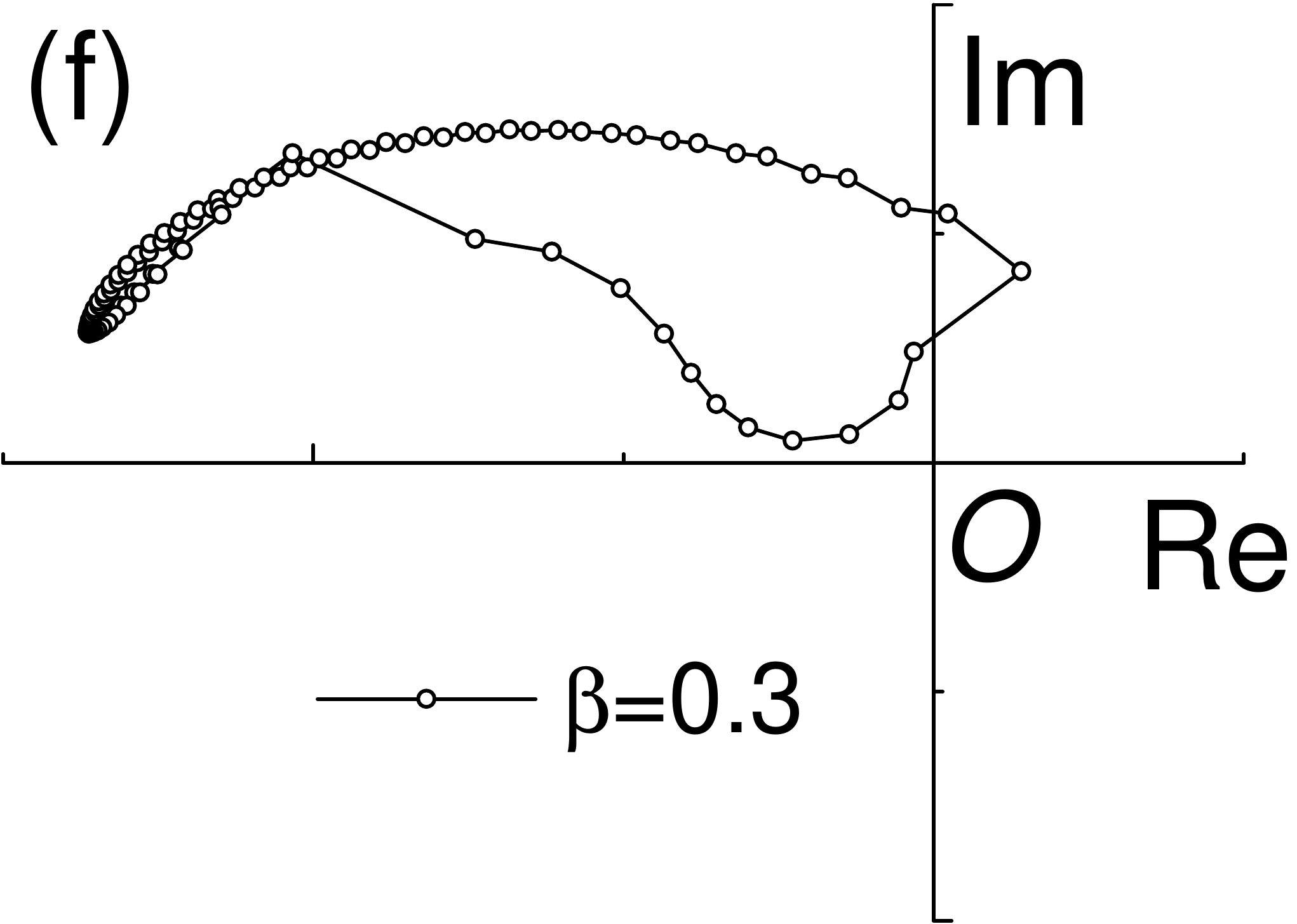}\label{modifiedwindingtransd5beta03}
	}
	\caption{The comparison of (a) real eigenfrequencies and (b) imaginary eigenfrequencies between the Bloch band structure (``Infinite") and the eigenmode distribution of a finite lattice with 100 NPs (``$N=100$"). (c) The localized bulk eigenmodes (mode numbers are 42 and 45) compared with a extended bulk eigenmode (mode number is 2). (d) Phase rigidity of the eigenmodes. System parameters are $d=5\mathrm{\mu m}$ and $\beta=0.7$. (e-f) The modified winding of $a_{12}(k_x)$ in the (e) $\beta=0.7$ and (f) $\beta=0.3$ cases for a finite chain with 100 NPs, by introducing complex wavenumbers $\tilde{k}_x$ for localized bulk eigenmodes (which are distributed near the divergences in the band structure).}
	\label{comparisond5}
\end{figure}

To demonstrate above consideration, we compare the eigenmode distribution of a finite chain (or band structure under OBC, Fig.\ref{d5transband}) with 100 NPs with the Bloch band structure of an infinitely long chain (Fig.\ref{bandstructuretransd5}), redrawing the real and imaginary parts of the band structures in Figs.\ref{comparisonreal} and \ref{comparisonimag}, respectively. Substantial deviations are observed near the divergences $k_x=\pm k$ (in this figure $k_x=2\pi/d-k$ is equivalent to $k_x=-k$), where the bulk eigenmodes can strongly couple with the free-space radiation. Those eigenmodes exhibit slightly larger IPRs than other bulk ones. In order to understand the behavior of them, we show, as an example, the dipole moments of the 42rd and 45th eigenmodes in Fig.\ref{edgebulkmode}. The momenta of these eigenmodes can be calculated from the eigenmode number according to Eq.(\ref{momentum}). It is surprisingly found that these bulk eigenmodes are weakly localized over the system edges, as compared to the second eigenmode that is extended over the chain (In Appendix \ref{longer_chain}, we provide more details of the localized bulk modes for much longer chains.). 

Here we would like to point out that this localization of bulk eigenmodes is actually a manifestation of the so-called ``non-Hermitian skin effect", which is recently intensively discussed and shown to be the key factor that breaks down the bulk-boundary correspondence in certain non-Hermitian systems \cite{yao2018edge,yao2018nonhermitian,lee2018anatomy,alvarez2018topologicalreview}. In its original definition, the non-Hermitian skin effect implies that all eigenmodes of certain non-Hermitian Hamiltonians are localized \cite{yao2018edge}. Nevertheless, in our system, the non-Hermiticity of an eigenmode depends on its momentum $k_x$, and only these eigenmodes with momenta near the singularities shows the strongest non-Hermiticity, leading to localized behavior near the edges. We call these eigenmodes as localized bulk modes. To quantify the non-Hermiticity of the eigenmodes, we calculate the measure of biorthogonality of an eigenmode $\phi$ of a finite chain, namely, the phase rigidity, defined as the ratio between biorthogonality and orthogonality \cite{alvarezPRB2018,eleuchPRA2016,wang2018topological,alvarez2018topologicalreview}
\begin{equation}
r_\phi=\frac{\langle \mathbf{p}_\phi^L|\mathbf{p}_\phi^R\rangle}{\langle \mathbf{p}_\phi^R|\mathbf{p}_\phi^R\rangle},
\end{equation}
where $\langle\mathbf{p}_\phi^L|$ is the left eigenvector of the  eigenmode. For Hermitian Hamiltonians, this quantity is exactly 1 for all eigenstates, while for a non-Hermitian Hamiltonian where $|\mathbf{p}_\phi^L\rangle\neq|\mathbf{p}_\phi^R\rangle$, $|r_\phi|$ is generally less than 1, and at and near an exceptional point (EP) it takes its minimum value $r_\phi\rightarrow0$. In Fig.\ref{phase_rigidity} we show the phase rigidity of the eigenmodes as a function of the momentum $k_x$. It can be regarded as a measure of Hermiticity, and a smaller $|r_\phi|$ indicates stronger non-Hermiticity of the eigenmode \cite{alvarezPRB2018,eleuchPRA2016,wang2018topological}. Evidently, those eigenmodes near the singularity exhibit very small phase rigidities and thus are strongly non-Hermitian.

So far, we have verified it is the non-Hermitian skin effect brought by the long-range interactions that leads to the breakdown of the conventional bulk-boundary correspondence. In order to re-establish the bulk-boundary correspondence for a finite lattice, here we resort to a discrete version of complex Zak phase. In particular, by first invoking the Bloch theorem, the discrete expression of the matrix elements in the Bloch Hamiltonian for a finite lattice can be obtained by replacing $\Phi(z,s,a)$ with $\sum_{n=0}^{N/2} z^n/(n+a)^s$ in Eqs.(\ref{a12T}) and (\ref{a21T}), where $N$ is the number of NPs. On the other hand, for the localized bulk eigenmodes, applying the Bloch theorem is not appropriate. To qualitatively fix this problem, following from the method proposed in Ref.\cite{yao2018edge}, we approximately introduce a complex momentum $\tilde{k}_x=k_x+ik_x''$ for the localized bulk modes near the singularities, where the imaginary part $k_x''$ accounts for the highly localized bulk eigenmodes. Note that owing to the power-law dipole-dipole interactions, all localized modes (including bulk and edge modes) are not rigorously exponentially localized but follow a power-law localization behavior (at least in the long range, see Ref.\cite{vodolaPRL2014} and Appendix \ref{longer_chain}). For the convenience of qualitatively demonstrating our idea, we use this simplified assumption. In Figs.\ref{modifiedwindingtransd5beta07} and \ref{modifiedwindingtransd5beta03}, we give the winding of $a_{12}(k_x)$ over the origin in the complex plane by simply introducing a small imaginary momentum $k_x''/(2\pi/d)=0.02$ for the localized bulk eigenmodes. It is hence found that the winding procedure eliminates the abrupt singularities shown in Figs.\ref{windingbeta07d5trans} and \ref{windingbeta03d5trans}. In this situation, the modified, or non-Bloch \cite{yao2018edge}, complex Zak phase $\tilde{\theta}_\mathrm{Z}^T$ corresponding to the modified $a_{12}(\tilde{k}_x)$ and $a_{21}(\tilde{k}_x)$ recovers the conventional SSH model behavior, i.e.,
\begin{equation}
\tilde{\theta}_\mathrm{Z}^T=\begin{cases}
\pi &{\beta=0.7, d=5\mathrm{\mu m}},\\
0 & {\beta=0.3, d=5\mathrm{\mu m}},
\end{cases}
\end{equation}
where the hat over the variables $\tilde{}$ denotes the modified quantities considering the non-Hermitian skin effect. This conclusion can be easily generalized to all $\beta>0.5$ and $\beta<0.5$ cases with a lattice constant beyond the topological phase transition point.

As a consequence, we have qualitatively rebuilt a non-Bloch bulk-boundary correspondence for the finite lattice with strong long-range interactions, although an exact construction still needs a detailed examination of the behaviors of localized bulk modes. A question not answered here is that whether a sufficiently long chain with $d=5\mathrm{\mu m}$ and $\beta=0.7$ can exhibit a topological phase transition from $\theta_\mathrm{Z}^T=\pi$ to $\theta_\mathrm{Z}^T=0$ as predicted by the Bloch band structure and corresponding complex Zak phase, because according to our numerical calculations performed over hundreds of thousands NPs no transition was found, which might be because the investigated chains are still not long enough, while longer chains are not accessible due to our limited computation resources \cite{pocockArxiv2017}. A more probable situation is that for arbitrarily long finite chains, the conventional bulk-boundary correspondence is always invalid since the localized bulk modes can always emerge and break this correspondence, as implied by the results in Appendix \ref{longer_chain}.

\section{Field enhancement and local density of states (LDOS)}
\begin{figure}[htbp]
	\flushleft
	\subfloat{
		\includegraphics[width=\linewidth]{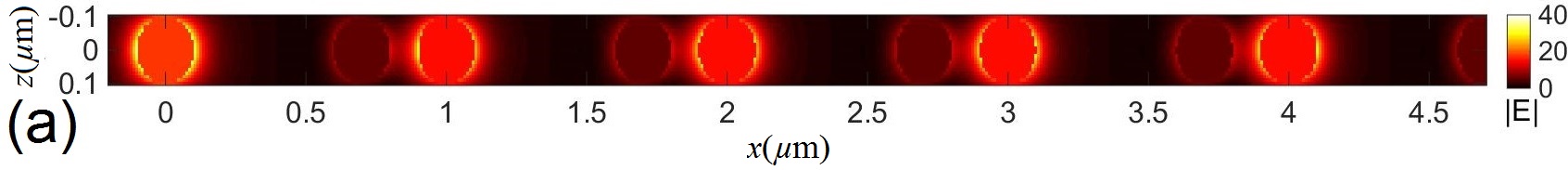}\label{Efield_top}
	}
	\hspace{0.01in}
	\subfloat{
		\includegraphics[width=\linewidth]{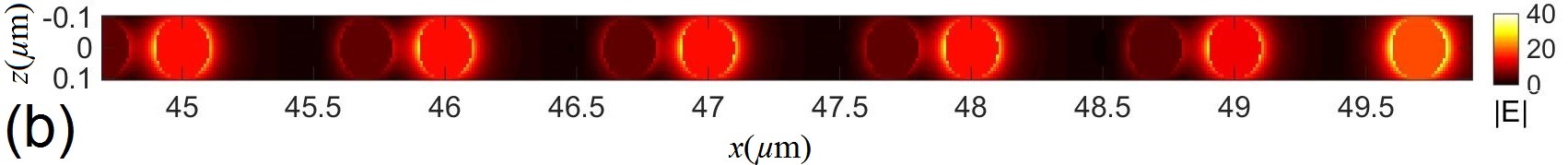}\label{Efield_end}
	}
	\hspace{0.01in}
	\subfloat{
		\includegraphics[width=0.45\linewidth]{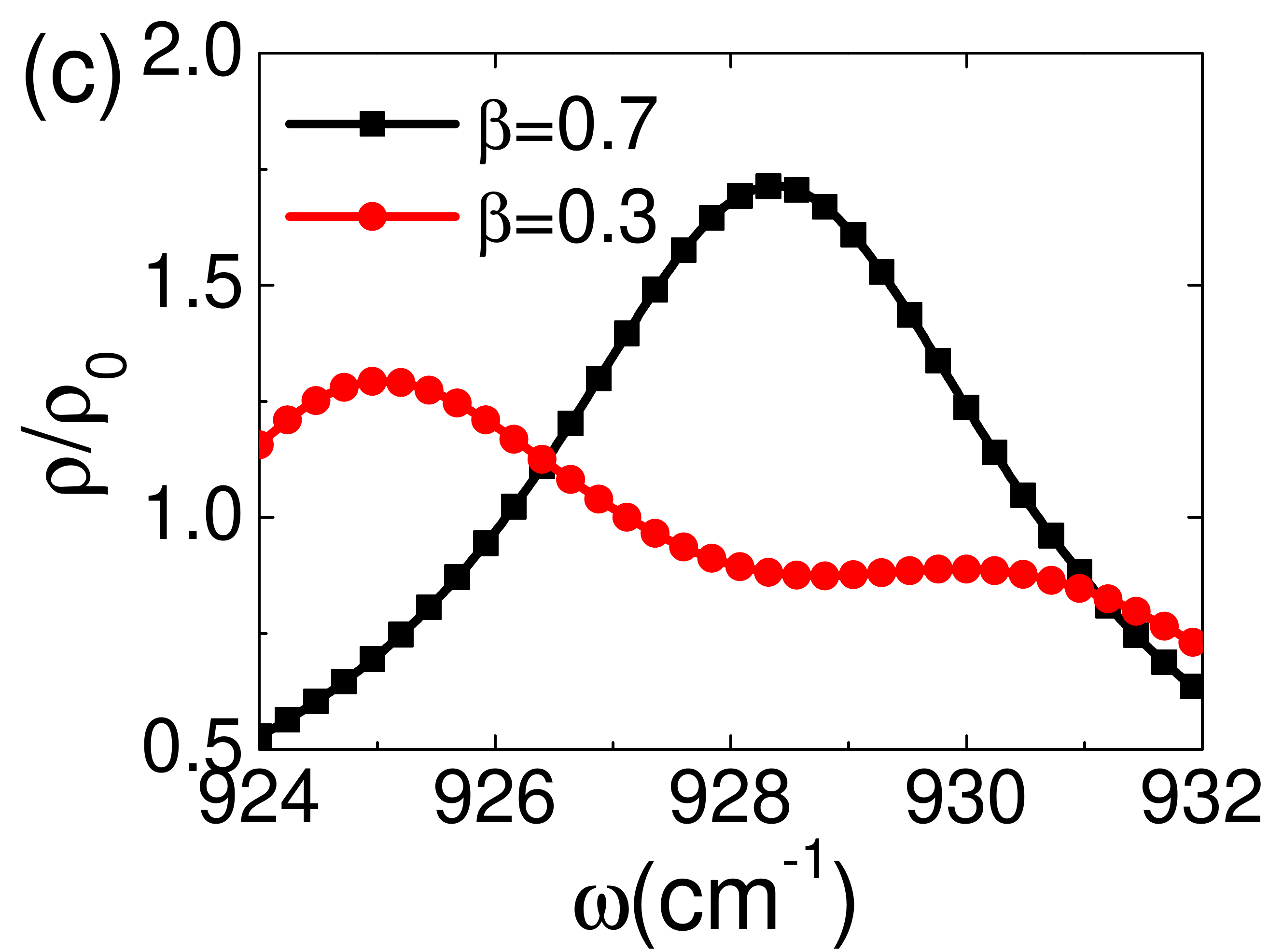}\label{LDOS}
	}
	\subfloat{
		\includegraphics[width=0.45\linewidth]{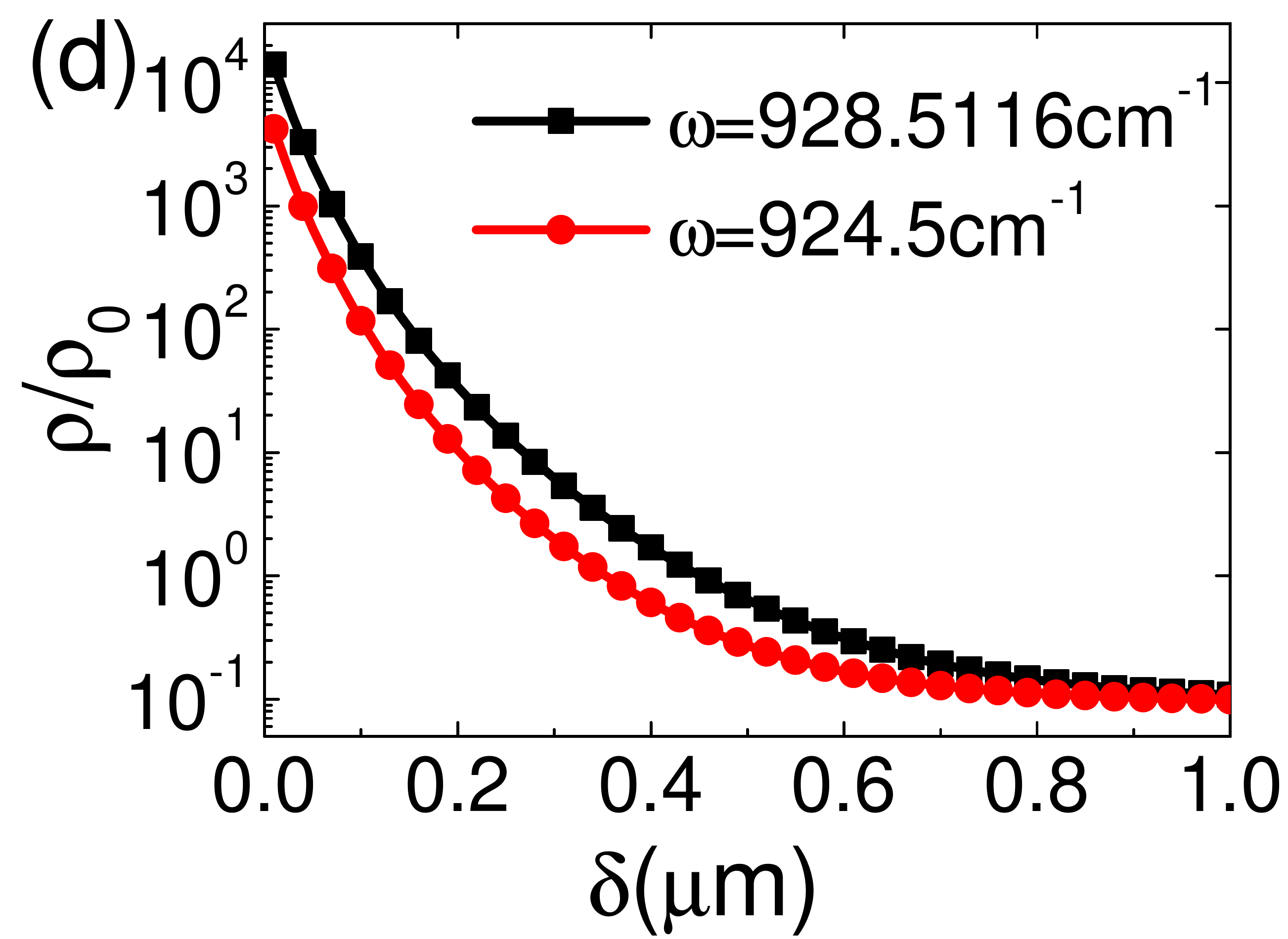}\label{LDOSwithd}
	}
	\caption{The distribution of the electric field $|\mathbf{E}|$ near the left (a) and right (b) edges of a dimerized SiC NP chain when a topological edge mode is excited using an evanescent plane wave with a wavevector $\mathbf{k}=\hat{x}\pi/d$ polarized along the $x$-axis, where the excitation frequency is $\omega=928.5116\mathrm{cm}^{-1}$. (c) Total LDOS at $x_s=-0.5\mathrm{\mu m}$ as a function of the angular frequency, for both topologically nontrivial ($\beta=0.7$) and trivial ($\beta=0.3$) chains, where $\rho_0=\omega^2/(\pi^2c^3)$ is the LDOS in the free space. (d) The distance dependence of the LDOS near the left-edge of the topologically nontrivial ($\beta=0.7$) chain (in logarithmic scale).}\label{figexcitation}
\end{figure}

Finally, we briefly investigate the excitation and the electromagnetic field enhancement of these TPhPs. In Figs.\ref{Efield_top} and \ref{Efield_end}, we show the electric field distribution near both edges of a dimerized SiC NP chain when a topological edge mode is excited using an evanescent plane wave with a wavevector $\mathbf{k}=\hat{x}\pi/d$, that is the wavenumber of the midgap mode, polarized along the $x$-axis, where the excitation frequency is $\omega=928.5116\mathrm{cm}^{-1}$, just lying in the bandgap. The SiC NP chain is composed of 100 identical NPs whose radius is $a=0.1\mathrm{\mu m}$, where the lattice constant is $d=1\mathrm{\mu m}$ and the dimerization parameter is $\beta=0.7$. The first NP is centered at $x=0\mathrm{\mu m}$, and the last NP is located at $x=50\mathrm{\mu m}$. It is clearly observed in Figs.\ref{Efield_top} and \ref{Efield_end} that the electric field is enormously enhanced in and near the NPs in the edges of the chain. In experiment, such an excitation method is available by using quantum dots \cite{sapienzaScience2010} as well scanning near-field optical microscopy (SNOM) tips \cite{alfaroNaturecomms2017} whose radiation field contains large momentum components .

Furthermore, we calculate the optical local density of states (LDOS) near the edge of the chain. The LDOS can also be obtained under the framework of the coupled-dipole equations, while the incident field is replaced by that emitted from a point source \cite{Pierrat2010,Caze2013}:
\begin{widetext}
\begin{equation}\label{coupled-dipole-LDOS}
\mathbf{p}_j(\omega)=\frac{\omega^2\alpha(\omega)}{c^2}\left[\mathbf{G}_0(\omega,\mathbf{r}_j,\mathbf{r}_s)\mathbf{p}_s+\sum_{i=1,i\neq j}^{N}\mathbf{G}_0(\omega,\mathbf{r}_j,\mathbf{r}_i)\mathbf{p}_i(\omega)\right],
\end{equation}
\end{widetext}
where $\mathbf{p}_s$ is the dipole moment of the emitting point source whose position is $\mathbf{r}_s$. After calculating the electromagnetic responses of all NPs based on Eq.(\ref{coupled-dipole-LDOS}), the total scattered field of the NP chain at an arbitrary position outside the NPs is computed as \cite{Pierrat2010,Caze2013}
\begin{equation}
\mathbf{E}_s(\mathbf{r})=\frac{\omega^2}{c^2}\sum_{i=1}^{N}\mathbf{G}_0(\omega,\mathbf{r},\mathbf{r}_i)\mathbf{p}_i(\omega).
\end{equation}
From the scattered field, it is straightforward to obtain the full Green's function with respect to the point source at $\mathbf{r}_s$ as $\mathbf{G}(\omega,\mathbf{r},\mathbf{r}_s)=\mathbf{G}_0(\omega,\mathbf{r},\mathbf{r}_s)+\mathbf{S}(\omega,\mathbf{r},\mathbf{r}_s)$ \cite{Pierrat2010,Caze2013} . Here the elements in scattering field tensor $\mathbf{S}(\omega,\mathbf{r},\mathbf{r}_s)$ can be calculated through the relation $\mathbf{E}_s(\mathbf{r})=\mathbf{S}(\omega,\mathbf{r},\mathbf{r}_s)\mathbf{p}_s$ by aligning the dipole moment of the point source along different axes. Afterwards LDOS is obtained from the full Green's function as
\begin{equation}
\rho(\mathbf{r}_s,\omega)=\frac{2\omega}{\pi c^2}\mathrm{Im}\left[\mathrm{Tr}\mathbf{G}(\omega,\mathbf{r}_s,\mathbf{r}_s)\right].
\end{equation}
Note this total LDOS contains all electromagnetic eigenmodes including both longitudinal and transverse ones. Fig.\ref{LDOS} shows the total LDOS at $x_s=-0.5\mathrm{\mu m}$ as a function of the angular frequency, where the point source is located along the negative $x$-axis and $x_s$ is its $x$-coordinate. It is not surprising that the LDOS is greatly enhanced in the bandgap frequency for the topologically nontrivial chain ($\beta=0.7$), while in the topologically trivial chain ($\beta=0.3$), the LDOS is suppressed to be lower than the vacuum value ($\rho_0=\omega^2/(\pi^2c^3)$). Hence this enhancement in LDOS is attributed to the topologically protected midgap TPhPs. In Fig.\ref{LDOSwithd}, we plot the distance dependence of the LDOS near the left-edge of the chain, where the distance is given by $\delta=-0.1\mathrm{\mu m}-x_s$. Two frequencies are investigated, including the frequency of the midgap edge mode ($\omega=928.5116\mathrm{cm}^{-1}$) and a frequency in the bulk band ($\omega=924.5\mathrm{cm}^{-1}$). The LDOS at $\omega=928.5116\mathrm{cm}^{-1}$ is always substantially larger thant that at $\omega=924.5\mathrm{cm}^{-1}$ all over the distance range (Note the logarithmic scale). As a result, we have verified that the existence of topological phonon polaritons gives rise to an appreciable enhancement to the electromagnetic field and LDOS. Moreover, this enhancement in return provides an indicator of TPhPs, and can aid us to experimentally locate and detect those modes.

\section{Conclusion}
In conclusion, we achieve topologically protected phonon polaritons by constructing 1D dimerized silicon SiC NP chains, which mimic the celebrated SSH model. However, different from the conventional SSH model, we carry out this study beyond the nearest-neighbor approximation and taking all near-field and far-field dipole-dipole interactions into account. For longitudinal modes, despite the consequences of non-Hermiticity and the breaking of chiral symmetry brought by this treatment, we show that such dimerized chains can still support topological protected midgap modes, i.e., TPhPs,  We reveal that in this non-Hermitian system, the band topology can be characterized by a quantized complex Zak phase, which indicates a topological phase transition point of $\beta=0.5$, like its Hermitian counterpart. By analyzing the eigenmodes of a finite chain as well as their inverse participation ratios (IPRs), we find topologically protected midgap modes and unequivocally verify the principle of bulk-boundary correspondence. 

For transverse modes, we further discover a topological phase transition in an infinitely long chain, brought by the increase of lattice constant due to the presence of strong long-range far-field dipole-dipole interactions decaying with the distance $r$ as $1/r$. However, for a finite chain, we show the non-Hermitian skin effect derived from the strong long-range far-field interactions leads to the breakdown of the bulk-boundary correspondence. In this sense, we actually give a realistic example for the non-Hermitian skin effect, which in our case is momentum-dependent as measured by the phase rigidity. Furthermore, by incorporating the effect of localized bulk eigenmodes and proposing a modified complex Zak phase for a finite lattice, we still reconstruct the bulk-boundary correspondence for finite chains, which recovers the topological  behavior of the conventional SSH model. Our systematic investigation provides profound implications to the study of non-Hermitian topological physics and quantum mechanical models with long-range interactions. Finally, we demonstrate the excitation of the topological phonon polaritons and show their enhancement to the photonic LDOS. These TPhPs offer an efficient tool for enhancing light-matter interaction in the mid-infrared. 

\begin{acknowledgments}
We thank the financial support from the National Natural Science Foundation of China (Nos. 51636004, 51476097), Shanghai Key Fundamental Research Grant (Nos. 18JC1413300, 16JC1403200), and the Foundation for Innovative Research Groups of the National Natural Science Foundation of China (No.51521004).
\end{acknowledgments}

\appendix
\section{Validation of the radiative correction of polarizability}\label{rad_correct_append}
\begin{figure}[htbp]
	\centering
	\includegraphics[width=0.6\linewidth]{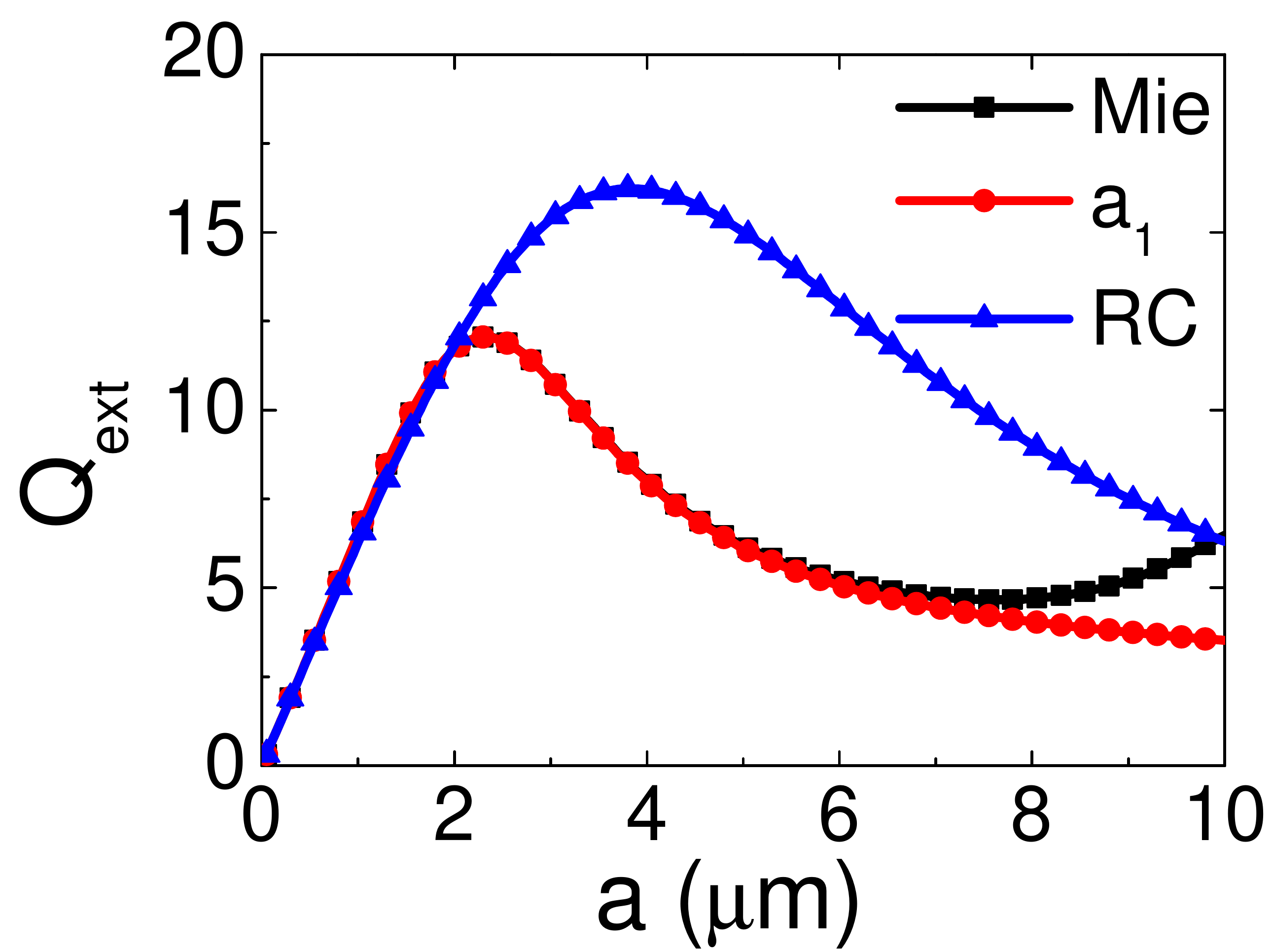}
	\caption{Extinction efficiency $Q_\mathrm{ext}$ of a single SiC NP as a function of radius $a$, the angular frequency is $\omega=928\mathrm{cm}^{-1}$. Results calculated from full Mie theory is compared with the electric dipole approximation using the polarizabilities from Eq.(\ref{alphamie}) (``$a_1$") and radiative correction Eq.(\ref{radiativecorrection}) (``RC").}\label{singlerc}
\end{figure}

The EM response of an individual SiC NP throughout this paper is described by the dipole polarizability with the radiative correction, which is given in Eq.(\ref{radiativecorrection}). To examine the validity of this approximation, we compare the result of this approximation with Mie theory and the result of electric dipole approximation using the first Mie coefficient as \cite{markelJPB2005} \begin{equation}\label{alphamie}
\alpha(\omega)=\frac{6\pi i}{k^3}\frac{m^2j_1(mx)[xj_1(x)]'-j_1(x)[mxj_1(mx)]'}{m^2j_1(mx)[xh_1(x)]'-h_1(x)[mxj_1(mx)]'},
\end{equation} 
where $k=\omega/c$ is the wavenumber in vacuum, $x=ka$ is the size parameter, and $m=\sqrt{\varepsilon_p}$ is the complex refractive index of SiC. $j_1$ and $h_1$ are first-order spherical Bessel functions and Hankel functions, respectively. The comparison is shown in Fig.\ref{singlerc} for the near-resonance angular frequency $\omega=928\mathrm{cm}^{-1}$. It is seen that for $a\lesssim2\mathrm{\mu m}$, the radiative correction agrees well with the full Mie solution.

\section{The effect of NP radius}\label{effect_of_radius}
\begin{figure}[htbp]
	\centering
	\subfloat{
		\includegraphics[width=0.46\linewidth]{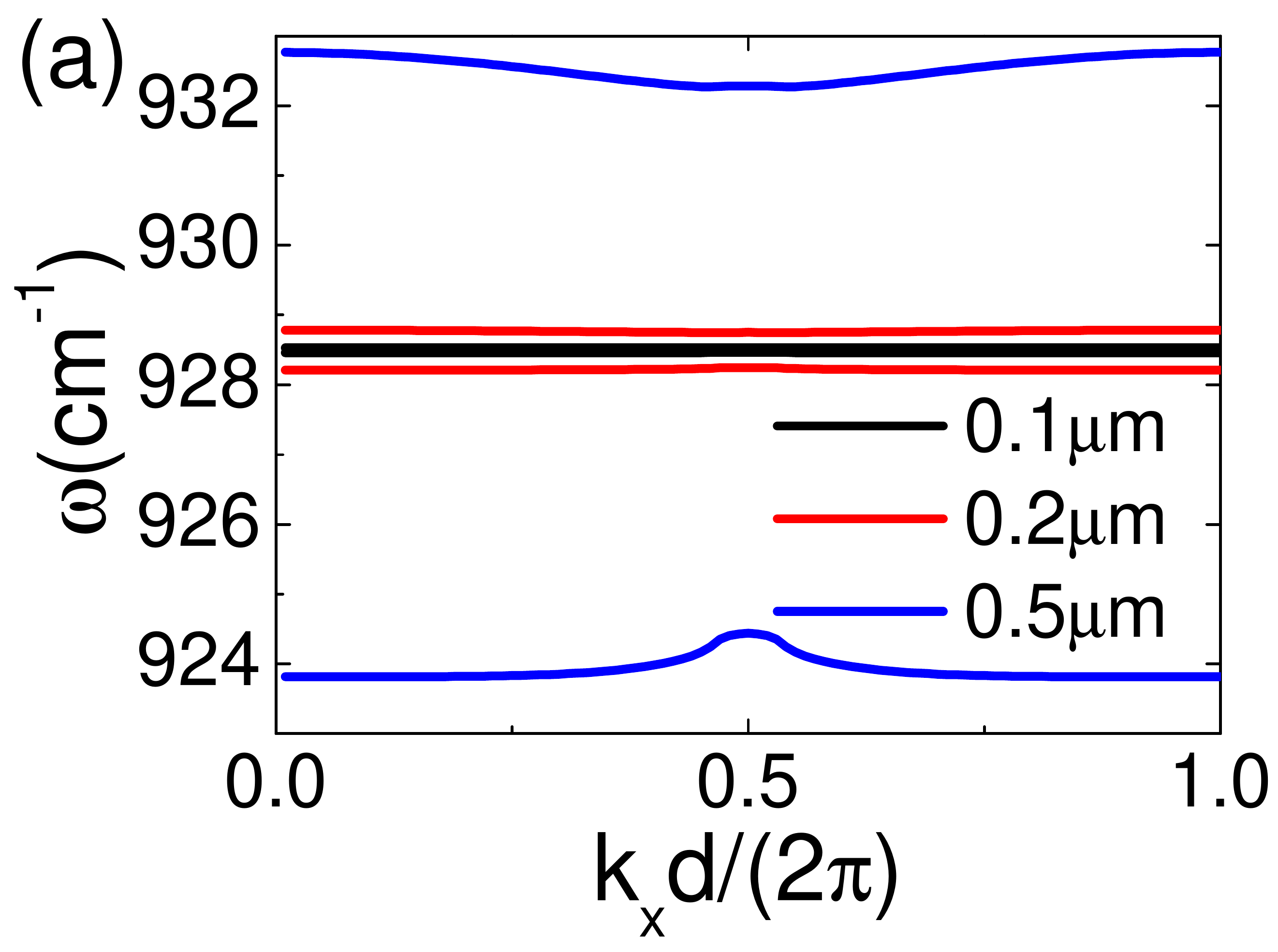}\label{sizeeffectlongreal}
	}
	\hspace{0.01in}
	\subfloat{
		\includegraphics[width=0.46\linewidth]{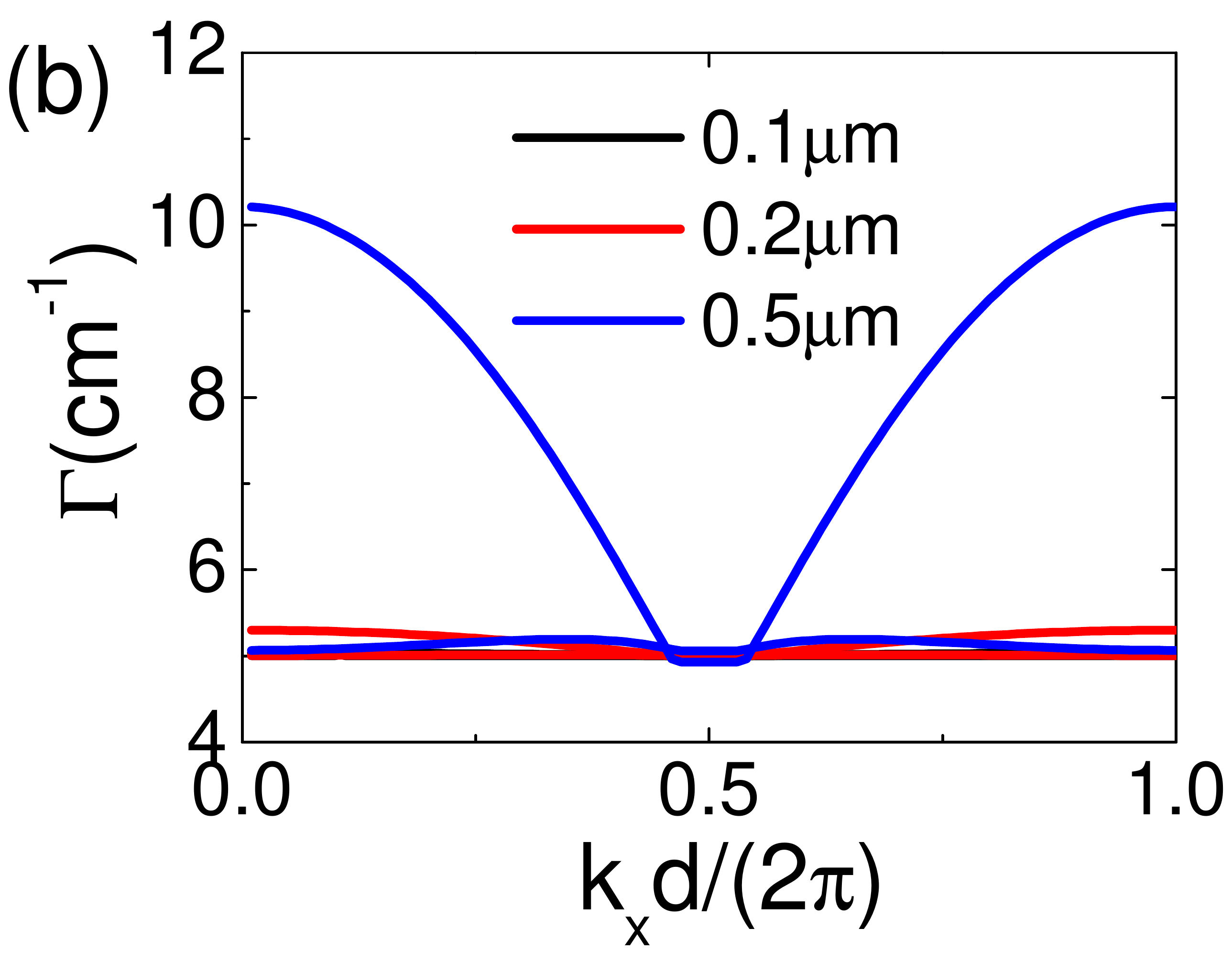}\label{sizeeffectlongimag}
	}
	\hspace{0.01in}
	\subfloat{
		\includegraphics[width=0.46\linewidth]{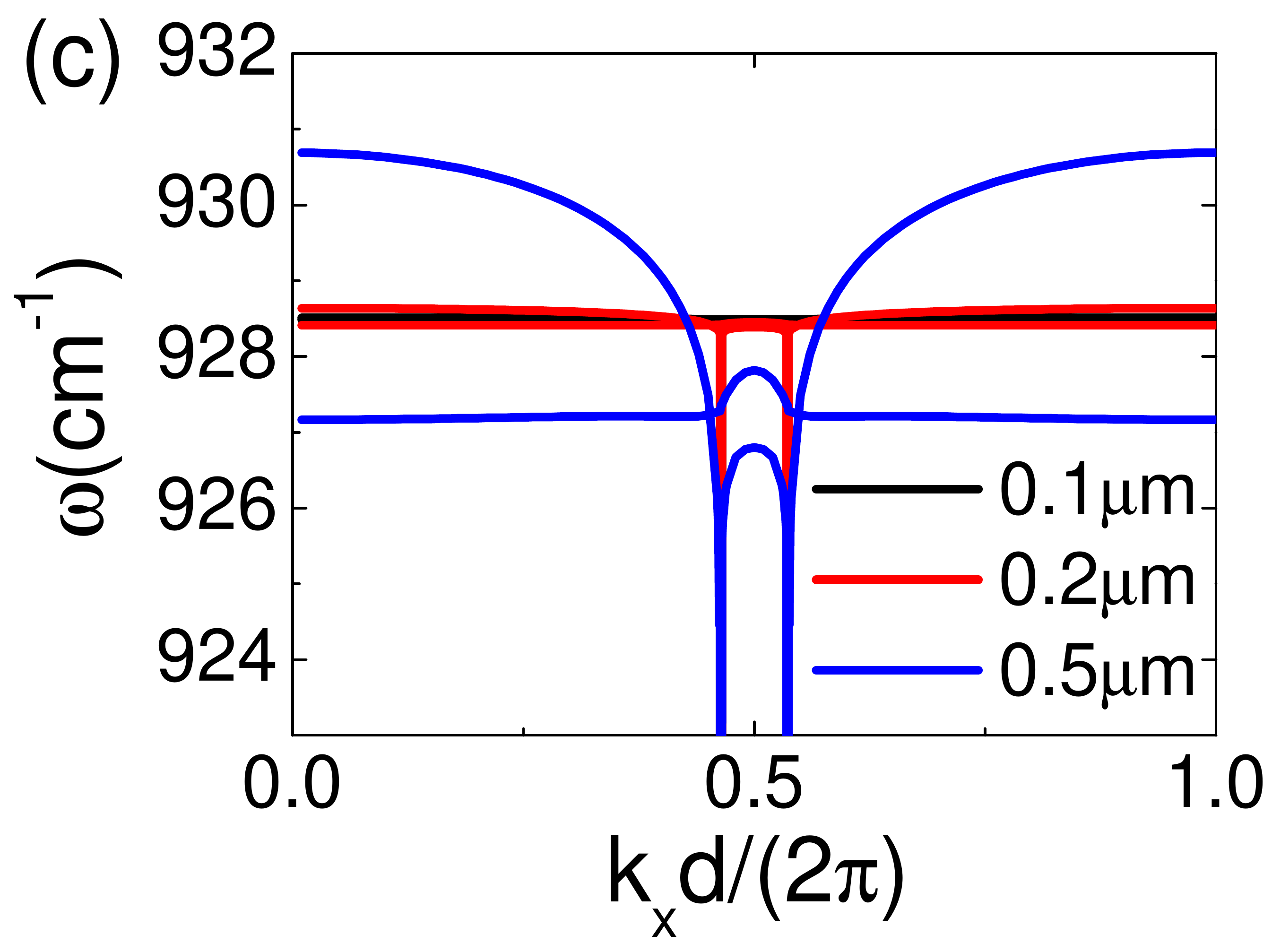}
		\label{sizeeffecttransreal}
	}
	\hspace{0.01in}
	\subfloat{
		\includegraphics[width=0.46\linewidth]{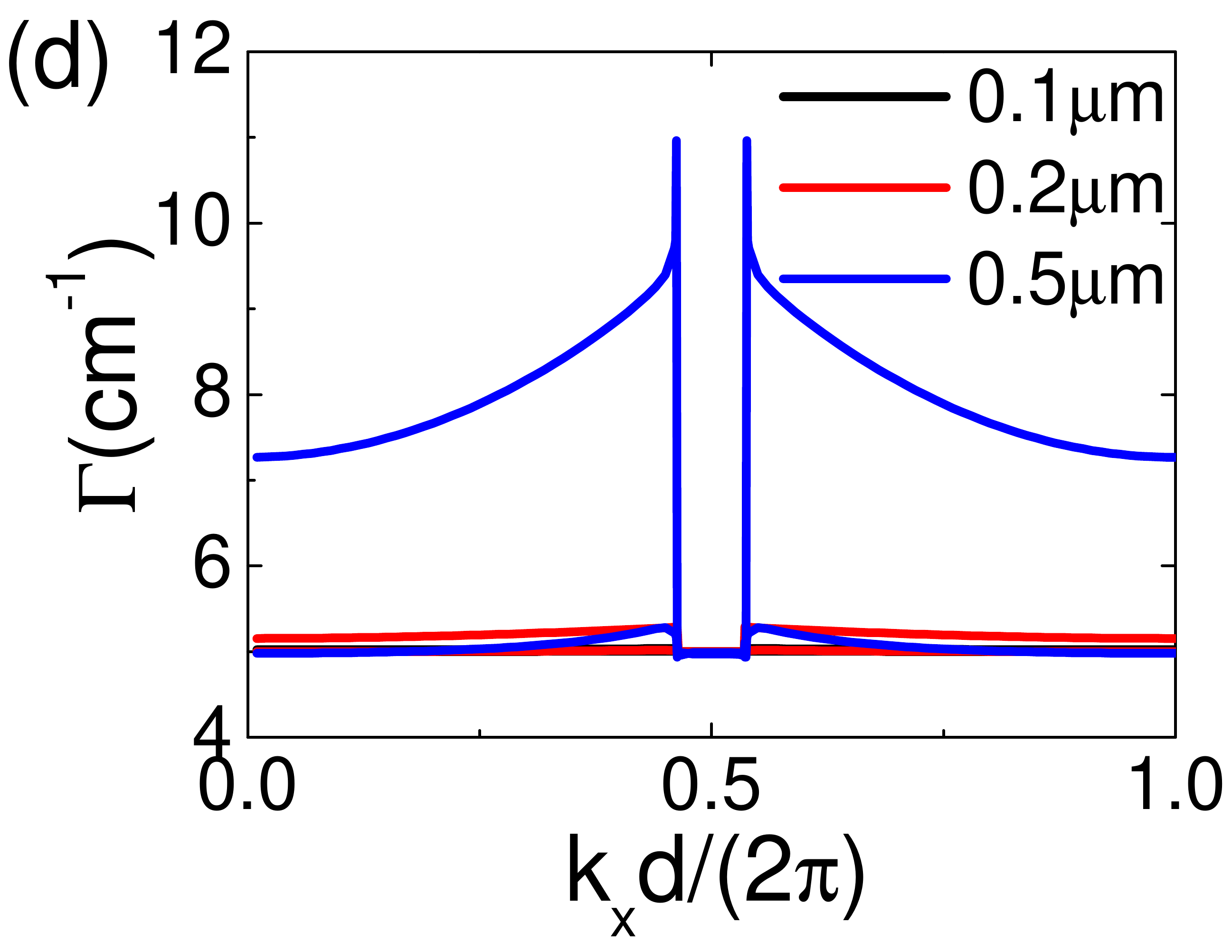}\label{sizeeffecttransimag}
	}
	\hspace{0.01in}
\subfloat{
	\includegraphics[width=0.46\linewidth]{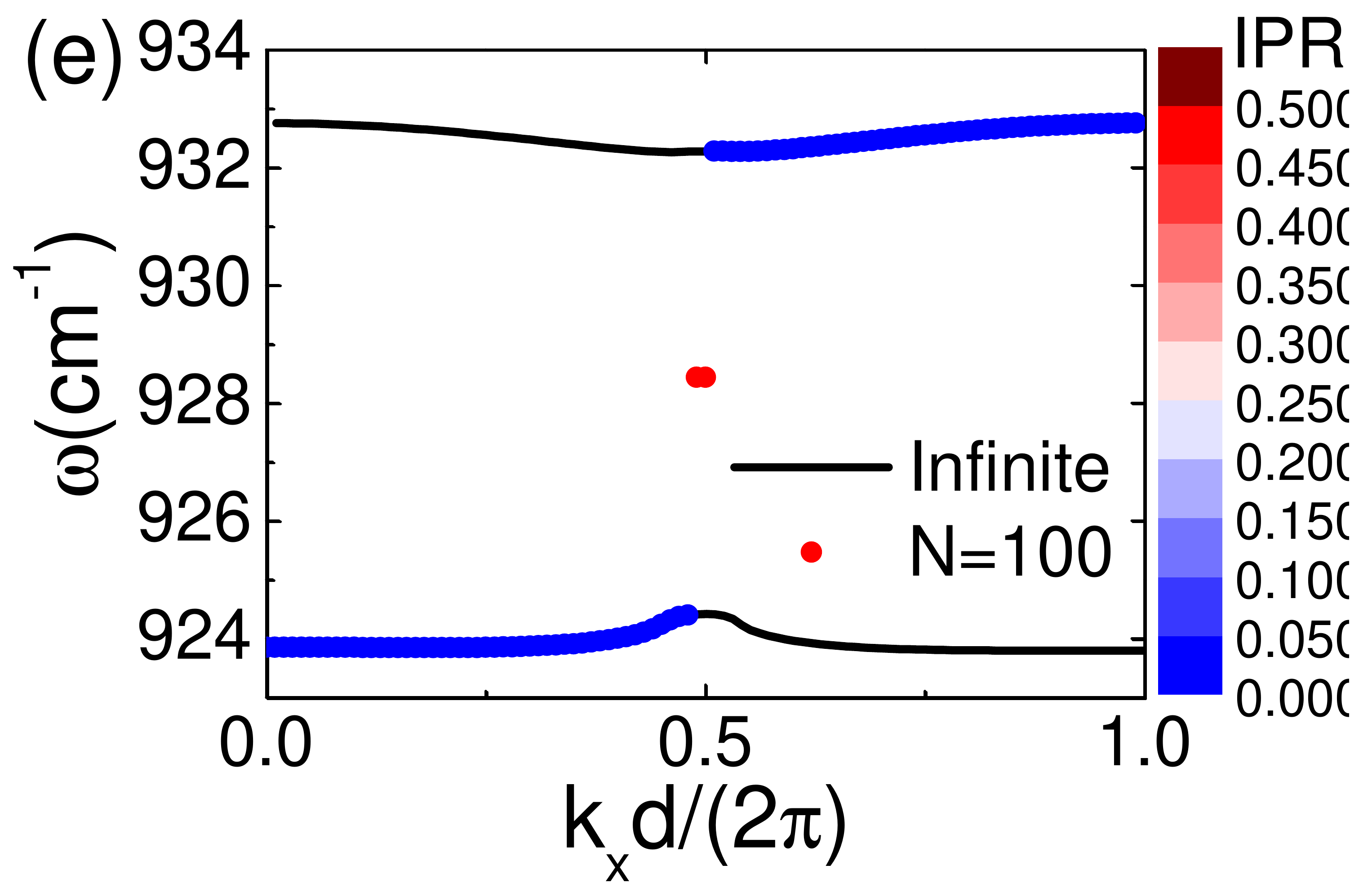}
	\label{comparisonreala05long}
}
\hspace{0.01in}
\subfloat{
	\includegraphics[width=0.46\linewidth]{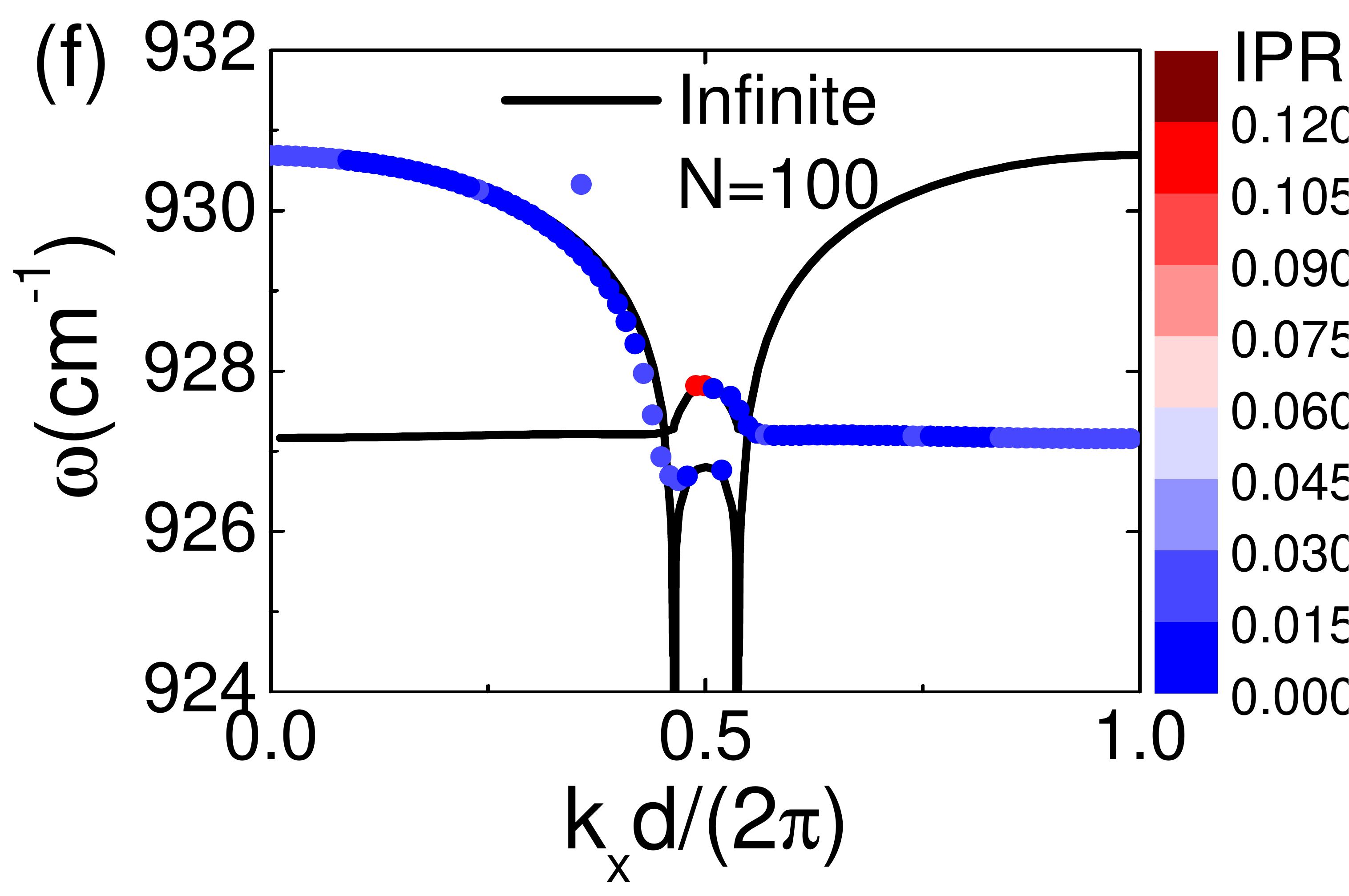}\label{comparisonreala05trans}
}
	\caption{(a-b) Real (a) and imaginary (b) parts of the longitudinal band structures of a dimerized chain with $d=5\mathrm{\mu m}$ and a dimerization parameter $\beta=0.7$ for different NP radii $a$ ($0.1, 0.2,0.5\mathrm{\mu m}$). (c-d) The same as (a-b) but for transverse modes.(e-f) The comparison of real eigenfrequencies between the longitudinal (e) and transverse (f) Bloch band structure (``Infinite") and corresponding eigenmode distributions of a finite chain with 100 NPs (``$N=100$"). System parameters are $d=5\mathrm{\mu m}$, $\beta=0.7$ and $a=0.5\mathrm{\mu m}$. }\label{sizeeffect}
\end{figure}

In this section, we briefly discuss the influence of NP size on the properties of the system. Here we focus on the parameter region where the electric dipole approximation using the radiative correction, as well as the coupled-dipole model, is valid (namely, $a\lesssim2\mu m$ and $\min{\{\beta d,(1-\beta)d\}}\geq3a$). The band structures for both longitudinal and transverse modes are shown in  Figs.\ref{sizeeffectlongreal}-\ref{sizeeffecttransimag}. Two features are recognized. The first is that the complex bandgap becomes wider when the size of the NP is increased, while the second is that the values of imaginary eigenfrequency (or decay rate) become substantially larger due to more significant radiative losses. Moreover, Figs.\ref{comparisonreala05long} and \ref{comparisonreala05trans} show the real parts of the longitudinal and transverse eigenmode distributions for a finite chain with $d=5\mathrm{\mu m}$, $\beta=0.7$ and $a=0.5\mathrm{\mu m}$, composed of 100 NPs. In a word, Fig.\ref{sizeeffect} tells us that despite the quantitative differences in the values of eigenfrequencies, no qualitative difference is observed. Moreover, for a fixed set of $d$, $\beta$ and number of NPs, we find the IPRs, dipole moment distributions and phase rigidities of the eigenmodes with the same mode number are exactly the same for different NP sizes. As a consequence, the topological properties are generally not affected by the size of the NP.

However, when the size of the NP continues to increase, the radiative correction becomes invalid. In the circumstance, Eq.(\ref{dispersionrelation}) will turn into a more complicated equation in which the values of $d$ and $a$ may cooperatively determine the band structure and topological properties because of possible internal resonances in the expression of polarizability (considering the structure of Eq.(\ref{alphamie})). This situation is much more complicated and needs further in-depth investigation.

\section{Localized bulk modes in longer chains}\label{longer_chain}
\begin{figure}[htbp]
	\centering
	\subfloat{
		\includegraphics[width=0.8\linewidth]{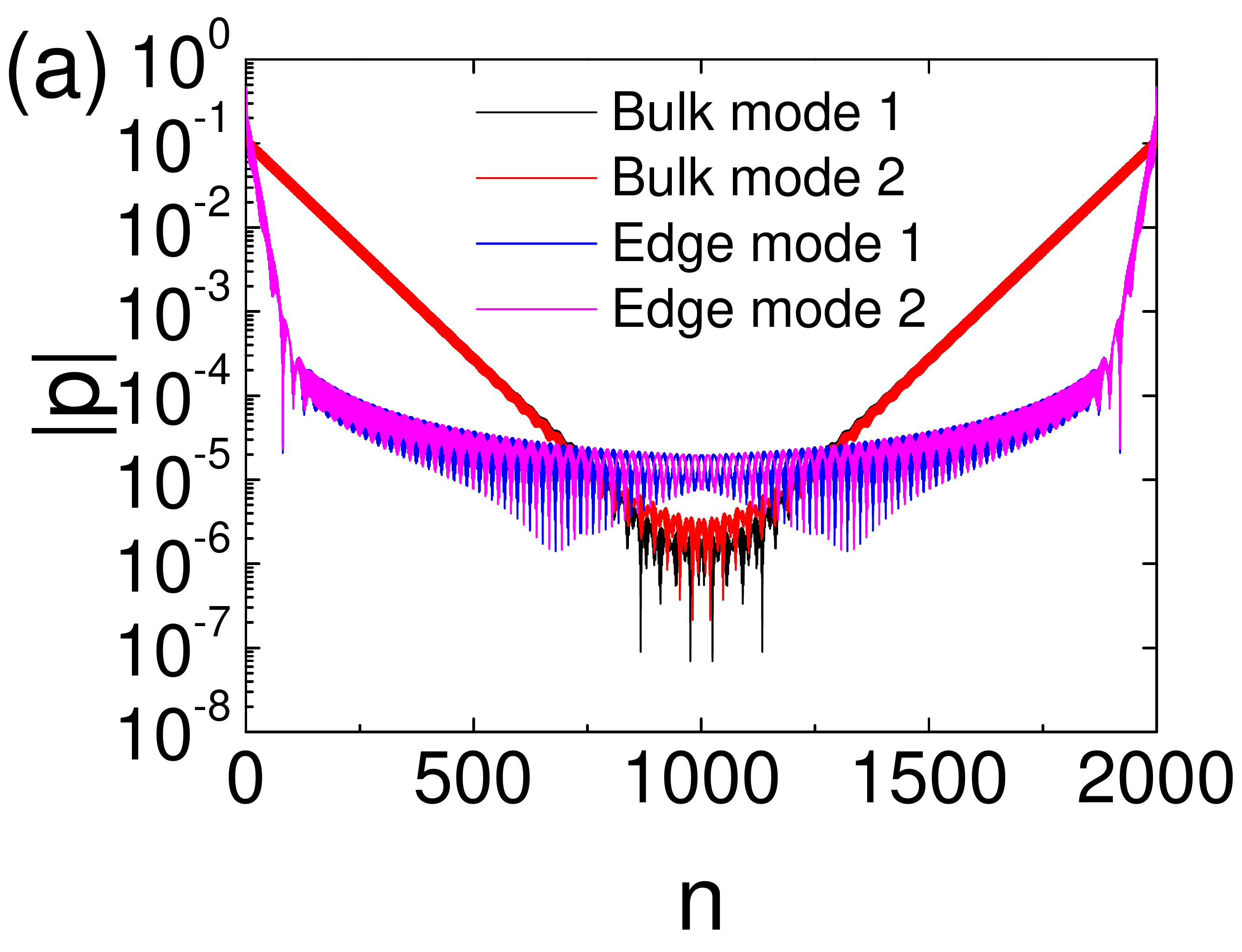}\label{NP2000}
	}
	\hspace{0.01in}
	\subfloat{
		\includegraphics[width=0.8\linewidth]{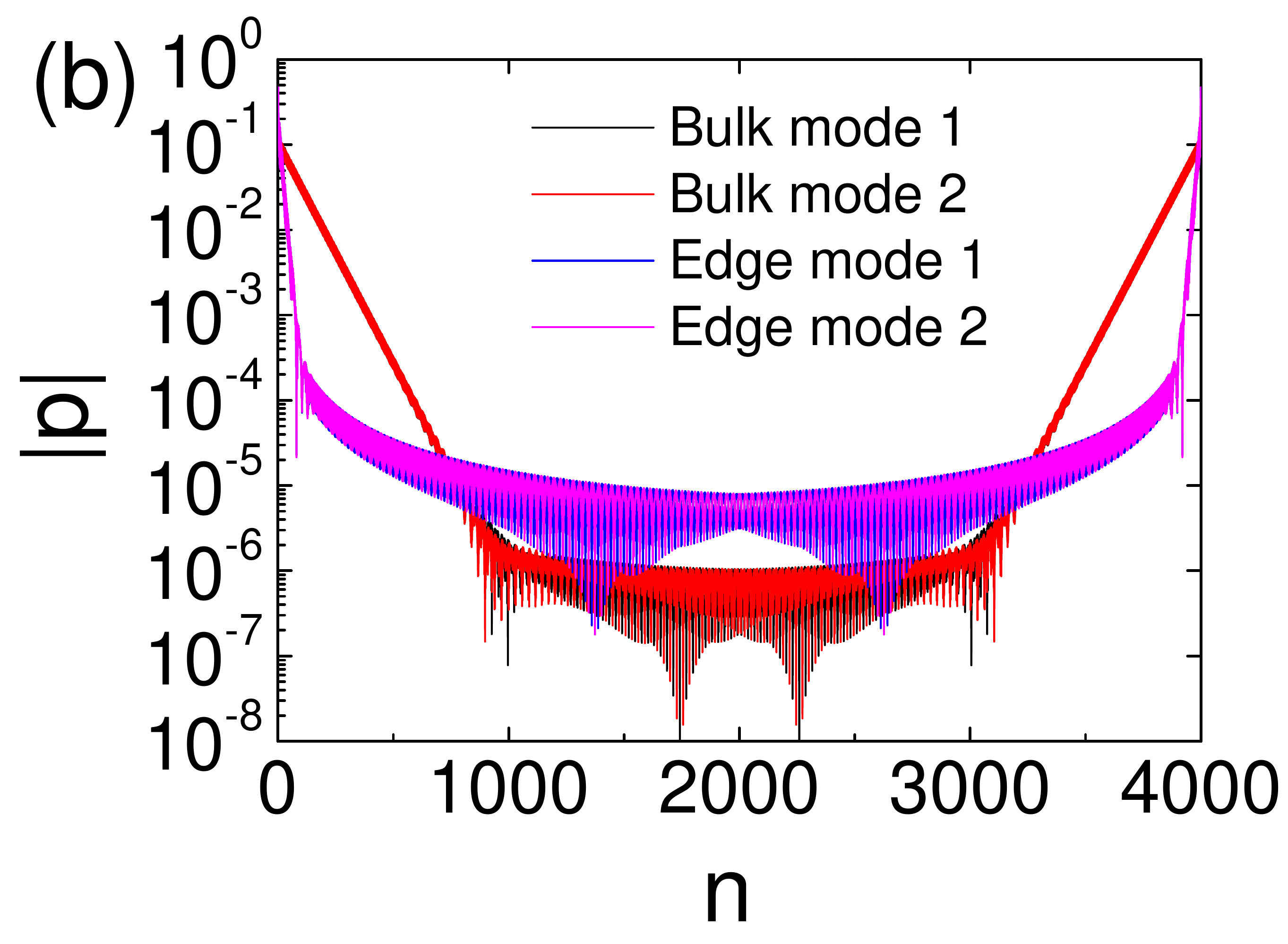}\label{NP4000}
	}
	\caption{The localized bulk modes and edge modes for very long chains with $\beta=0.7$, $d=5\mathrm{\mu m}$ and $a=0.1\mathrm{\mu m}$. (a) $N=2000$. (b) $N=4000$.}\label{longerchains}
\end{figure}

In the section, we investigate the non-Hermitian skin effect of transverse modes in longer chains, because the chain with 100 NPs in the main text is too short to display the long-range decaying behavior of localized bulk modes arising from strong long-range interactions. In Figs.\ref{NP2000} and \ref{NP4000} we show the dipole moment distributions of typical localized bulk modes for much longer chains containing 2000 and 4000 NPs respectively, in comparison with the localized topological edge modes. It is identified that the topologically protected edge modes decay exponentially from the edge with a very short localization length while decay algebraically in the long range. In the meanwhile, the exponential decaying behavior of localized bulk modes can persist over a very long range (nearly 1000NPs). For the chain composed of 2000 NPs these bulk modes seem to be entirely exponentially localized, and the algebraic decaying behavior in the long range \cite{vodolaPRL2014} only emerges in the 4000-NP chain. This observation further corroborates our modified method assuming exponential localization of certain bulk modes for finite chains not too long.

\bibliography{ssh_sphp}
\end{document}